\PassOptionsToPackage{table}{xcolor}
\documentclass{article}
\usepackage{iclr2027_conference,times}
\iclrfinalcopy
\usepackage{amsmath,amsfonts,bm}

\def\eqref#1{equation~\ref{#1}}

\def\1{\bm{1}}

\DeclareMathAlphabet{\mathsfit}{\encodingdefault}{\sfdefault}{m}{sl}
\SetMathAlphabet{\mathsfit}{bold}{\encodingdefault}{\sfdefault}{bx}{n}

\usepackage[compact]{titlesec}

\usepackage{graphicx}
\usepackage{amsmath}
\usepackage{amssymb}
\usepackage{amsfonts}
\usepackage{booktabs}
\usepackage{microtype}
\usepackage{hyperref}
\usepackage{xurl}
\usepackage{xcolor}
\usepackage{array}
\usepackage{tabularx}
\usepackage{makecell}
\usepackage{multirow}
\usepackage{longtable}
\usepackage{pdflscape}
\usepackage{subcaption}
\usepackage{tikz}
\usepackage{pgfplots}
\usepackage{pgfplotstable}
\usepackage{svg}
\usepackage{stfloats}
\usepackage{adjustbox}
\usepackage{colortbl}
\usepackage{enumitem}
\usepackage{xspace}
\usepackage{pifont}
\usepackage{algorithm}
\usepackage{algpseudocode}
\usepackage{listings}
\usepackage[most]{tcolorbox}
\definecolor{sqlbg}{RGB}{248,248,248}
\definecolor{sqlframe}{RGB}{220,220,220}
\definecolor{sqlkeyword}{RGB}{0,92,175}
\definecolor{sqlstring}{RGB}{120,70,0}
\definecolor{sqlcomment}{RGB}{100,100,100}

\lstdefinelanguage{SQL}{
  morekeywords={
    SELECT,FROM,WHERE,GROUP,BY,ORDER,HAVING,AS,AND,OR,NOT,
    BETWEEN,IN,IS,NULL,COUNT,SUM,AVG,MIN,MAX,CAST,REAL,
    INTEGER,TEXT,CASE,WHEN,THEN,ELSE,END,DISTINCT,LIMIT
  },
  sensitive=false,
  morecomment=[l]{--},
  morestring=[b]',
  morestring=[b]"
}

\lstdefinestyle{sqlstyle}{
  language=SQL,
  basicstyle=\ttfamily\scriptsize,
  keywordstyle=\color{sqlkeyword}\bfseries,
  stringstyle=\color{sqlstring},
  commentstyle=\color{sqlcomment}\itshape,
  showstringspaces=false,
  columns=fullflexible,
  keepspaces=true,
  breaklines=true,
  breakatwhitespace=true,
  tabsize=2,
  frame=none,
  numbers=none,
  xleftmargin=0pt,
  xrightmargin=0pt
}

\newtcblisting{sqlbox}{
  listing only,
  listing style=sqlstyle,
  colback=sqlbg,
  colframe=sqlframe,
  boxrule=0.4pt,
  arc=1.5mm,
  left=1mm,
  right=1mm,
  top=1mm,
  bottom=1mm,
  enhanced,
  breakable
}

\usetikzlibrary{arrows.meta,positioning,fit,calc,patterns,patterns.meta}
\usepgfplotslibrary{groupplots}
\pgfplotsset{compat=1.18}
\pgfplotsset{bar width/.style={/pgf/bar width={#1}}}

\newcommand{\vyas}[1]{}
\newcommand{\lesley}[1]{}
\newcommand{\shinan}[1]{}
\newcommand{\jialin}[1]{}
\newcommand{\steven}[1]{}

\definecolor{FirstPlace}{HTML}{1397B8}
\definecolor{SecondPlace}{HTML}{7B45E5}
\definecolor{ThirdPlace}{HTML}{F28E2B}
\definecolor{OverallTint}{HTML}{F8F1DA}
\definecolor{RuleGray}{HTML}{C8CDD3}
\definecolor{modelarf}{HTML}{EE7733}
\definecolor{modelbayesnet}{HTML}{666666}
\definecolor{modelctgan}{HTML}{CC3311}
\definecolor{modelforestdiffusion}{HTML}{66CCEE}
\definecolor{modelrealtabformer}{HTML}{6F4C9B}
\definecolor{modeltabbyflow}{HTML}{CC6677}
\definecolor{modeltabddpm}{HTML}{004488}
\definecolor{modeltabdiff}{HTML}{009988}
\definecolor{modeltabpfgen}{HTML}{DDAA33}
\definecolor{modeltabsyn}{HTML}{4477AA}
\definecolor{modeltvae}{HTML}{882255}
\definecolor{modelreal}{HTML}{000000}

\newcolumntype{L}[1]{>{\raggedright\arraybackslash}p{#1}}
\newcolumntype{C}[1]{>{\centering\arraybackslash}p{#1}}

\newcommand{\NumDatasets}{49}
\newcommand{\NumModels}{11}

\DeclareRobustCommand{\BenchmarkName}{\textsc{TabQueryBench}\xspace}

\title{\BenchmarkName: A Query-Centric Benchmark \\for Synthetic Tabular Data}

\makeatletter
\renewcommand{\@maketitle}{\vbox{\hsize\textwidth
  {\LARGE\scshape\@title\par}
  \vskip 0.15in
  \@author\par
  \vskip 0.3in minus 0.1in
}}
\makeatother

\author{%
\begin{minipage}{\textwidth}
\raggedright
\textbf{Jialin Zhang}\textsuperscript{1,2}\quad
\textbf{Fenghao Dong}\textsuperscript{3}\quad
\textbf{Yajie Zhou}\textsuperscript{4}\quad
\textbf{Vyas Sekar}\textsuperscript{3}\quad
\textbf{Shinan Liu}\textsuperscript{1}\\[6pt]
{\small
\textsuperscript{1}University of Hong Kong\quad
\textsuperscript{2}Tongji University\\
\textsuperscript{3}Carnegie Mellon University\quad
\textsuperscript{4}University of Maryland, College Park\endgraf}
\vspace{4pt}
{\footnotesize\ttfamily
fredzhang@tongji.edu.cn, fenghaod@andrew.cmu.edu, leszhou@umd.edu\\
vsekar@andrew.cmu.edu, shinan6@hku.hk\endgraf}
\end{minipage}%
}
\date{}
\hypersetup{
  pdftitle={TabQueryBench: A Query-Centric Benchmark for Synthetic Tabular Data},
  pdfauthor={Jialin Zhang, Fenghao Dong, Yajie Zhou, Vyas Sekar, Shinan Liu},
  pdfsubject={A query-centric benchmark for synthetic tabular data},
  hidelinks
}

\begin{document}

\maketitle

\begin{abstract}
Synthetic tabular data support use cases like data sharing, model development under access restrictions, and rapid prototyping of analytical workflows. Modern generative models are evaluated by their statistical similarity, correlation structure, privacy, and downstream machine-learning utility. However, such evaluations leave a gap: they rarely test the structure that matters for analytical queries. 
We present \BenchmarkName{}, a query-centric benchmark for evaluating synthetic tabular data with SQL-shaped analytical queries. \BenchmarkName{} distills 12 public query sources into 44 reusable query templates and grounds them to each dataset through a policy-guided template-to-SQL pipeline. Across 49 datasets and 11 generative models, it produces more than 100 executable queries per dataset while preserving cross-model comparability.
Our evaluation reveals a substantial gap between conventional and query-centric fidelity. Even the strongest model, RealTabFormer, reaches only $0.76\pm0.23$ query-centric fidelity compared with $1.00$ for real data. Failures are most pronounced for high-cardinality attributes and local or tail queries, while model quality also trades off against generation cost. These results show that synthetic data can appear statistically realistic yet still fail to preserve the analytical behavior of real data, motivating query-centric fidelity as a necessary complement to existing evaluation metrics.
We open source \BenchmarkName{} to facilitate research in this direction.\footnote{Code: \url{https://github.com/TabQueryBench/TabQueryBench}. Docker images and data: \url{https://huggingface.co/datasets/TabQueryBench2026/TabQueryBench/tree/main}.}
\end{abstract}

\section{Introduction}

Synthetic tabular data are now used for data sharing \citep{bates2019synthetic,pdpc2024synthetic,jiang2024netdiffusion}, model development under access restrictions \citep{bates2019synthetic,pdpc2024synthetic,chu2024feasibility}, and rapid algorithm or system prototyping \citep{patki2016sdv,walonoski2018synthea,kannan2025highfidelity,gupta2025generative}. As these use cases mature, benchmarking matters more, because what makes synthetic data useful in these settings is not distributional resemblance alone, but the preservation of the analytical properties that the database community has long treated as the core measure of data quality \citep{poess2000new,nambiar2006making,poess2002tpcdsnext,ding2021dsb,tpchspec,tpcdsspec,jiang2023generative,chu2026netssm}. 

Synthetic data consumers often care less about synthetic records as standalone samples. They care more about the functions that those records make possible. \textit{A useful synthetic dataset does not merely approximate marginal distributions. It should ideally expose a controlled approximation of hidden dataset attributes}, such as table structure, column relationships, valid joins, and executable queries \citep{walonoski2018synthea,kannan2025highfidelity}. This structural view is important for (1) collaboration, where a partner wants to preview the structure and quality of a sensitive dataset before accessing the real data \citep{bates2019synthetic,pdpc2024synthetic}; (2) software and data-system testing, where engineers need realistic schemas, constraints, edge cases, and relational consistency \citep{walonoski2018synthea}; and (3) query-centric analytics, where the target is not an individual row but the answer to aggregates, joins, SQL functions, and text-to-SQL workloads \citep{kannan2025highfidelity,caferoglu2025sing}.

\begin{figure*}[t]
    \centering
    \includegraphics[
        width=0.95\textwidth,
        height=0.72\textheight,
        keepaspectratio
    ]{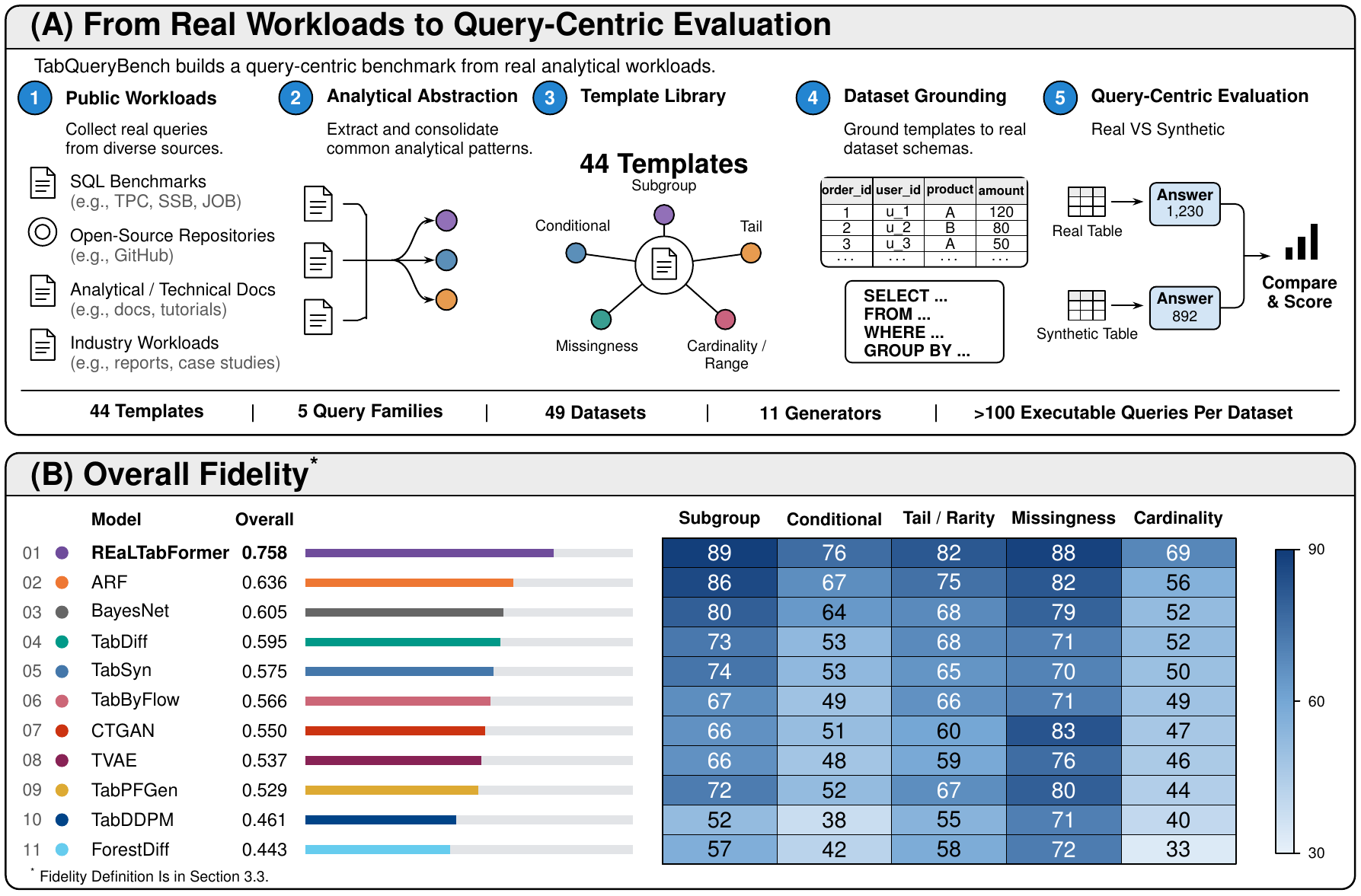}
    \captionsetup{skip=2pt}
    \caption{Overview of \BenchmarkName{}.}
    \label{fig:intro_project_overview}
\end{figure*}

\textit{Our key observation is that existing synthetic tabular benchmarks have limited coverage of the analytical structures that occur in real analytical queries.} Current benchmarks score column-level distance, predictive ML utility, or task-specific utility proxies~\citep{lautrup2024syntheval,snoke2018general,elemam2022utilitymetrics,hernandez2023tripartite,kaabachi2025scoping}, none of which directly assesses the analytical properties that diverse downstream workflows care about. We also find that picking generative models based on conventional distance-based metrics may even be misleading for practitioners~\citep{vanbreugel2023realerrors}.
A useful contrast comes from the database community, which has long treated occurring query sets as first-class evaluation objects: benchmarks such as TPC-H and TPC-DS encode representative decision-support SQL workloads designed to reflect recurring analytical patterns in industry practice~\citep{poess2000new,nambiar2006making,ding2021dsb,poess2002tpcdsnext}. Although designed to compare database engines, such recurring query patterns also provide natural \emph{structural assessors} for synthetic data. In contrast, existing synthetic-data benchmarks do not yet use them in this role, which shows a natural reality/generation target mismatch.

\textit{Our answer: \BenchmarkName{}} (Figure~\ref{fig:intro_project_overview}).
We argue for a shift from the conventional benchmarking used for tabular generative models to
\emph{query-centric} benchmarking, where {\em domain-relevant queries} serve as structural assessors of synthetic-data fidelity.
The key design challenge is to make these assessors both realistic and generic.
\BenchmarkName{} addresses this challenge with a two-stage construction pipeline. First, it taxonomizes recurring analytical logic from public sources of analytical queries into reusable five-family templates (i.e., subgroup, conditional, tail/rarity, cardinality/range, and missingness). Second, it grounds these templates to each benchmark dataset through a constrained realization policy that binds template roles to schema-appropriate columns and validates the resulting queries. This design preserves schema-level realism while enabling comparable evaluation across datasets, domains, and generative models. \BenchmarkName{} also provides fine-grained feedback on specific queries and query families rather than only an aggregate score.

In this work, we identify the following key findings: (1) Current tabular generative models can have good distance-based fidelity, but they still fall short on query-centric fidelity; (2) Tabular generative models fail to synthesize columns with very high-cardinality discrete support; (3) SOTA generative models preserve good global conditional query-centric fidelity, but fail more on local queries; (4) Tail fidelity deteriorates as queries move toward the extreme tail, even for SOTA generative models; (5) There is a clear cost-fidelity tradeoff in tabular data generation and BayesNet achieves the best balance.
We make three main contributions:
\begin{itemize}[leftmargin=*, topsep=2pt, itemsep=1pt, parsep=0pt, partopsep=0pt]
    \item \BenchmarkName{} is one of the first tabular data benchmarks that focus on workload-level fidelity evaluation.

    \item \BenchmarkName{} provides an automated and practical framework for operators to generate realistic queries without dataset-specific handcrafting.

    \item \BenchmarkName{} reveals actionable diagnosis to operators, not just one fidelity score.
    Across 49 datasets and 11 generators, it exposes failures in high-cardinality support, local conditional structure, and extreme tails, while also showing fidelity--cost tradeoffs.
\end{itemize}

\section{Related Work and Motivation}
\label{sec:current_benchmarks}

\subsection{Related Works}
\label{sec:related_works}

\textbf{Tabular generative models.}
\label{subsec:rw-tabular-generative-models}
Tabular generative models learn distributions over heterogeneous records.
Approaches range from Bayesian networks and tree-based synthesis to
adversarial, variational, diffusion, transformer-based, and flow-matching
models~\citep{ping2017datasynthesizer,watson2023arf,xu2019ctgan,kotelnikov2023tabddpm,solatorio2023realtabformer,guzmancordero2025tabbyflow}.
These approaches differ in how they represent dependencies; our focus is
whether their synthetic tables preserve the answers to analytical queries.
Appendix~\ref{app:related_tabular_models} reviews the model families.

\newcommand{\cmark}{\textcolor{green!45!black}{\ding{51}}}
\newcommand{\xmark}{\textcolor{red!75!black}{\ding{55}}}
\newcommand{\pmark}{\textcolor{orange!80!black}{\textbf{\raisebox{0.15ex}{\scriptsize$\triangle$}}}}
\newcommand{\uspec}{\ensuremath{+\mathit{u}}}
\newcommand{\unspec}{\ensuremath{\mathit{u}}}
\newcommand{\qfam}[1]{\raisebox{0.05ex}{\textcircled{\raisebox{-0.10ex}{\tiny #1}}}}

\newcolumntype{Y}{>{\centering\arraybackslash}X}
\newcolumntype{Z}{>{\columncolor{tqbblue}\centering\arraybackslash}X}
\newcolumntype{Q}[1]{>{\centering\arraybackslash}m{#1}}
\newcolumntype{B}[1]{>{\columncolor{tqbblue}\centering\arraybackslash}m{#1}}
\definecolor{tqbgray}{gray}{0.92}
\definecolor{tqbblue}{RGB}{226,238,250}

\begin{table*}[!t]
\centering
\scriptsize
\setlength{\tabcolsep}{0.25pt}
\renewcommand{\arraystretch}{1.08}

\begin{tabular}{@{}
>{\raggedright\arraybackslash}m{1.64in}
>{\raggedright\arraybackslash}m{0.76in}
>{\raggedright\arraybackslash}m{0.88in}
Q{0.28in}
Q{0.26in}
Q{0.20in}
Q{0.20in}
Q{0.20in}
Q{0.20in}
Q{0.20in}
B{0.25in}
B{0.25in}
@{}}

\toprule
\textbf{Benchmark} &
\textbf{Target setting} &
\textbf{Eval. focus} &
\textbf{Data} &
\textbf{Model} &
\textbf{\qfam{1}} &
\textbf{\qfam{2}} &
\textbf{\qfam{3}} &
\textbf{\qfam{4}} &
\textbf{\qfam{5}} &
\textbf{Dist.} &
\textbf{Cost} \\
\midrule

Synthcity~\citep{qian2023synthcity}
& Generic tabular
& Metric suite
& 18\uspec
& 5\uspec
& \xmark
& \xmark
& \xmark
& \pmark
& \xmark
& \cmark
& \xmark \\

SDGym~\citep{sdgym2026}
& Generic tabular
& Global fidelity
& 22\uspec
& \unspec
& \xmark
& \xmark
& \xmark
& \xmark
& \xmark
& \cmark
& \cmark \\

SyntheRela~\citep{hudovernik2024benchmarking}
& Multi-table
& Relational fidelity
& 39
& 6
& \xmark
& \xmark
& \xmark
& \xmark
& \xmark
& \pmark
& \pmark \\

SynthRO~\citep{santangelo2025synthro}
& Health domain
& Domain validation
& \unspec
& \unspec
& \xmark
& \xmark
& \xmark
& \xmark
& \xmark
& \cmark
& \xmark \\

SynthEval~\citep{lautrup2024syntheval}
& Generic tabular
& Metric suite
& \unspec
& \unspec
& \xmark
& \xmark
& \xmark
& \xmark
& \xmark
& \cmark
& \xmark \\

DP benchmark~\citep{tao2021dpbenchmark}
& Private tabular
& Privacy--utility
& 7
& 12
& \xmark
& \xmark
& \xmark
& \xmark
& \xmark
& \cmark
& \pmark \\

TabStruct~\citep{jiang2026tabstruct}
& Generic tabular
& Structural metrics
& 29
& 13
& \xmark
& \pmark
& \xmark
& \xmark
& \xmark
& \cmark
& \cmark \\

\midrule
\rowcolor{tqbgray}
\textbf{\BenchmarkName{} (ours)}
& Generic tabular
& Query-centric fidelity
& \NumDatasets{}
& \NumModels{}
& \cmark
& \cmark
& \cmark
& \cmark
& \cmark
& \cellcolor{tqbblue}\cmark
& \cellcolor{tqbblue}\cmark \\

\bottomrule
\end{tabular}

\vspace{-2pt}
\captionsetup{skip=2pt}
\caption{Comparison of related benchmarks for synthetic tabular data and adjacent tabular evaluation tasks.
\qfam{1} subgroup, \qfam{2} conditional, \qfam{3} tail,
\qfam{4} missingness, and \qfam{5} high-cardinality.
\cmark{} = direct, \pmark{} = partial, \xmark{} = absent,
\unspec{} = user-specified.}
\label{tab:benchmark_comparison}
\end{table*}

\textbf{Tabular generative model benchmarks.}
Existing frameworks assess distributional fidelity, privacy, and
predictive utility~\citep{qian2023synthcity,lautrup2024syntheval,hansen2023reimagining}.
TabStruct further evaluates structural fidelity through conditional
independence and \emph{global utility}, which aggregates predictive utility
across variables without requiring a ground-truth causal graph
\citep{jiang2026tabstruct}.
\BenchmarkName{} complements these perspectives by directly comparing
dataset-grounded SQL answers across five analytical query families.
Appendix~\ref{app:related_synthetic_benchmarks} details the distinctions
and compares the evaluation settings.

\textbf{Database query benchmarks.}
TPC-H, TPC-DS, ClickBench, and H2O db-benchmark use recurring analytical
queries to evaluate database or data-processing systems
\citep{tpchspec,tpcdsspec,clickbench2022,h2odbbenchmark}.
We use public SQL patterns as sources of reusable templates, adapting
them to assess answer preservation in synthetic tables.
Appendix~\ref{app:related_query_benchmarks} explains this connection.

\subsection{Motivation: Distance--Query Fidelity Mismatch}
\label{sec:mismatch-two-types}

Distance-based metrics summarize global resemblance between real and synthetic tables, but analytical queries often depend on query-relevant subsets, group identities, conditional rankings, or rare supports. 
This mismatch motivates evaluating synthetic tables with query-centric fidelity in addition to classical distance-based metrics~\citep{swanberg2025api}.
Figure~\ref{fig:distance-query-raw-scatter} compares each model's
Wasserstein distance fidelity with its official query score. Both axes are
therefore higher-is-better. Appendix~\ref{app:related_fidelity_gap} gives more details of this example.

\begin{figure*}[t]
  \centering
  \begin{tikzpicture}
\begin{groupplot}[
  group style={group size=4 by 1, horizontal sep=0.045\linewidth},
  width=0.27\textwidth,
  height=0.22\textwidth,
  xmin=0.3, xmax=1.0,
  ymin=0.3, ymax=1.0,
  xtick={0.3,0.5,0.7,0.9,1.0},
  ytick={0.3,0.5,0.7,0.9,1.0},
  grid=both,
  minor tick num=1,
  grid style={draw=black!8},
  major grid style={draw=black!14},
  axis line style={draw=black!55},
  tick style={draw=black!55},
  ticklabel style={font=\fontsize{8}{6}\selectfont},
  label style={font=\fontsize{8}{7}\selectfont},
  xlabel={$1-\mathrm{Wasserstein}$},
  ylabel={Query score},
  clip=false,
  legend style={at={(2.0,-0.52)}, anchor=north, draw=none, font=\fontsize{8}{6.5}\selectfont, legend columns=11, transpose legend=false, column sep=3pt, cells={anchor=west}},
  legend image post style={scale=0.5}
]
\nextgroupplot[title={ClickBench}, ylabel={Query score}]
\addplot[dashed, line width=0.5pt, draw=black!50, forget plot] coordinates {(0.3,0.3) (1.0,1.0)};
\addplot[only marks, mark=*, mark size=1.55pt, draw=modelarf, fill=modelarf] coordinates {(0.98755029,0.69613933)}; \addlegendentry{ARF}
\addplot[only marks, mark=*, mark size=1.55pt, draw=modelbayesnet, fill=modelbayesnet] coordinates {(0.75697153,0.46851861)}; \addlegendentry{Bayes}
\addplot[only marks, mark=*, mark size=1.55pt, draw=modelctgan, fill=modelctgan] coordinates {(0.99302047,0.61757501)}; \addlegendentry{CTGAN}
\addplot[only marks, mark=*, mark size=1.55pt, draw=modelrealtabformer, fill=modelrealtabformer] coordinates {(0.98654911,0.61264928)}; \addlegendentry{RTF}
\addplot[only marks, mark=*, mark size=1.55pt, draw=modeltabbyflow, fill=modeltabbyflow] coordinates {(0.99492538,0.69635765)}; \addlegendentry{T-Flow}
\addplot[only marks, mark=*, mark size=1.55pt, draw=modeltabddpm, fill=modeltabddpm] coordinates {(0.91010723,0.58481925)}; \addlegendentry{T-DDPM}
\addplot[only marks, mark=*, mark size=1.55pt, draw=modeltabdiff, fill=modeltabdiff] coordinates {(0.99861144,0.74814656)}; \addlegendentry{T-Diff}
\addplot[only marks, mark=*, mark size=1.55pt, draw=modeltabpfgen, fill=modeltabpfgen] coordinates {(0.99627788,0.65524627)}; \addlegendentry{TPF}
\addplot[only marks, mark=*, mark size=1.55pt, draw=modeltabsyn, fill=modeltabsyn] coordinates {(0.96945665,0.54593217)}; \addlegendentry{T-Syn}
\addplot[only marks, mark=*, mark size=1.55pt, draw=modeltvae, fill=modeltvae] coordinates {(0.98769997,0.65340160)}; \addlegendentry{TVAE}
\addlegendimage{only marks, mark=*, mark size=1.55pt, draw=modelforestdiffusion, fill=modelforestdiffusion} \addlegendentry{F-Diff}
\nextgroupplot[title={c8}, ylabel={}]
\addplot[dashed, line width=0.5pt, draw=black!50, forget plot] coordinates {(0.3,0.3) (1.0,1.0)};
\addplot[only marks, mark=*, mark size=1.55pt, draw=modelarf, fill=modelarf] coordinates {(0.99128800,0.83809200)};
\addplot[only marks, mark=*, mark size=1.55pt, draw=modelbayesnet, fill=modelbayesnet] coordinates {(0.99218800,0.91587300)};
\addplot[only marks, mark=*, mark size=1.55pt, draw=modelctgan, fill=modelctgan] coordinates {(0.93494100,0.77797400)};
\addplot[only marks, mark=*, mark size=1.55pt, draw=modelrealtabformer, fill=modelrealtabformer] coordinates {(0.99656500,0.99602400)};
\addplot[only marks, mark=*, mark size=1.55pt, draw=modeltabbyflow, fill=modeltabbyflow] coordinates {(0.96915700,0.81771300)};
\addplot[only marks, mark=*, mark size=1.55pt, draw=modeltabddpm, fill=modeltabddpm] coordinates {(0.65772200,0.41061800)};
\addplot[only marks, mark=*, mark size=1.55pt, draw=modeltabdiff, fill=modeltabdiff] coordinates {(0.95087200,0.80284000)};
\addplot[only marks, mark=*, mark size=1.55pt, draw=modeltabpfgen, fill=modeltabpfgen] coordinates {(0.99448600,0.99170600)};
\addplot[only marks, mark=*, mark size=1.55pt, draw=modeltabsyn, fill=modeltabsyn] coordinates {(0.99158100,0.95926000)};
\addplot[only marks, mark=*, mark size=1.55pt, draw=modeltvae, fill=modeltvae] coordinates {(0.92668400,0.85869500)};
\addplot[only marks, mark=*, mark size=1.55pt, draw=modelforestdiffusion, fill=modelforestdiffusion] coordinates {(0.98574400,0.95600100)};
\nextgroupplot[title={m4}, ylabel={}]
\addplot[dashed, line width=0.5pt, draw=black!50, forget plot] coordinates {(0.3,0.3) (1.0,1.0)};
\addplot[only marks, mark=*, mark size=1.55pt, draw=modelarf, fill=modelarf] coordinates {(0.99053800,0.75971900)};
\addplot[only marks, mark=*, mark size=1.55pt, draw=modelbayesnet, fill=modelbayesnet] coordinates {(0.99242100,0.67548300)};
\addplot[only marks, mark=*, mark size=1.55pt, draw=modelctgan, fill=modelctgan] coordinates {(0.94923400,0.68094800)};
\addplot[only marks, mark=*, mark size=1.55pt, draw=modelrealtabformer, fill=modelrealtabformer] coordinates {(0.99226300,0.88715900)};
\addplot[only marks, mark=*, mark size=1.55pt, draw=modeltabbyflow, fill=modeltabbyflow] coordinates {(0.95993000,0.58652700)};
\addplot[only marks, mark=*, mark size=1.55pt, draw=modeltabddpm, fill=modeltabddpm] coordinates {(0.50772900,0.35212200)};
\addplot[only marks, mark=*, mark size=1.55pt, draw=modeltabdiff, fill=modeltabdiff] coordinates {(0.97813000,0.71552100)};
\addplot[only marks, mark=*, mark size=1.55pt, draw=modeltabpfgen, fill=modeltabpfgen] coordinates {(0.98716200,0.66708300)};
\addplot[only marks, mark=*, mark size=1.55pt, draw=modeltabsyn, fill=modeltabsyn] coordinates {(0.99081900,0.73630100)};
\addplot[only marks, mark=*, mark size=1.55pt, draw=modeltvae, fill=modeltvae] coordinates {(0.95653500,0.64642500)};
\addplot[only marks, mark=*, mark size=1.55pt, draw=modelforestdiffusion, fill=modelforestdiffusion] coordinates {(0.93786500,0.58389300)};
\nextgroupplot[title={n2}, ylabel={}]
\addplot[dashed, line width=0.5pt, draw=black!50, forget plot] coordinates {(0.3,0.3) (1.0,1.0)};
\addplot[only marks, mark=*, mark size=1.55pt, draw=modelarf, fill=modelarf] coordinates {(0.97044025,0.48433131)};
\addplot[only marks, mark=*, mark size=1.55pt, draw=modelbayesnet, fill=modelbayesnet] coordinates {(0.99024100,0.41709261)};
\addplot[only marks, mark=*, mark size=1.55pt, draw=modelctgan, fill=modelctgan] coordinates {(0.92734810,0.41412269)};
\addplot[only marks, mark=*, mark size=1.55pt, draw=modelrealtabformer, fill=modelrealtabformer] coordinates {(0.98965835,0.80484525)};
\addplot[only marks, mark=*, mark size=1.55pt, draw=modeltabbyflow, fill=modeltabbyflow] coordinates {(0.77825744,0.42085385)};
\addplot[only marks, mark=*, mark size=1.55pt, draw=modeltabddpm, fill=modeltabddpm] coordinates {(0.82530379,0.57292657)};
\addplot[only marks, mark=*, mark size=1.55pt, draw=modeltabpfgen, fill=modeltabpfgen] coordinates {(0.98192046,0.45441514)};
\addplot[only marks, mark=*, mark size=1.55pt, draw=modeltabsyn, fill=modeltabsyn] coordinates {(0.98630928,0.51271054)};
\addplot[only marks, mark=*, mark size=1.55pt, draw=modeltvae, fill=modeltvae] coordinates {(0.95240142,0.40685118)};
\addplot[only marks, mark=*, mark size=1.55pt, draw=modelforestdiffusion, fill=modelforestdiffusion] coordinates {(0.82373332,0.35950926)};
\end{groupplot}
\end{tikzpicture}%
  \captionsetup{skip=2pt}
  \caption{Distance-based and query scores do not align across generative models. The four panels show four datasets, respectively. Colors identify the models in the shared legend.}
  \label{fig:distance-query-raw-scatter}
\end{figure*}

\section{Benchmark Scope and Design}
\label{sec:design}
\subsection{Scope}
\label{sec:scope}

\BenchmarkName{} evaluates whether synthetic tables preserve the answers
to reusable analytical queries. The current benchmark focuses on
single-table analysis, drawing on query patterns from public OLAP and
decision-support collections
\citep{poess2000new,nambiar2006making,poess2007workload}.
These patterns capture analytical structure that can be assessed across
datasets through schema-aware grounding. Queries tied to specific row
identifiers or private business rules, temporal forecasting beyond simple
time-bucket aggregates, and relational joins are outside the scope of
the current evaluation.

We consider \emph{mimicry-style} synthetic-data generation: each
generative model learns from a real training split and produces a
synthetic table under the same schema, with the aim of preserving its
analytical structure. We do not evaluate differentially private
synthesis, generation under task-specific conditions, or text-to-table
generation. The evaluation reports query-centric fidelity by analytical
family, allowing different aspects of structure preservation to be
examined separately.

\subsection{Dataset and Generative Model Suites}
\label{sec:suites}

\textbf{Dataset suite.}
\label{sec:datasets}
We curate 49 single-table datasets from UCI, Kaggle, OpenML, and
HuggingFace~\citep{uci,kaggle,openml,huggingface}, comprising
19 categorical-dominant, 19 numerical-dominant, and 11 mixed-type datasets. These regimes probe discrete support and subgroup structure,
numerical ranges and tail behavior, and their interactions, respectively.
The suite spans multiple application domains and includes high-cardinality and missingness-rich tables to test these structural challenges.
Details of these datasets are provided in
Appendix~\ref{app:dataset_catalog},
Table~\ref{tab:appendix_dataset_catalog_all}.

\textbf{Generative model suite.}
\label{sec:models}
We evaluate 11 tabular generative models spanning Bayesian networks,
tree-based methods, adversarial and variational models, diffusion,
transformers, and flow matching
\citep{ping2017datasynthesizer,watson2023arf,xu2019ctgan,kotelnikov2023tabddpm,zhang2024tabsyn,shi2025tabdiff,jolicoeurmartineau2023forestdiffusion,solatorio2023realtabformer,ma2024tabpfgen,guzmancordero2025tabbyflow}.
This selection enables comparisons across different modeling approaches.

For each dataset, all generative models use the same training split,
schema, and synthetic sample-size requirement. The full model roster
and supporting references are provided in Appendix
Table~\ref{tab:model_taxonomy}. Detailed training, runtime, and
hyperparameter settings are provided in
Appendix~\ref{app:evaluation_settings_details}.

\subsection{\BenchmarkName{} Design}
\label{sec:query_protocol}

\BenchmarkName{} constructs analytical queries by extracting and
consolidating reusable templates from public SQL sources, grounding them
to individual datasets, and scoring query answers.

\textbf{Stage 1: Template construction.}
\label{sec:query_goal}
We extract recurring analytical query patterns from public SQL sources,
including TPC-H, TPC-DS, ClickBench, and H2O db-benchmark
\citep{tpchspec,tpcdsspec,clickbench2022,h2odbbenchmark},
with additional patterns from curated SQL repositories and database
engine documentation. We consolidate these patterns into 44 reusable
templates and deduplicate overlapping templates across sources.
Appendix~\ref{app:query_template_construction} provides the detailed
construction procedure and source-to-template attribution.

\begin{figure*}[t]
\centering
\includegraphics[width=\textwidth]{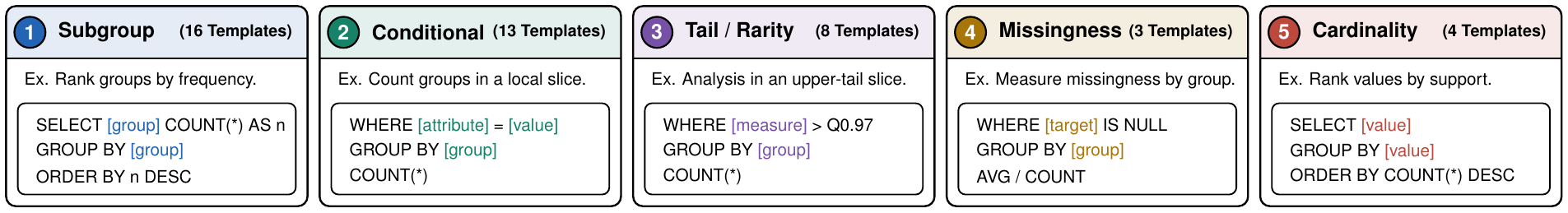}
  \vspace{-12pt}
  \captionsetup{skip=2pt}
\caption{The five query families in \BenchmarkName{} and their representative examples.
}
\label{fig:query-family-examples}
\end{figure*}

\textbf{Query families.}
\label{sec:query_families_templates}
The resulting library contains 44 templates organized into five families,
each probing a distinct aspect of analytical structure. 
The five families are derived from recurring single-table analytical patterns in our source registry and are kept separate because they probe distinct properties that synthetic data must preserve.
\textbf{Subgroup} (16 templates) measures grouped summary statistics and
subgroup sizes;
\textbf{Conditional} (13 templates) measures conditional summaries at full grouped
scope and within filtered slices;
\textbf{Tail / rarity} (8 templates) measures rare-member coverage and tail
magnitudes or sizes;
\textbf{Missingness} (3 templates) measures marginal and bucket-conditioned
missingness patterns; and
\textbf{Cardinality / range} (4 templates) measures discrete support and rank
behavior, together with numerical ranges.
These families target structural properties whose preservation may not
be captured by aggregate statistical distances.
Figure~\ref{fig:query-family-examples} summarizes the five-family taxonomy
and its representative analytical questions, while
Appendix~\ref{app:template_taxonomy_catalog} lists the complete template
inventory, and Table~\ref{tab:five-family-grounded-sql} gives a grounded
question and SQL example for each family.

\textbf{Stage 2: Dataset grounding and SQL realization.}
\label{sec:query_grounding}
\label{sec:query_coverage}
For each dataset, the agent uses its schema and data profile to select
applicable templates and eligible columns under fixed template policies,
skipping templates without valid bindings. The agent performs constrained
SQL realization while preserving the analytical structure specified by each template. 
Accepted SQL queries are accompanied by natural-language questions for readability. 
Across the 49 benchmark datasets, each dataset activates at least 10 templates and yields more than 100 executable SQL queries. Appendix~\ref{app:query_grounding} describes the grounding and
SQL realization procedure.

\textbf{Query scoring.}
We execute the grounded SQL queries on real and synthetic data and compare
their answers using scoring rules tailored to the different SQL result
output types. 
The detailed scoring standards are provided in
Appendix~\ref{app:query_scoring}.

\section{Evaluation}
\subsection{Setup}
\label{sec:evaluation_settings}

We evaluate \BenchmarkName{} on \NumDatasets{} datasets and \NumModels{} synthetic tabular generative models. For each dataset, we split the real table into a training split and a held-out test split (4:1). Each generative model is trained only on the training split and is then asked to synthesize a table under the same schema. The synthetic table is generated with the same number of rows as the training split.

We then execute the same grounded SQL workload on both the real training table and the synthetic table, and compute query fidelity by comparing their query answers. The held-out test split is retained for future downstream evaluation and is not used in the current scoring pipeline. This setup ensures that all generators are evaluated under the same input schema, output contract, and workload.

The agent selects applicable templates and dataset columns. The profiler,
binder, and validator are deterministic. Only the SQL realizer
calls an LLM, and it does so only inside the constrained
template-to-SQL realization step. We package each tabular generative model in
its own Docker image and use the recommended configuration from the original
implementation when available, otherwise tuning within a bounded search range
chosen to balance quality and runtime. Full details on SQL grounding,
validation, infrastructure, API cost, and hyperparameter ranges are provided
in Appendix~\ref{app:evaluation_settings_details}.

\newcommand{\findingbox}[2]{%
\begin{tcolorbox}[
  enhanced,
  colback=blue!3,
  colframe=black,
  boxrule=0.9pt,
  arc=2pt,
  left=4pt,
  right=4pt,
  top=6pt,
  bottom=4pt,
  boxsep=2pt,
  before skip=8pt,
  after skip=8pt,
  title={#1},
  coltitle=white,
  fonttitle=\bfseries,
  attach boxed title to top left={xshift=10pt,yshift*=-\tcboxedtitleheight/2},
  boxed title style={
    colback=black,
    colframe=black,
    boxrule=0pt,
    arc=2pt,
    left=5pt,
    right=5pt,
    top=2pt,
    bottom=2pt
  }
]
#2
\end{tcolorbox}%
}

\subsection{How Well Do State-of-the-Art Tabular Generative Models Perform under Query-Centric Evaluation?}

\findingbox{}{\textbf{Finding 1: }Current state-of-the-art tabular generative models often appear faithful under distance-based metrics, but still fall well short on query-centric fidelity (Figure~\ref{fig:distance_query_overall_scatter}).}

To provide an overview of \BenchmarkName{} results, we evaluate whether current tabular generative models preserve the analytical behavior exposed by dataset-grounded analytical queries. We are also curious about whether this query-centric approach provides a different view compared to conventional distance-based fidelity.

\begin{figure*}[t]
\centering
\begin{minipage}[t]{0.485\textwidth}
  \vspace{0pt}
  \centering
  \begin{tikzpicture}
\definecolor{modelarf}{HTML}{EE7733}
\definecolor{modelbayesnet}{HTML}{666666}
\definecolor{modelctgan}{HTML}{CC3311}
\definecolor{modelforestdiffusion}{HTML}{66CCEE}
\definecolor{modelrealtabformer}{HTML}{6F4C9B}
\definecolor{modeltabbyflow}{HTML}{CC6677}
\definecolor{modeltabddpm}{HTML}{004488}
\definecolor{modeltabdiff}{HTML}{009988}
\definecolor{modeltabpfgen}{HTML}{DDAA33}
\definecolor{modeltabsyn}{HTML}{4477AA}
\definecolor{modeltvae}{HTML}{882255}
\begin{axis}[
width=0.79\linewidth,
height=0.57\linewidth,
scale only axis,
xmin=0.4, xmax=1.0, ymin=0.4, ymax=1.0,
xtick={0.4,0.5,0.6,0.7,0.8,0.9,1.0},
ytick={0.4,0.5,0.6,0.7,0.8,0.9,1.0},
grid=both, grid style={draw=black!9}, major grid style={draw=black!12},
axis line style={draw=black!55}, tick style={draw=black!55},
xlabel={49 Datasets' Distance-based Fidelity},
ylabel={49 Datasets' Query Fidelity},
ticklabel style={font=\scriptsize},
label style={font=\footnotesize},
clip=false,
trim axis left,
trim axis right
]
\addplot[domain=0.4:1.0,samples=2,draw=black!55,dashed,line width=0.75pt] {x};
\addplot[only marks,mark=*,mark size=3.0pt,draw=modelrealtabformer,fill=modelrealtabformer] coordinates {(0.908,0.766)}
node[anchor=east, xshift=-3pt, font=\scriptsize, text=modelrealtabformer, inner sep=0.9pt] {RTF};
\addplot[only marks,mark=*,mark size=2.7pt,draw=modeltabdiff,fill=modeltabdiff] coordinates {(0.864,0.678)}
node[anchor=east, xshift=-3pt, font=\scriptsize, text=modeltabdiff, inner sep=0.9pt] {T-Diff};
\addplot[only marks,mark=*,mark size=2.7pt,draw=modelarf,fill=modelarf] coordinates {(0.911,0.653)}
node[anchor=west, xshift=3pt, font=\scriptsize, text=modelarf, inner sep=0.9pt] {ARF};
\addplot[only marks,mark=*,mark size=2.7pt,draw=modeltabbyflow,fill=modeltabbyflow] coordinates {(0.793,0.646)}
node[anchor=west, xshift=3pt, font=\scriptsize, text=modeltabbyflow, inner sep=0.9pt] {T-Flow};
\addplot[only marks,mark=*,mark size=2.7pt,draw=modeltabsyn,fill=modeltabsyn] coordinates {(0.787,0.630)}
node[anchor=east, xshift=-3pt, yshift=-2pt, font=\scriptsize, text=modeltabsyn, inner sep=0.9pt] {T-Syn};
\addplot[only marks,mark=*,mark size=2.7pt,draw=modelbayesnet,fill=modelbayesnet] coordinates {(0.871,0.618)}
node[anchor=west, xshift=3pt, font=\scriptsize, text=modelbayesnet, inner sep=0.9pt] {Bayes};
\addplot[only marks,mark=*,mark size=2.7pt,draw=modelforestdiffusion,fill=modelforestdiffusion] coordinates {(0.611,0.588)}
node[anchor=east, xshift=23pt, font=\scriptsize, text=modelforestdiffusion, inner sep=0.9pt] {F-Diff};
\addplot[only marks,mark=*,mark size=2.7pt,draw=modeltabpfgen,fill=modeltabpfgen] coordinates {(0.859,0.551)}
node[anchor=west, xshift=3pt, font=\scriptsize, text=modeltabpfgen, inner sep=0.9pt] {TPF};
\addplot[only marks,mark=*,mark size=2.7pt,draw=modeltvae,fill=modeltvae] coordinates {(0.804,0.549)}
node[anchor=south east, xshift=-2pt, yshift=3pt, font=\scriptsize, text=modeltvae, inner sep=0.9pt] {TVAE};
\addplot[only marks,mark=*,mark size=2.7pt,draw=modelctgan,fill=modelctgan] coordinates {(0.812,0.542)}
node[anchor=north west, xshift=2pt, yshift=-3pt, font=\scriptsize, text=modelctgan, inner sep=0.9pt] {CTGAN};
\addplot[only marks,mark=*,mark size=2.7pt,draw=modeltabddpm,fill=modeltabddpm] coordinates {(0.560,0.487)}
node[anchor=east, xshift=30pt, font=\scriptsize, text=modeltabddpm, inner sep=0.9pt] {T-DDPM};
\end{axis}
\end{tikzpicture}
  \captionsetup{skip=2pt}
  \caption{Distance-based and query overall scores do not align across generative models.
  Detailed values appear in Appendix Table~\ref{tab:benchmark_overall_real_main}.
  }
  \label{fig:distance_query_overall_scatter}
\end{minipage}\hfill
\begin{minipage}[t]{0.485\textwidth}
  \vspace{0pt}
  \centering
  \definecolor{modelcolorarf}{HTML}{EE7733}
\definecolor{modelcolorbayesnet}{HTML}{666666}
\definecolor{modelcolorctgan}{HTML}{CC3311}
\definecolor{modelcolorforestdiffusion}{HTML}{66CCEE}
\definecolor{modelcolorrealtabformer}{HTML}{6F4C9B}
\definecolor{modelcolortabbyflow}{HTML}{CC6677}
\definecolor{modelcolortabdiff}{HTML}{009988}
\definecolor{modelcolortabpfgen}{HTML}{DDAA33}
\definecolor{modelcolortabsyn}{HTML}{4477AA}
\definecolor{modelcolortvae}{HTML}{882255}
\begin{tikzpicture}
\begin{loglogaxis}[
width=0.79\linewidth,
height=0.57\linewidth,
scale only axis,
xlabel={Real distinct count},
ylabel={Synthetic distinct count},
grid=none,
legend style={draw=none, fill=none, font=\scriptsize, at={(0.03,0.97)}, anchor=north west, row sep=0pt, /tikz/every even column/.append style={column sep=0.18cm}},
legend columns=2,
tick label style={font=\scriptsize},
label style={font=\footnotesize},
xmin=35, xmax=110000,
ymin=35, ymax=110000,
clip=false,
trim axis left,
trim axis right,
]
\addplot[dashed, line width=0.8pt, color=blue!70!black, forget plot] coordinates {(35,35) (110000,110000)};
\addplot[only marks, mark=*, mark size=0.95pt, draw=modelcolorarf, fill=modelcolorarf, fill opacity=0.75, draw opacity=0.85] coordinates {(1219,1217) (2215,2212) (11927,11735) (1220,1220) (1518,1516) (2218,2216) (484,424) (1634,1366) (97504,63087) (33068,24476) (1179,1114) (96777,62598) (15786,13729) (2541,1649) (2426,1533) (2181,2086) (1480,952)};
\addlegendentry{ARF}
\addplot[only marks, mark=*, mark size=0.95pt, draw=modelcolorbayesnet, fill=modelcolorbayesnet, fill opacity=0.75, draw opacity=0.85] coordinates {(1219,64) (2215,64) (11927,64) (1220,1220) (1518,1518) (2218,2216) (484,421) (1634,1375) (97504,224) (33068,255) (1179,255) (96777,242) (15786,256) (2541,1572) (2426,1562) (2181,256) (1480,951)};
\addlegendentry{Bayes}
\addplot[only marks, mark=*, mark size=0.95pt, draw=modelcolorctgan, fill=modelcolorctgan, fill opacity=0.75, draw opacity=0.85] coordinates {(1219,1219) (2215,2215) (11927,11927) (1220,1220) (1518,1518) (2218,2218) (484,474) (1634,1491) (97504,19643) (33068,9679) (1179,848) (96777,19630) (15786,8207) (2541,1581) (2426,1559) (2181,2163) (1480,972)};
\addlegendentry{CTGAN}
\addplot[only marks, mark=*, mark size=0.95pt, draw=modelcolorforestdiffusion, fill=modelcolorforestdiffusion, fill opacity=0.75, draw opacity=0.85] coordinates {(1219,629) (2215,752) (11927,963) (1220,636) (1518,678) (2218,751) (484,239) (1634,612) (97504,1022) (33068,74) (1179,293) (96777,1022) (15786,915) (2181,607)};
\addlegendentry{F-Diff}
\addplot[only marks, mark=*, mark size=0.95pt, draw=modelcolorrealtabformer, fill=modelcolorrealtabformer, fill opacity=0.75, draw opacity=0.85] coordinates {(1219,1212) (2215,2174) (11927,10907) (1220,1099) (1518,1508) (2218,2191) (484,45) (1634,48) (97504,62197) (33068,24057) (1179,1097) (96777,61865) (15786,13594) (2541,150) (2426,146) (2181,59) (1480,648)};
\addlegendentry{RTF}
\addplot[only marks, mark=*, mark size=0.95pt, draw=modelcolortabbyflow, fill=modelcolortabbyflow, fill opacity=0.75, draw opacity=0.85] coordinates {(1219,1219) (2215,2215) (11927,11927) (1220,1221) (1518,1519) (2218,2219) (484,467) (1634,1562) (97504,63933) (33068,9944) (1179,50) (96777,63708) (15786,15730) (2541,1609) (2426,1564) (2181,2181) (1480,570)};
\addlegendentry{T-Flow}
\addplot[only marks, mark=*, mark size=0.95pt, draw=modelcolortabdiff, fill=modelcolortabdiff, fill opacity=0.75, draw opacity=0.85] coordinates {(1219,1219) (2215,2214) (11927,11733) (1220,1220) (1518,1518) (2218,2217) (484,411) (1634,1355) (2541,1579) (2426,1532) (2181,1834) (1480,949)};
\addlegendentry{T-Diff}
\addplot[only marks, mark=*, mark size=0.95pt, draw=modelcolortabpfgen, fill=modelcolortabpfgen, fill opacity=0.75, draw opacity=0.85] coordinates {(1219,1219) (2215,2215) (11927,11926) (1220,1220) (1518,1518) (2218,2218) (484,434) (1634,1459) (2541,1535) (2426,1547) (2181,2168) (1480,949)};
\addlegendentry{TPF}
\addplot[only marks, mark=*, mark size=0.95pt, draw=modelcolortabsyn, fill=modelcolortabsyn, fill opacity=0.75, draw opacity=0.85] coordinates {(1219,932) (2215,1115) (11927,723) (1220,1221) (1518,1519) (2218,2203) (484,433) (1634,1016) (2541,2550) (2426,1462) (2181,1699) (1480,846)};
\addlegendentry{T-Syn}
\addplot[only marks, mark=*, mark size=0.95pt, draw=modelcolortvae, fill=modelcolortvae, fill opacity=0.75, draw opacity=0.85] coordinates {(1219,1049) (2215,1952) (11927,8388) (1220,39) (1518,62) (2218,81) (484,324) (1634,924) (2541,83) (2426,84) (2181,1774) (1480,532)};
\addlegendentry{TVAE}
\end{loglogaxis}
\end{tikzpicture}
  \vspace{-10pt}
  \captionsetup{skip=2pt}
  \caption{Real vs.\ synthetic distinct counts on high-cardinality columns. Current generative models still struggle to preserve high-cardinality support. Details in Table~\ref{tab:cardinality-range-five-case-raw}.}
  \label{fig:cardinality-scatter-v1-local-preview}
\end{minipage}
\end{figure*}

Figure~\ref{fig:distance_query_overall_scatter} compares each generative model's distance-based overall score with its query overall score, with both scores computed against the same real training split and the same-size synthetic output. Every generative model lies below the diagonal, so the query-centric view is uniformly stricter than the distance-based one rather than reflecting a mismatch in reference splits. The ordering also shifts. RealTabFormer remains strongest overall, but ARF and TabPFGen look especially strong under distance-based fidelity relative to their query-side scores, while BayesNet remains comparatively stronger once we ask whether the same synthetic tables preserve the structure that matters for analytical queries. The full per-generative-model summary is deferred to Appendix Table~\ref{tab:benchmark_overall_real_main}.

The appendix table also shows that the failure is not uniform across query families. RealTabFormer is strongest on the main query families, reaching $0.83_{\pm 0.20}$ on Tail/Rarity, $0.75_{\pm 0.26}$ on Conditional, and $0.81_{\pm 0.26}$ on Missingness, but weaker generators still drop sharply in these same families. Thus, distance-based fidelity can give an overly optimistic picture: it hides the subgroup, conditional, and rare-event failures that matter for real analytical use.

\subsection{Where Do Failures Under Query-Centric Fidelity Concentrate?}

A major goal for \BenchmarkName{} is to localize failure modes rather than reporting only an aggregate query score. We study two complementary axes: query families, which reveal difficult structures to preserve, and data regimes, which reveal how categorical, numerical, and mixed schemas change the failure profile.

\subsubsection{Breaking Down By Query Family}

We break down query-centric fidelity across the query families and demonstrate analysis and insights. We first localize failures by query family. This view asks which kinds of analytical query objects are most fragile under synthetic generation, rather than averaging all query behavior together. In the main text, we focus on three representative hard regimes. Conditional queries test whether grouped structure remains reliable after the analysis is restricted to a filtered local slice. Tail and rarity queries test whether rare regions remain both identifiable and queryable as the support becomes more extreme. High-cardinality queries test whether synthetic data preserves the large discrete support that matters for analytical queries. For each family, we follow the same progression: we first define the query object, then show one grounded example, then summarize the benchmark evidence, and finally explain what the result implies about current synthetic tabular generative models.

\vspace{-0.5em}
\findingbox{}{\textbf{Finding 2: }Current SOTA tabular generative models often fail to preserve high-cardinality discrete support (Figure~\ref{fig:cardinality-scatter-v1-local-preview} and Table~\ref{tab:cardinality-range-five-case-raw}).}

We study \emph{cardinality} because it determines whether the synthetic table preserves the state space needed by downstream analytical queries. For discrete columns, cardinality measures whether the generated table retains the distinct values present in the real data. This is especially important for high-cardinality attributes, where a column may contain hundreds, thousands, or even more distinct states. If a generative model drops a large portion of these values, then later filtering, grouping, and subgroup queries no longer operate over the same support as the real table. In this case, the synthetic table may appear reasonable under aggregate distributional metrics while still failing to represent the set of values that downstream analyses depend on.

Figure~\ref{fig:cardinality-scatter-v1-local-preview} shows that this failure is common on high-cardinality discrete columns. 
Many points fall far below this line, especially when the real column contains thousands to tens of thousands of distinct values. This means that the synthetic table often preserves only a fraction of the real discrete support. The problem is also generative-model-dependent: some generative models remain close to the diagonal for several columns, while others collapse high-cardinality support to only hundreds of synthetic values.

Table~\ref{tab:cardinality-range-five-case-raw} shows the support loss at the raw column level. For columns with relatively moderate cardinality, some generative models can preserve the distinct count almost exactly. 
Several generative models collapse the same columns to only a small fraction of the real support: BayesNet repeatedly generates only $64$ or $256$ distinct values, ForestDiffusion often reduces thousands of real states to hundreds, and TVAE can collapse high-cardinality columns even more severely.

\begin{table*}
\centering
\footnotesize
\setlength{\tabcolsep}{2.5pt}
\begin{adjustbox}{max width=\textwidth}
  \begin{tabular}{lllcclccccccccccc}
    \toprule
    Dataset & Rows & Column Name & Kind & Real & ARF & Bayes & CTGAN & F-Diff$^{*}$ & RTF & T-Flow & T-DDPM$^{*}$ & T-Diff$^{*}$ & TPF$^{*}$ & T-Syn$^{*}$ & TVAE$^{*}$ \\
    \midrule
    c14 & 300,000 & 7 & discrete & 1219 & {\color{SecondPlace}\textbf{1217}} & 64 & {\color{FirstPlace}\textbf{1219}} & 629 & {\color{ThirdPlace}\textbf{1212}} & {\color{FirstPlace}\textbf{1219}} & - & {\color{FirstPlace}\textbf{1219}} & {\color{FirstPlace}\textbf{1219}} & 932 & 1049 \\
    c15 & 600,000 & nom\_9 & discrete & 2218 & {\color{ThirdPlace}\textbf{2216}} & {\color{ThirdPlace}\textbf{2216}} & {\color{FirstPlace}\textbf{2218}} & 751 & 2191 & {\color{SecondPlace}\textbf{2219}} & - & {\color{SecondPlace}\textbf{2217}} & {\color{FirstPlace}\textbf{2218}} & 2203 & 81 \\
    c17 & 8,809 & listed\_in & discrete & 484 & 424 & 421 & {\color{FirstPlace}\textbf{474}} & 239 & 45 & {\color{SecondPlace}\textbf{467}} & - & 411 & {\color{ThirdPlace}\textbf{434}} & 433 & 324 \\
    c17 & 8,809 & date\_added & discrete & 1634 & 1366 & 1375 & {\color{SecondPlace}\textbf{1491}} & 612 & 48 & {\color{FirstPlace}\textbf{1562}} & - & 1355 & {\color{ThirdPlace}\textbf{1459}} & 1016 & 924 \\
    c18 & 129,975 & region\_1 & discrete & 1179 & {\color{FirstPlace}\textbf{1114}} & 255 & {\color{ThirdPlace}\textbf{848}} & 293 & {\color{SecondPlace}\textbf{1097}} & 50 & - & - & - & - & - \\
    c18 & 129,975 & title & discrete & 96777 & {\color{SecondPlace}\textbf{62598}} & 242 & 19630 & 1022 & {\color{ThirdPlace}\textbf{61865}} & {\color{FirstPlace}\textbf{63708}} & - & - & - & - & - \\
    c18 & 129,975 & winery & discrete & 15786 & {\color{SecondPlace}\textbf{13729}} & 256 & 8207 & 915 & {\color{ThirdPlace}\textbf{13594}} & {\color{FirstPlace}\textbf{15730}} & - & - & - & - & - \\
    c3 & 2,551 & ATRINS-DONOR-521 & discrete & 2541 & {\color{SecondPlace}\textbf{1649}} & 1572 & 1581 & - & 150 & {\color{ThirdPlace}\textbf{1609}} & - & 1579 & 1535 & {\color{FirstPlace}\textbf{2550}} & 83 \\
    c3 & 2,551 & CCAGC & discrete & 2426 & 1533 & {\color{SecondPlace}\textbf{1562}} & {\color{ThirdPlace}\textbf{1559}} & - & 146 & {\color{FirstPlace}\textbf{1564}} & - & 1532 & 1547 & 1462 & 84 \\
    c19 & 48,697 & channel\_title & discrete & 2181 & 2086 & 256 & {\color{ThirdPlace}\textbf{2163}} & 607 & 59 & {\color{FirstPlace}\textbf{2181}} & - & 1834 & {\color{SecondPlace}\textbf{2168}} & 1699 & 1774 \\
    \bottomrule
  \end{tabular}
\end{adjustbox}
  \vspace{-2pt}
  \captionsetup{skip=2pt}
\caption{Representative column-level cases for cardinality fidelity. Each generative-model entry is a distinct-value count. Entries for generative models are highlighted as {\color{FirstPlace}\textbf{First}}, {\color{SecondPlace}\textbf{Second}}, and {\color{ThirdPlace}\textbf{Third}} within each row by closeness to the real statistics.
Models marked with $^{*}$ have unavailable results due to distinct-value limitations or exceeding the 72-hour time limit for training or generation.
}
\label{tab:cardinality-range-five-case-raw}
\end{table*}

\findingbox{}{\textbf{Finding 3: }Current SOTA generative models preserve global conditional structure more reliably than local-slice conditional structure (Figure~\ref{fig:local-vs-defiltered-scatter}).}
\vspace{0.2em}

For the local--global comparison, we define a conditional query
containing a \texttt{WHERE} predicate as a \emph{local} query. Its
paired \emph{global} counterpart is obtained by deleting only the
\texttt{WHERE} clause, while keeping the remaining
\texttt{SELECT}, \texttt{FROM}, \texttt{GROUP BY}, and
\texttt{ORDER BY} structure unchanged. For example,
\texttt{... WHERE Subtopic = 'Linear Transformations' GROUP BY ...}
is paired with the same query without the \texttt{WHERE} clause.
Thus, the comparison isolates the effect of evaluating the same
conditional structure within a filtered slice rather than over the
full table; see Appendix~\ref{app:experiment_conditional} for the
formal construction.

Figure~\ref{fig:local-vs-defiltered-scatter} shows that local-slice conditional queries are often less faithfully preserved than their global counterparts. Each point represents one generative model on one dataset under one paired conditional query. 
The points below the diagonal correspond to cases where the same query structure becomes less faithful after restricting the analysis to a subpopulation.
This result shows that high global conditional fidelity is not sufficient: a synthetic table may answer broad conditional queries accurately while still failing on the local slices that analysts use to study specific subgroups.

\findingbox{}{\textbf{Finding 4: }Current SOTA generative models fail to preserve rare cases and tail distributions, which leads to poor fidelity on tail-focused analytical queries (Figure~\ref{fig:tail_threshold_model_curves_v2_pair}).}

We first ask whether current generative models preserve the rare regions of individual columns, before evaluating full tail queries. The tail-region definition and the two complementary scores, \emph{tail size} and \emph{tail coverage}, are documented in Appendix~\ref{app:tail_rarity_diagnostics}.

Figure~\ref{fig:tail_threshold_model_curves_v2_pair} shows that current generative models have limited ability to preserve rare regions, and this limitation becomes more severe as the tail threshold tightens.
Figure~\ref{fig:tail_threshold_model_curves_v2_pair} shows that this trend is not caused by a single weak generative model. 
However, the rare-region failure is broadly shared across current SOTA generative models.

\begin{figure*}[t]
\centering

\begin{minipage}[t]{0.485\textwidth}
  \vspace{0pt}
  \centering

  \begin{minipage}[t][0.72\linewidth][t]{\linewidth}
    \centering

    \centering
\begingroup
\definecolor{tplL1}{HTML}{1F77B4}
\definecolor{tplL2}{HTML}{FF7F0E}
\definecolor{tplL3}{HTML}{2CA02C}
\definecolor{tplL4}{HTML}{D62728}
\definecolor{tplL5}{HTML}{9467BD}
\begin{tikzpicture}
\begin{axis}[
width=\linewidth,
height=0.62\linewidth,
xmin=0.0, xmax=1.0,
ymin=0.0, ymax=1.0,
xlabel={Global counterpart score},
ylabel={Local-slice score},
tick label style={font=\normalsize},
label style={font=\normalsize},
xtick={0,0.2,0.4,0.6,0.8,1.0},
ytick={0,0.2,0.4,0.6,0.8,1.0},
xmajorgrids=false, ymajorgrids=false,
grid=none,
axis line style={draw=black!70},
tick style={draw=black!70},
clip=false,
legend columns=1,
legend cell align=left,
legend style={draw=none, fill=none, font=\normalsize, at={(0.05,0.95)}, anchor=north west, row sep=0.04cm},
]
\addplot[black, dashed, line width=1.0pt, forget plot] coordinates {(0,0) (1,1)};
\addplot+[only marks, mark=*, mark size=0.95pt, draw=black!45, fill=black!45, fill opacity=0.32, draw opacity=0.45] coordinates { (1.000000,1.000000) (1.000000,1.000000) (1.000000,1.000000) (1.000000,1.000000) (1.000000,1.000000) (1.000000,1.000000) (0.400000,0.400000) (1.000000,1.000000) (1.000000,1.000000) (1.000000,1.000000) (1.000000,1.000000) (1.000000,1.000000) (1.000000,1.000000) (1.000000,1.000000) (1.000000,1.000000) (1.000000,1.000000) (0.000000,0.000000) (1.000000,1.000000) (1.000000,1.000000) (1.000000,1.000000) (1.000000,1.000000) (0.056592,0.048502) (0.001968,0.003377) (0.000196,0.000000) (0.000000,0.000000) (0.059032,0.037397) (0.000000,0.000000) (0.000000,0.000000) (0.000000,0.000000) (0.005772,0.002547) (0.000000,0.000000) (0.019376,0.022343) (1.000000,1.000000) (1.000000,1.000000) (1.000000,1.000000) (1.000000,1.000000) (1.000000,1.000000) (1.000000,1.000000) (1.000000,1.000000) (1.000000,1.000000) (0.500000,0.500000) (1.000000,1.000000) (1.000000,1.000000) (0.617647,0.500000) (0.600000,0.470588) (0.512195,0.390244) (0.500000,0.380952) (0.954545,0.888889) (1.000000,0.941176) (0.875000,0.666667) (0.525000,0.410256) (1.000000,1.000000) (0.677419,0.571429) (0.600000,0.457143) (1.000000,0.866667) (0.937500,0.812500) (0.937500,0.866667) (0.937500,0.812500) (1.000000,1.000000) (1.000000,0.923077) (0.933333,0.928571) (0.937500,0.812500) (1.000000,1.000000) (0.937500,1.000000) (1.000000,0.923077) (1.000000,1.000000) (1.000000,1.000000) (1.000000,1.000000) (0.000000,0.000000) (1.000000,1.000000) (1.000000,1.000000) (1.000000,1.000000) (1.000000,1.000000) (1.000000,1.000000) (0.000000,0.000000) (1.000000,1.000000) (1.000000,1.000000) (1.000000,1.000000) (1.000000,1.000000) (1.000000,1.000000) (1.000000,1.000000) (1.000000,1.000000) (1.000000,0.833333) (1.000000,1.000000) (1.000000,1.000000) (1.000000,1.000000) (0.666667,0.500000) (0.000000,0.000000) (0.001875,0.001871) (0.000207,0.000122) (0.000000,0.000000) (0.887403,0.861965) (0.000000,0.000000) (0.000000,0.000000) (0.000000,0.000000) (0.000000,0.000000) (0.000000,0.000000) (0.000069,0.000044) (1.000000,1.000000) (1.000000,1.000000) (1.000000,1.000000) (1.000000,1.000000) (1.000000,1.000000) (1.000000,1.000000) (1.000000,1.000000) (1.000000,1.000000) (1.000000,1.000000) (1.000000,1.000000) (1.000000,1.000000) (0.000000,0.000000) (0.000000,0.000000) (0.000000,0.000000) (0.000000,0.000000) (0.000000,0.000000) (0.000000,0.000000) (0.000000,0.000000) (0.000000,0.000000) (0.000000,0.000000) (1.000000,1.000000) (1.000000,1.000000) (1.000000,1.000000) (1.000000,1.000000) (1.000000,1.000000) (1.000000,1.000000) (0.055556,0.000000) (1.000000,1.000000) (1.000000,1.000000) (1.000000,1.000000) (0.277778,0.166667) (0.685083,0.680000) (0.631250,0.620155) (0.604762,0.556757) (0.323024,0.250000) (0.855172,0.817391) (0.046358,0.040650) (0.418719,0.387640) (0.393782,0.353293) (0.239726,0.279661) (0.864060,0.829787) (1.000000,0.991525) (0.952991,0.882096) (0.722939,0.717770) (0.991525,1.000000) (0.952586,0.892377) (0.617936,0.596479) (1.000000,0.991525) (0.919149,0.672566) (0.439920,0.429293) (0.000000,0.000000) (0.000000,0.000000) (0.035785,0.026713) (0.177966,0.119658) (0.116379,0.062780) (0.000994,0.001161) (0.000000,0.000000) (0.000000,0.000000) (0.000000,0.000000) (0.000000,0.000000) (0.000000,0.000000) (0.719273,0.718584) (0.000000,0.000000) (0.000000,0.000000) (0.845703,0.789352) (0.991525,0.991453) (0.982759,0.888889) (0.797998,0.785106) (0.000000,0.000000) (0.000000,0.000000) (0.348907,0.324042) (0.983051,0.914530) (0.892241,0.855895) (0.904762,1.000000) (0.078125,0.454545) (0.825397,0.500000) (0.000000,0.000000) (0.047619,0.000000) (0.000000,0.000000) (0.000000,0.000000) (0.857143,0.500000) (0.000000,0.000000) (0.412698,0.900000) (1.000000,1.000000) (1.000000,1.000000) (1.000000,1.000000) (0.000000,0.000000) (1.000000,1.000000) (0.000000,0.000000) (1.000000,1.000000) (0.000000,0.000000) (1.000000,1.000000) (0.000000,0.000000) (1.000000,1.000000) (1.000000,1.000000) (1.000000,1.000000) (1.000000,1.000000) (0.000000,0.000000) (1.000000,1.000000) (0.000000,0.000000) (0.000000,0.000000) (1.000000,1.000000) (0.000000,0.000000) (1.000000,1.000000) (1.000000,0.970588) (0.970588,0.911765) (0.823529,0.848485) (0.058824,0.060606) (1.000000,0.970588) (1.000000,0.970588) (0.058824,0.060606) (1.000000,0.970588) (1.000000,1.000000) (0.970588,0.941176) (0.676471,0.666667) (1.000000,1.000000) (0.793103,0.807692) (0.766667,0.777778) (0.000000,0.000000) (0.956522,1.000000) (0.956522,0.809524) (0.724138,0.760000) (0.766667,0.800000) (0.956522,0.952381) (0.846154,1.000000) (0.521739,0.380952) (1.000000,1.000000) (1.000000,1.000000) (1.000000,1.000000) (1.000000,1.000000) (1.000000,1.000000) (1.000000,1.000000) (1.000000,1.000000) (1.000000,1.000000) (1.000000,1.000000) (1.000000,1.000000) (1.000000,1.000000) (1.000000,1.000000) (1.000000,1.000000) (1.000000,1.000000) (1.000000,1.000000) (1.000000,1.000000) (1.000000,1.000000) (0.000000,0.000000) (0.000000,0.000000) (1.000000,1.000000) (0.000000,0.000000) (1.000000,0.916667) (1.000000,1.000000) (1.000000,1.000000) (1.000000,1.000000) (0.714286,0.714286) (1.000000,1.000000) (0.714286,0.714286) (0.714286,0.714286) (0.714286,0.714286) (0.833333,0.833333) (0.714286,0.714286) (1.000000,1.000000) (1.000000,1.000000) (1.000000,1.000000) (1.000000,1.000000) (1.000000,1.000000) (1.000000,1.000000) (1.000000,1.000000) (1.000000,1.000000) (1.000000,1.000000) (1.000000,1.000000) (1.000000,1.000000) (1.000000,1.000000) (1.000000,1.000000) (0.687500,0.593750) (1.000000,1.000000) (1.000000,1.000000) (1.000000,1.000000) (1.000000,1.000000) (1.000000,1.000000) (1.000000,1.000000) (1.000000,1.000000) (1.000000,1.000000) (1.000000,1.000000) (1.000000,1.000000) (1.000000,1.000000) (1.000000,1.000000) (1.000000,1.000000) (1.000000,1.000000) (1.000000,1.000000) (0.250000,0.250000) (0.250000,0.250000) (1.000000,1.000000) (1.000000,1.000000) (0.250000,0.250000) (0.250000,0.250000) (0.250000,0.250000) (0.250000,0.250000) (0.250000,0.250000) (0.250000,0.250000) (1.000000,1.000000) (1.000000,1.000000) (0.143713,0.172662) (0.088889,0.400000) (1.000000,1.000000) (1.000000,1.000000) (0.003585,0.003881) (0.000000,0.000000) (0.000000,0.000000) (0.000238,0.000000) (0.588060,0.579661) (0.029955,0.029208) (0.004839,0.005435) (0.036439,0.049470) (0.000000,0.000000) (0.071315,0.097540) (0.000000,0.000000) (0.762712,0.741379) (1.000000,0.976744) (0.740000,0.714286) (0.846154,0.840000) (0.888889,0.930233) (0.891304,0.866667) (0.020000,0.021739) (0.833333,0.811321) (1.000000,1.000000) (0.000000,0.000000) (0.734694,0.744681) (0.000000,0.000000) (0.733990,0.391144) (0.226837,0.000000) (0.000000,0.000000) (0.800595,0.808333) (0.000000,0.000000) (0.000000,0.000000) (0.000000,0.000000) (0.000000,0.000000) (0.000000,0.000000) (0.778711,0.758333) (0.333333,1.000000) (0.666667,1.000000) (0.200000,0.000000) (0.666667,0.500000) (0.333333,1.000000) (0.333333,0.000000) (0.666667,1.000000) (0.000000,0.000000) (0.666667,1.000000) (0.333333,0.000000) (0.000000,0.000000) (0.802187,0.280320) (0.722391,0.154321) (0.000000,0.000000) (0.906103,0.490667) (0.000000,0.000000) (0.000000,0.000000) (0.000000,0.000000) (0.000000,0.000000) (0.000000,0.000000) (0.781962,0.293624) (0.000000,0.000000) (0.000000,0.000000) (0.000000,0.000000) (0.000000,0.000000) (0.000000,0.000000) (0.000000,0.000000) (0.000000,0.000000) (0.000000,0.000000) (0.000000,0.000000) (0.000000,0.000000) (0.000000,0.000000) (0.000000,0.000000) (0.000103,0.000074) (0.000000,0.000000) (0.000000,0.000000) (0.762476,0.704814) (0.000257,0.000155) (0.000000,0.000000) (0.000051,0.000000) (0.000103,0.000164) (0.000051,0.000074) (0.000000,0.000000) (0.000000,0.000000) (0.000000,0.000000) (0.000000,0.000000) (0.000000,0.000000) (0.000000,0.000000) (0.000000,0.000000) (0.000000,0.000000) (0.000000,0.000000) (0.000000,0.000000) (0.007686,0.027692) (0.000618,0.001439) (0.000000,0.000000) (0.000000,0.000000) (0.097561,0.090909) (0.000000,0.000000) (0.022005,0.000000) (0.000000,0.000000) (0.023041,0.036765) (0.000000,0.000000) (0.000000,1.000000) (0.755102,1.000000) (0.043210,1.000000) (0.000000,1.000000) (0.712500,1.000000) (0.000000,1.000000) (0.000000,1.000000) (0.000000,1.000000) (0.000000,1.000000) (0.000000,1.000000) (0.618557,1.000000) (0.000000,0.000000) (0.759494,0.652778) (0.708861,0.671053) (0.000000,0.000000) (0.647059,0.524590) (0.000000,0.000000) (0.000000,0.000000) (0.000000,0.000000) (0.000000,0.000000) (0.746667,0.765625) };
\end{axis}
\end{tikzpicture}
\endgroup%

  \end{minipage}
  \vspace{-30pt}
  \captionsetup{skip=2pt}
  \caption{
    Local-slice vs.\ global-counterpart conditional scores.
    Most points lie below the diagonal, indicating that
    local-slice conditions are generally harder to preserve
    than global-counterpart conditions.
    Details appear in Appendix~\ref{app:experiment_conditional}.
  }
  \label{fig:local-vs-defiltered-scatter}

\end{minipage}
\hfill
\begin{minipage}[t]{0.485\textwidth}
  \vspace{0pt}
  \centering

  \begin{minipage}[t][0.72\linewidth][t]{\linewidth}
    \centering

    \begin{tikzpicture}
\begin{axis}[
width=\linewidth,
height=0.62\linewidth,
xmin=0.7, xmax=12.5,
ymin=0.360, ymax=0.828,
xtick={1,2,3,4,5,6,7,8,9},
xticklabels={10\%,8\%,6\%,4\%,3\%,2\%,1\%,0.5\%,0.1\%},
xticklabel style={rotate=28, anchor=base, yshift=-10pt, font=\scriptsize},
yticklabel style={font=\scriptsize},
ylabel={Mean tail overall score},
xlabel={Tail threshold},
xlabel style={yshift=5pt},
axis line style={draw=black!70}, tick style={draw=black!70},
grid=major, grid style={dashed, gray!35}, major grid style={dashed, gray!35},
clip=false,
]
\addplot[line width=1.05pt, color=modelrealtabformer, mark=*, mark size=1.7pt, mark options={fill=modelrealtabformer,draw=modelrealtabformer}] coordinates { (1,0.782161) (2,0.775523) (3,0.770375) (4,0.758953) (5,0.752320) (6,0.737424) (7,0.722322) (8,0.706187) (9,0.661999) };
\node[font=\tiny\bfseries, text=modelrealtabformer, anchor=west, xshift=2pt, inner sep=0.6pt] at (axis cs:7.95,0.755) {RTF};

\addplot[line width=1.05pt, color=modelarf, mark=*, mark size=1.7pt, mark options={fill=modelarf,draw=modelarf}] coordinates { (1,0.754092) (2,0.750489) (3,0.745419) (4,0.736274) (5,0.730255) (6,0.724344) (7,0.712112) (8,0.698102) (9,0.656889) };
\node[font=\tiny\bfseries, text=modelarf, anchor=west, xshift=2pt, inner sep=0.6pt] at (axis cs:9.35,0.700) {ARF};

\addplot[line width=1.05pt, color=modelbayesnet, mark=*, mark size=1.7pt, mark options={fill=modelbayesnet,draw=modelbayesnet}] coordinates { (1,0.780770) (2,0.770578) (3,0.755194) (4,0.737679) (5,0.724831) (6,0.711987) (7,0.685941) (8,0.664673) (9,0.631392) };
\node[font=\tiny\bfseries, text=modelbayesnet, anchor=west, xshift=2pt, inner sep=0.6pt] at (axis cs:9.35,0.665) {Bayes};

\addplot[line width=1.05pt, color=modelctgan, mark=*, mark size=1.7pt, mark options={fill=modelctgan,draw=modelctgan}] coordinates { (1,0.525823) (2,0.519525) (3,0.507454) (4,0.494605) (5,0.486095) (6,0.476324) (7,0.464886) (8,0.453414) (9,0.428820) };
\node[font=\tiny\bfseries, text=modelctgan, anchor=west, xshift=2pt, inner sep=0.6pt] at (axis cs:9.35,0.445) {CTGAN};

\addplot[line width=1.05pt, color=modeltabpfgen, mark=*, mark size=1.7pt, mark options={fill=modeltabpfgen,draw=modeltabpfgen}] coordinates { (1,0.746187) (2,0.737553) (3,0.731742) (4,0.715759) (5,0.707689) (6,0.694600) (7,0.660587) (8,0.633046) (9,0.590798) };
\node[font=\tiny\bfseries, text=modeltabpfgen, anchor=west, xshift=2pt, inner sep=0.6pt] at (axis cs:4.25,0.660) {TPF};

\addplot[line width=1.05pt, color=modeltabdiff, mark=*, mark size=1.7pt, mark options={fill=modeltabdiff,draw=modeltabdiff}] coordinates { (1,0.743302) (2,0.741009) (3,0.734557) (4,0.727142) (5,0.721946) (6,0.715542) (7,0.693300) (8,0.668193) (9,0.623938) };
\node[font=\tiny\bfseries, text=modeltabdiff, anchor=west, xshift=2pt, inner sep=0.6pt] at (axis cs:9.35,0.625) {T-Diff};

\addplot[line width=1.05pt, color=modeltabbyflow, mark=*, mark size=1.7pt, mark options={fill=modeltabbyflow,draw=modeltabbyflow}] coordinates { (1,0.604670) (2,0.601787) (3,0.594199) (4,0.583869) (5,0.577485) (6,0.569420) (7,0.555166) (8,0.546439) (9,0.530879) };
\node[font=\tiny\bfseries, text=modeltabbyflow, anchor=west, xshift=2pt, inner sep=0.6pt] at (axis cs:5.15,0.605) {T-Flow};

\addplot[line width=1.05pt, color=modeltabsyn, mark=*, mark size=1.7pt, mark options={fill=modeltabsyn,draw=modeltabsyn}] coordinates { (1,0.606044) (2,0.599632) (3,0.592535) (4,0.581741) (5,0.577569) (6,0.568421) (7,0.559301) (8,0.548823) (9,0.531092) };
\node[font=\tiny\bfseries, text=modeltabsyn, anchor=west, xshift=2pt, inner sep=0.6pt] at (axis cs:7.05,0.585) {T-Syn};

\addplot[line width=1.05pt, color=modeltvae, mark=*, mark size=1.7pt, mark options={fill=modeltvae,draw=modeltvae}] coordinates { (1,0.499544) (2,0.489979) (3,0.478889) (4,0.472907) (5,0.462540) (6,0.447176) (7,0.430706) (8,0.420786) (9,0.394207) };
\node[font=\tiny\bfseries, text=modeltvae, anchor=west, xshift=2pt, inner sep=0.6pt] at (axis cs:9.35,0.395) {TVAE};

\addplot[line width=1.05pt, color=modelforestdiffusion, mark=*, mark size=1.7pt, mark options={fill=modelforestdiffusion,draw=modelforestdiffusion}] coordinates { (1,0.534175) (2,0.526027) (3,0.520377) (4,0.501063) (5,0.487275) (6,0.481329) (7,0.471872) (8,0.468176) (9,0.506445) };
\node[font=\tiny\bfseries, text=modelforestdiffusion, anchor=west, xshift=2pt, inner sep=0.6pt] at (axis cs:9.35,0.500) {F-Diff};

\addplot[line width=1.05pt, color=modeltabddpm, mark=*, mark size=1.7pt, mark options={fill=modeltabddpm,draw=modeltabddpm}] coordinates { (1,0.558615) (2,0.558634) (3,0.558003) (4,0.553818) (5,0.547115) (6,0.538674) (7,0.529426) (8,0.519719) (9,0.518104) };
\node[font=\tiny\bfseries, text=modeltabddpm, anchor=west, xshift=2pt, inner sep=0.6pt] at (axis cs:9.35,0.545) {T-DDPM};

\end{axis}
\end{tikzpicture}%

  \end{minipage}

  \vspace{-27pt}
  \captionsetup{skip=2pt}
  \caption{
    The broadly consistent decline across models shows that
    tail degradation is a general pattern, not one driven by
    only a few models.
  }
  \label{fig:tail_threshold_model_curves_v2_pair}

\end{minipage}

\end{figure*}

\textbf{Missingness.}
Missingness provides a useful contrast. Generative models often preserve the overall missing rate, but this does not guarantee that missingness remains associated with the same conditioning attributes. 
We therefore evaluate both marginal missingness and broad co-missingness. 
For example, one grounded query on dataset \texttt{c5} groups rows by \texttt{odor} and measures the missing rate of \texttt{stalk-root} within each group.
A synthetic table may match the global missing fraction while still failing this grouped query if the relationship between the conditioning attribute and the missingness pattern is distorted. 
Detailed query definitions are provided in Appendix~\ref{app:experiment_missingness}.

\subsubsection{Breaking Down by Categorical vs. Numerical vs. Mixed}

We further separate results across categorical, numerical, and mixed datasets, which we define by the dominant column type in each schema. This analysis tests whether the same query family fails differently depending on the schema regime, rather than treating all tabular datasets as one homogeneous benchmark pool.

Table~\ref{tab:benchmark_query_category_main} shows that queries on categorical datasets are often about preserving discrete support, subgroup identity, and rare-state membership, whereas queries on numerical datasets more often test range structure, continuous conditional behavior, and tail sizes. 

\begin{table*}
\centering

\definecolor{FirstPlace}{HTML}{1397B8}%
\definecolor{SecondPlace}{HTML}{7B45E5}%
\definecolor{ThirdPlace}{HTML}{000000}%
\definecolor{OverallTint}{HTML}{F8F1DA}%
\definecolor{RuleGray}{HTML}{C8CDD3}%
\arrayrulecolor{RuleGray}%
\setlength{\tabcolsep}{2.2pt}%
\renewcommand{\arraystretch}{1.18}%
\scriptsize
\newcommand{\RegimeScore}[3]{{\color{#1}\textbf{\makecell[c]{#2\\[-1pt]$\pm #3$}}}}%
\begin{tabular}{@{}l c c c c c c c c c c c@{}}
\toprule
\multirow{2}{*}{\textbf{Category}} & \multicolumn{1}{c}{\textbf{Overall}} & \multicolumn{2}{c}{\textbf{Subgroup}} & \multicolumn{2}{c}{\textbf{Conditional}} & \multicolumn{2}{c}{\textbf{Tail / Rarity}} & \multicolumn{2}{c}{\textbf{Missingness}} & \multicolumn{2}{c}{\textbf{Cardinality / Range}} \\
\cmidrule(lr){2-2}
\cmidrule(lr){3-4}
\cmidrule(lr){5-6}
\cmidrule(lr){7-8}
\cmidrule(lr){9-10}
\cmidrule(lr){11-12}
 & \cellcolor{OverallTint}\makecell[c]{Query\\overall} & \makecell[c]{Structure} & \makecell[c]{Size} & \makecell[c]{Global\\structure} & \makecell[c]{Local\\slice} & \makecell[c]{Tail\\coverage} & \makecell[c]{Tail\\size} & \makecell[c]{Marginal} & \makecell[c]{Broad\\co-missing} & \makecell[c]{Discrete} & \makecell[c]{Continuous} \\
\midrule
Categorical & \cellcolor{OverallTint}\RegimeScore{FirstPlace}{0.68}{0.07} & \RegimeScore{FirstPlace}{0.77}{0.06} & \RegimeScore{FirstPlace}{0.78}{0.06} & \RegimeScore{FirstPlace}{0.64}{0.06} & \RegimeScore{FirstPlace}{0.63}{0.10} & \RegimeScore{FirstPlace}{0.14}{0.12} & \RegimeScore{SecondPlace}{0.31}{0.24} & \RegimeScore{ThirdPlace}{0.93}{0.01} & \RegimeScore{ThirdPlace}{0.93}{0.01} & \RegimeScore{SecondPlace}{0.66}{0.21} & \RegimeScore{ThirdPlace}{0.88}{0.16} \\
Numerical & \cellcolor{OverallTint}\RegimeScore{ThirdPlace}{0.59}{0.06} & \RegimeScore{ThirdPlace}{0.52}{0.08} & \RegimeScore{ThirdPlace}{0.54}{0.10} & \RegimeScore{ThirdPlace}{0.32}{0.06} & \RegimeScore{ThirdPlace}{0.15}{0.09} & \RegimeScore{ThirdPlace}{0.07}{0.07} & \RegimeScore{FirstPlace}{0.60}{0.11} & \RegimeScore{FirstPlace}{0.99}{0.01} & \RegimeScore{FirstPlace}{0.99}{0.01} & \RegimeScore{ThirdPlace}{0.58}{0.25} & \RegimeScore{SecondPlace}{0.92}{0.08} \\
Mix & \cellcolor{OverallTint}\RegimeScore{SecondPlace}{0.63}{0.06} & \RegimeScore{SecondPlace}{0.55}{0.05} & \RegimeScore{SecondPlace}{0.63}{0.05} & \RegimeScore{SecondPlace}{0.51}{0.06} & \RegimeScore{SecondPlace}{0.56}{0.09} & \RegimeScore{SecondPlace}{0.10}{0.12} & \RegimeScore{ThirdPlace}{0.23}{0.18} & \RegimeScore{SecondPlace}{0.94}{0.02} & \RegimeScore{SecondPlace}{0.94}{0.02} & \RegimeScore{FirstPlace}{0.85}{0.14} & \RegimeScore{FirstPlace}{0.94}{0.08} \\
\bottomrule
\end{tabular}

\captionsetup{skip=4pt}
\caption{\textbf{Query taxonomy by data regime.} Rows are ranked as {\color{FirstPlace}\textbf{First}}, {\color{SecondPlace}\textbf{Second}} and Third. 
}
\label{tab:benchmark_query_category_main}
\end{table*}

\subsection{Fidelity-cost Pareto tradeoff}

\findingbox{}{\textbf{Finding 5: }There is a clear cost-fidelity tradeoff in tabular data generation, with different tradeoff regimes for training and generation cost (Figures~\ref{fig:query_train_time_tradeoff} and Table~\ref{tab:benchmark_overall_real_main}).}

Figure~\ref{fig:query_train_time_tradeoff} shows that query-centric fidelity and runtime separate into distinct practical regimes once training and generation cost are considered separately. RealTabFormer achieves the highest mean Query overall score across the 49-dataset benchmark, but it is also by far the most expensive generative model to train. BayesNet lies on the low-training-cost/strong-fidelity end of the training frontier, while ARF offers a stronger generation-time tradeoff because it has high query fidelity and sub-second median generation cost per 1k rows. Details are shown in Appendix Table~\ref{tab:benchmark_overall_real_main}.

\begin{figure*}[t]
\centering
\begin{minipage}[t]{0.65\textwidth}
\begin{minipage}[t]{0.49\linewidth}
  \vspace{0pt}
  \centering
  \begin{tikzpicture}
\definecolor{modelrealtabformer}{HTML}{6F4C9B}
\definecolor{modeltvae}{HTML}{882255}
\definecolor{modelforestdiffusion}{HTML}{66CCEE}
\definecolor{modeltabddpm}{HTML}{004488}
\definecolor{modeltabsyn}{HTML}{4477AA}
\definecolor{modeltabdiff}{HTML}{009988}
\definecolor{modelctgan}{HTML}{CC3311}
\definecolor{modelarf}{HTML}{EE7733}
\definecolor{modelbayesnet}{HTML}{666666}
\definecolor{modeltabpfgen}{HTML}{DDAA33}
\definecolor{modeltabbyflow}{HTML}{CC6677}
\begin{axis}[
width=\linewidth,
height=4.35cm,
xmode=log,
xmin=0.05, xmax=80,
ymin=0.38, ymax=0.80,
xtick={0.1,1,10,50},
xticklabels={0.1,1,10,50},
minor tick num=0,
grid=both, grid style={draw=black!9}, major grid style={draw=black!12},
axis line style={draw=black!55}, tick style={draw=black!55},
xlabel={Median train time (min)},
ylabel={Mean query score},
ticklabel style={font=\scriptsize},
label style={font=\footnotesize},
clip=false
]
\addplot[only marks, mark=*, mark size=2.9pt, draw=modelarf, fill=modelarf] coordinates {(9.22,0.64)}
node[anchor=west, xshift=2pt, yshift=2pt, font=\scriptsize, text=modelarf, fill=white, fill opacity=0.78, text opacity=1, inner sep=0.9pt] {7};
\addplot[only marks, mark=*, mark size=2.9pt, draw=modelbayesnet, fill=modelbayesnet] coordinates {(1.10,0.60)}
node[anchor=west, xshift=2pt, yshift=3pt, font=\scriptsize, text=modelbayesnet, fill=white, fill opacity=0.78, text opacity=1, inner sep=0.9pt] {11};
\addplot[only marks, mark=*, mark size=2.6pt, draw=modelctgan, fill=modelctgan] coordinates {(9.08,0.55)}
node[anchor=south east, xshift=3pt, yshift=-9pt, font=\scriptsize, text=modelctgan, fill=white, fill opacity=0.78, text opacity=1, inner sep=0.9pt] {6};
\addplot[only marks, mark=*, mark size=2.6pt, draw=modelforestdiffusion, fill=modelforestdiffusion] coordinates {(7.25,0.44)}
node[anchor=west, xshift=2pt, yshift=-1pt, font=\scriptsize, text=modelforestdiffusion, fill=white, fill opacity=0.78, text opacity=1, inner sep=0.9pt] {4};
\addplot[only marks, mark=*, mark size=2.9pt, draw=modelrealtabformer, fill=modelrealtabformer] coordinates {(54.93,0.76)}
node[anchor=east, xshift=-2pt, font=\scriptsize, text=modelrealtabformer, fill=white, fill opacity=0.78, text opacity=1, inner sep=0.9pt] {1};
\addplot[only marks, mark=*, mark size=2.6pt, draw=modeltabbyflow, fill=modeltabbyflow] coordinates {(6.67,0.57)}
node[anchor=north east, xshift=-3pt, yshift=5pt, font=\scriptsize, text=modeltabbyflow, fill=white, fill opacity=0.78, text opacity=1, inner sep=0.9pt] {10};
\addplot[only marks, mark=*, mark size=2.6pt, draw=modeltabddpm, fill=modeltabddpm] coordinates {(2.80,0.46)}
node[anchor=north east, xshift=-2pt, yshift=-2pt, font=\scriptsize, text=modeltabddpm, fill=white, fill opacity=0.78, text opacity=1, inner sep=0.9pt] {2};
\addplot[only marks, mark=*, mark size=2.6pt, draw=modeltabdiff, fill=modeltabdiff] coordinates {(7.56,0.59)}
node[anchor=west, xshift=2pt, yshift=3pt, font=\scriptsize, text=modeltabdiff, fill=white, fill opacity=0.78, text opacity=1, inner sep=0.9pt] {5};
\addplot[only marks, mark=*, mark size=2.6pt, draw=modeltabpfgen, fill=modeltabpfgen] coordinates {(0.05,0.53)}
node[anchor=west, xshift=2pt, yshift=-4pt, font=\scriptsize, text=modeltabpfgen, fill=white, fill opacity=0.78, text opacity=1, inner sep=0.9pt] {8};
\addplot[only marks, mark=*, mark size=2.6pt, draw=modeltabsyn, fill=modeltabsyn] coordinates {(23.07,0.57)}
node[anchor=west, xshift=2pt, yshift=3pt, font=\scriptsize, text=modeltabsyn, fill=white, fill opacity=0.78, text opacity=1, inner sep=0.9pt] {3};
\addplot[only marks, mark=*, mark size=2.6pt, draw=modeltvae, fill=modeltvae] coordinates {(4.12,0.54)}
node[anchor=north west, xshift=-3pt, yshift=-3pt, font=\scriptsize, text=modeltvae, fill=white, fill opacity=0.78, text opacity=1, inner sep=0.9pt] {9};
\end{axis}
\end{tikzpicture}
  \par\vspace{-0.25cm}
\end{minipage}\hspace*{0.006\textwidth}%
\begin{minipage}[t]{0.49\linewidth}
  \vspace{0pt}
  \centering
  \begin{tikzpicture}
\definecolor{modelrealtabformer}{HTML}{6F4C9B}
\definecolor{modeltvae}{HTML}{882255}
\definecolor{modelforestdiffusion}{HTML}{66CCEE}
\definecolor{modeltabddpm}{HTML}{004488}
\definecolor{modeltabsyn}{HTML}{4477AA}
\definecolor{modeltabdiff}{HTML}{009988}
\definecolor{modelctgan}{HTML}{CC3311}
\definecolor{modelarf}{HTML}{EE7733}
\definecolor{modelbayesnet}{HTML}{666666}
\definecolor{modeltabpfgen}{HTML}{DDAA33}
\definecolor{modeltabbyflow}{HTML}{CC6677}
\begin{axis}[
width=\linewidth,
height=4.35cm,
xmode=log,
xmin=0.5, xmax=70,
ymin=0.38, ymax=0.80,
xtick={0.5,1,2,5,10,50},
xticklabels={0.5,1,2,5,10,50},
minor tick num=0,
grid=both, grid style={draw=black!9}, major grid style={draw=black!12},
axis line style={draw=black!55}, tick style={draw=black!55},
xlabel={Median generation time (s/1k rows)},
ylabel={},
ticklabel style={font=\scriptsize},
label style={font=\footnotesize},
clip=false
]
\addplot[only marks, mark=*, mark size=2.9pt, draw=modelarf, fill=modelarf] coordinates {(0.84,0.64)}
node[anchor=west, xshift=2pt, yshift=3pt, font=\scriptsize, text=modelarf, fill=white, fill opacity=0.78, text opacity=1, inner sep=0.9pt] {7};
\addplot[only marks, mark=*, mark size=2.7pt, draw=modelbayesnet, fill=modelbayesnet] coordinates {(1.92,0.60)}
node[anchor=south west, xshift=2pt, yshift=0pt, font=\scriptsize, text=modelbayesnet, fill=white, fill opacity=0.78, text opacity=1, inner sep=0.9pt] {11};
\addplot[only marks, mark=*, mark size=2.6pt, draw=modelctgan, fill=modelctgan] coordinates {(0.94,0.55)}
node[anchor=north west, xshift=-2pt, yshift=-3pt, font=\scriptsize, text=modelctgan, fill=white, fill opacity=0.78, text opacity=1, inner sep=0.9pt] {6};
\addplot[only marks, mark=*, mark size=2.6pt, draw=modelforestdiffusion, fill=modelforestdiffusion] coordinates {(0.72,0.44)}
node[anchor=west, xshift=2pt, yshift=-2pt, font=\scriptsize, text=modelforestdiffusion, fill=white, fill opacity=0.78, text opacity=1, inner sep=0.9pt] {4};
\addplot[only marks, mark=*, mark size=2.9pt, draw=modelrealtabformer, fill=modelrealtabformer] coordinates {(10.45,0.76)}
node[anchor=west, xshift=2pt, font=\scriptsize, text=modelrealtabformer, fill=white, fill opacity=0.78, text opacity=1, inner sep=0.9pt] {1};
\addplot[only marks, mark=*, mark size=2.6pt, draw=modeltabbyflow, fill=modeltabbyflow] coordinates {(3.19,0.57)}
node[anchor=west, xshift=2pt, yshift=-1pt, font=\scriptsize, text=modeltabbyflow, fill=white, fill opacity=0.78, text opacity=1, inner sep=0.9pt] {10};
\addplot[only marks, mark=*, mark size=2.6pt, draw=modeltabddpm, fill=modeltabddpm] coordinates {(6.22,0.46)}
node[anchor=west, xshift=2pt, yshift=-1pt, font=\scriptsize, text=modeltabddpm, fill=white, fill opacity=0.78, text opacity=1, inner sep=0.9pt] {2};
\addplot[only marks, mark=*, mark size=2.6pt, draw=modeltabdiff, fill=modeltabdiff] coordinates {(1.60,0.59)}
node[anchor=south west, xshift=-8pt, yshift=-2pt, font=\scriptsize, text=modeltabdiff, fill=white, fill opacity=0.78, text opacity=1, inner sep=0.9pt] {5};
\addplot[only marks, mark=*, mark size=2.7pt, draw=modeltabpfgen, fill=modeltabpfgen] coordinates {(45.29,0.53)}
node[anchor=east, xshift=-2pt, font=\scriptsize, text=modeltabpfgen, fill=white, fill opacity=0.78, text opacity=1, inner sep=0.9pt] {8};
\addplot[only marks, mark=*, mark size=2.6pt, draw=modeltabsyn, fill=modeltabsyn] coordinates {(0.71,0.57)}
node[anchor=east, xshift=0pt, yshift=-6pt, font=\scriptsize, text=modeltabsyn, fill=white, fill opacity=0.78, text opacity=1, inner sep=0.9pt] {3};
\addplot[only marks, mark=*, mark size=2.6pt, draw=modeltvae, fill=modeltvae] coordinates {(1.89,0.54)}
node[anchor=north west, xshift=2pt, yshift=2pt, font=\scriptsize, text=modeltvae, fill=white, fill opacity=0.78, text opacity=1, inner sep=0.9pt] {9};
\end{axis}
\end{tikzpicture}
  \par\vspace{-0.25cm}
\end{minipage}
\captionsetup{skip=2pt,justification=raggedright,singlelinecheck=false}

\caption{Query-centric fidelity versus runtime. The left panel shows median
training time on a log scale, and the middle panel shows median generation
time per 1k rows. Numbers mean different models align with Figure~\ref{fig:query_rank_stability_ai_model_lines}. 
Details in Table~\ref{tab:benchmark_overall_real_main}. 
}
\label{fig:query_train_time_tradeoff}\label{fig:query_generation_time_tradeoff}
\end{minipage}\hspace*{0.006\textwidth}%
\begin{minipage}[t]{0.348\textwidth}
  \vspace{0pt}
  \centering
  \makebox[\linewidth][l]{%
  \hspace*{-0.2\textwidth}%
  \begin{tikzpicture}
\definecolor{modelrealtabformer}{HTML}{6F4C9B}
\definecolor{modelarf}{HTML}{EE7733}
\definecolor{modelbayesnet}{HTML}{666666}
\definecolor{modeltabdiff}{HTML}{009988}
\definecolor{modeltabsyn}{HTML}{4477AA}
\definecolor{modeltabbyflow}{HTML}{CC6677}
\definecolor{modelctgan}{HTML}{CC3311}
\definecolor{modeltvae}{HTML}{882255}
\definecolor{modeltabpfgen}{HTML}{DDAA33}
\definecolor{modeltabddpm}{HTML}{004488}
\definecolor{modelforestdiffusion}{HTML}{66CCEE}
\begin{axis}[
  width=\linewidth, height=4.35cm,
  xmin=.65, xmax=11.35, ymin=.35, ymax=11.65,
  xtick={1,3,5,7,9,11}, ytick={1,...,11},
  yticklabels={\textcolor{modelrealtabformer}{RTF (1)},\textcolor{modelarf}{ARF (2)},\textcolor{modelbayesnet}{Bayes (3)},\textcolor{modeltabdiff}{T-Diff (4)},\textcolor{modeltabsyn}{T-Syn (5)},\textcolor{modeltabbyflow}{T-Flow (6)},\textcolor{modelctgan}{CTGAN (7)},\textcolor{modeltvae}{TVAE (8)},\textcolor{modeltabpfgen}{TPF (9)},\textcolor{modeltabddpm}{T-DDPM (10)},\textcolor{modelforestdiffusion}{F-Diff (11)}},
  y dir=reverse, xlabel={Rank (1 = best)},
  xticklabel style={font=\scriptsize},
  yticklabel style={font=\tiny,text width=1.65cm,align=right},
  label style={font=\footnotesize}, tick style={draw=black!55},
  axis line style={draw=black!55}, grid=major,
  major grid style={draw=black!10}, clip=false, enlargelimits=false,
]
\foreach \y/\lo/\hi/\med/\col/\delta/\a/\b/\c/\d/\e/\f/\g in {
  1/1/1/1/modelrealtabformer/0/1/1/1/1/1/1/1,
  2/2/2/2/modelarf/0/2/2/2/2/2/2/2,
  3/3/3/3/modelbayesnet/0/3/3/3/3/3/3/3,
  4/4/4/4/modeltabdiff/0/4/4/4/4/4/4/4,
  5/5/6/5/modeltabsyn/1/6/6/5/5/5/5/5,
  6/5/6/6/modeltabbyflow/1/5/5/6/6/6/6/6,
  7/7/8/7/modelctgan/1/7/8/7/7/8/7/7,
  8/7/9/8/modeltvae/2/9/7/8/8/7/8/8,
  9/8/9/9/modeltabpfgen/1/8/9/9/9/9/9/9,
  10/10/11/10/modeltabddpm/1/11/10/10/10/10/10/10,
  11/10/11/11/modelforestdiffusion/1/10/11/11/11/11/11/11}
  {
   \pgfmathparse{\y-0.035}\let\ya\pgfmathresult
   \pgfmathparse{\y-0.018}\let\yb\pgfmathresult
   \pgfmathparse{\y+0.018}\let\yc\pgfmathresult
   \pgfmathparse{\y+0.035}\let\yd\pgfmathresult
   \pgfmathparse{\y+0.052}\let\ye\pgfmathresult
   \pgfmathparse{\y+0.069}\let\yf\pgfmathresult
   \pgfmathparse{\y+0.086}\let\yg\pgfmathresult
   \addplot[thick,draw=black!45] coordinates {(\lo,\y) (\hi,\y)};
   \edef\smallstyle{only marks,mark=*,mark size=1.25pt,draw=white,line width=.25pt,fill=\col}
   \expandafter\addplot\expandafter[\smallstyle]
     coordinates {(\a,\ya) (\b,\yb) (\c,\y) (\d,\yc) (\e,\yd) (\f,\ye) (\g,\yf)};
   \edef\medstyle{only marks,mark=*,mark size=2.05pt,draw=white,line width=.35pt,fill=\col}
   \expandafter\addplot\expandafter[\medstyle] coordinates {(\med,\y)};
   }
\end{axis}
\end{tikzpicture}%
}
  \par\vspace{0.3cm}
  \captionsetup{skip=2pt,justification=raggedright,singlelinecheck=false}
  \caption{Generator ranks across seven AI workload builders.
  Segments show min--max rank; large dots show medians.}
  \label{fig:query_rank_stability_ai_model_lines}
\end{minipage}
\end{figure*}

\subsection{The Stability of Query Generation Pipeline}

We assess whether the choice of LLMs used for SQL realization affects the benchmark's comparative conclusions. 
We regenerate the query sets using three AI builders, including Claude Haiku-4.5, GLM-5.3 and GLM-5.3Flash, running three times for the two GLM models, under the same template library and dataset schemas.
Figure~\ref{fig:query_rank_stability_ai_model_lines} shows that the same four generative models occupy the top four positions across
all seven query sets. The overall query-centric fidelity score,
averaged across generative models, ranges from 0.572 to 0.588,
an absolute spread of 0.016. These results indicate that the
identification of the leading generative models is insensitive to
the choice of SQL builder within the tested configurations.
Complete scores are reported in
Appendix Table~\ref{tab:appendix_query_rank_stability_v11_ai_model_scores}.

\subsection{Structural Coverage of the Generated Query Workload}
\label{sec:structural_query_coverage}

Beyond ranking stability, we audit whether the complete generated workload
covers the structural space represented by public analytical SQL workloads.
We retain all $2{,}763$ valid generated queries in the default query version. 
After schema-specific normalization and SQL parsing with SQLGlot~\citep{sqlglot}, these queries
contain $44$ unique query shapes, while the source corpus contains $286$ unique shapes from $320$ valid SQL queries. For each source shape, we find its closest generated shape using
Tree Similarity of Edit Distance (TSED)~\citep{song-etal-2024-revisiting}, an AST-based metric that
normalizes tree-edit distance into a similarity score in [0, 1], with higher values indicating greater
structural similarity. The resulting
Generated$\to$Source TSED score is $0.702$, indicating substantial structural overlap between the generated and public workloads, while still leaving non-trivial source-side structures uncovered.

Complementing this structural view, we
compare the query-score distributions of the official ClickBench workload ($n=39$) and TQBClick
($n=144$). We first normalize query scores to $[0,1]$ and express query ranks as percentiles, making
the two workloads comparable despite their different numbers of queries. We then compute the
normalized 1-Wasserstein distance as the area of the absolute difference between their empirical
quantile functions over the full percentile range. The resulting distance is 0.1702, corresponding to
an average normalized-score separation of 17.02\% points across ranked quantiles. More details lie in Appendix~\ref{app:tsed_tree_construction}.

\section{Conclusions}

\BenchmarkName{} reframes synthetic tabular data evaluation around analytical query answers. Instead of asking only whether a synthetic table matches the real table under aggregate statistical distances, it asks whether the table preserves the answers to queries that downstream users are likely to run. We instantiate this idea with \NumDatasets{} datasets and \NumModels{} generative models. 
More concretely, \BenchmarkName{} contributes a workload-grounded evaluation framework that
distills recurring analytical structures into 44 reusable templates across five query families,
automatically grounds them to diverse dataset schemas, and evaluates synthetic data through
the resulting query answers at both query and family levels.

Our evaluation shows that distance-based fidelity does not reliably predict query-centric fidelity. 
Beyond an aggregate score, \BenchmarkName{} localizes where synthetic data fails to preserve
analytical behavior, revealing failures in high-cardinality support, local conditional structure,
and extreme tails.
In our released repository, users can also invoke the \BenchmarkName{} API following the provided requirements to automatically generate an executable query workload for a target dataset. This enables workload-aware evaluation, model selection, and failure diagnosis beyond a single global resemblance score.

\subsection*{AI use statement}

During research and manuscript preparation, we used LLM-based assistants and coding agents to assist with software implementation and debugging, language editing and writing refinement, and the preparation or refinement of selected conceptual figures. These tools were used as assistants rather than as autonomous decision makers: all generated code, text, figures, and technical suggestions were reviewed and, where necessary, revised by the authors.

AI is also used as a constrained component of the benchmark construction pipeline. Within the benchmark, an LLM agent receives a dataset profile and a fixed template library, selects applicable templates and dataset columns, and produces the structured query specification under predefined constraints. The LLM does not freely generate the final SQL. The specification is validated against the template contract and deterministically rendered into executable SQL; invalid, non-executable, or contract-violating outputs are rejected and regenerated under the same constraints. Profiling, binding, validation, SQL rendering, scoring, and benchmark aggregation are deterministic and are not delegated to the LLM.

To evaluate the robustness and external validity of the agent-grounded query generation pipeline, we conduct three dedicated analyses. First, we run the same grounding pipeline with multiple AI models to examine whether its behavior and downstream conclusions are robust to the choice of underlying model. Second, we apply queries produced by our agent to ClickBench and compare their workload characteristics with those of the real analytical SQL workload provided by ClickBench. Third, we measure structural similarity between our generated queries and all available SQL queries collected from the 12 source workloads using Tree Similarity of Edit Distance (TSED), providing an external assessment of how closely the generated workload reflects the structural space of real-world analytical SQL.

\subsection*{Ethics statement}

This study evaluates synthetic tabular-data generators using publicly available datasets and does not recruit human participants or collect new personal data. The benchmark is intended to measure analytical fidelity and does not establish privacy protection. Synthetic data that achieves high fidelity may still retain sensitive or identifying information, so benchmark scores should not be interpreted as evidence of privacy, anonymity, or regulatory compliance. Users of the released artifacts should follow the licenses, terms, and usage restrictions associated with the original data sources and should conduct separate privacy and risk assessments before applying the benchmark or its outputs to sensitive data.

\subsection*{Reproducibility statement}

We provide the benchmark code, query-template specifications, grounding and validation procedures, model adapters, experiment configurations, and the scripts and data used to generate the reported figures and tables in the accompanying public repositories: \href{https://github.com/TabQueryBench/TabQueryBench}{https://github.com/TabQueryBench/TabQueryBench} and \href{https://huggingface.co/datasets/TabQueryBench2026/TabQueryBench/tree/main}{https://huggingface.co/datasets/TabQueryBench2026/TabQueryBench}. The evaluation protocol specifies the dataset splits, model interfaces, sampling budgets, query-grounding constraints, scoring rules, runtime measurements, and aggregation procedures. Each generator is isolated in a Docker environment and evaluated through a common fit--sample--restore--validate interface. We record random seeds where supported, together with model versions, prompts, inference settings, and other configurations needed to reproduce the experiments. The appendix documents the dataset roster, model settings, API usage, query-generation protocol, and scoring standard required to reproduce the reported results.

\begingroup
\raggedright
\bibliographystyle{iclr2027_conference}
\bibliography{tabquerybench_references}
\endgroup

\clearpage
\appendix

\noindent{\LARGE\bfseries Appendix}

\vspace{1.4em}
\noindent
{\large\bfseries \BenchmarkName: A Query-Centric Benchmark for Synthetic Tabular Data}

\vspace{1.2em}
\noindent{\Large\bfseries Table of Contents}
\par\vspace{-0.45em}
\noindent\rule{\textwidth}{0.4pt}
\vspace{0.35em}

\begingroup
\footnotesize
\setlength{\parindent}{0pt}
\setlength{\parskip}{0.14em}
\newcommand{\appToCEntry}[4]{%
  \noindent\makebox[#1][l]{#2}%
  \hyperref[#3]{#4}\dotfill\hyperref[#3]{\pageref*{#3}}\par}
\newcommand{\appToCMain}[3]{%
  \vspace{0.25em}%
  \appToCEntry{2.3em}{\bfseries #1}{#2}{\bfseries #3}}
\newcommand{\appToCSub}[3]{%
  \appToCEntry{4.0em}{\hspace*{1.5em}#1}{#2}{#3}}
\appToCMain{A}{app:related_works}{Related Works}
\appToCSub{A.1}{app:related_tabular_models}{Tabular Generative Models}
\appToCSub{A.2}{app:related_synthetic_benchmarks}{Tabular Generative Model Benchmarks}
\appToCSub{A.3}{app:related_query_benchmarks}{Database Query Benchmarks}
\appToCSub{A.4}{app:related_fidelity_gap}{Distance-Based vs. Query-Centric Fidelity}
\appToCMain{B}{app:experiment}{Synthetic Data Experiment}
\appToCSub{B.1}{app:evaluation_scope}{Benchmark Scope}
\appToCSub{B.2}{app:evaluation_settings_details}{Experimental Settings Details}
\appToCSub{B.3}{app:dataset_catalog}{Dataset Catalog}
\appToCSub{B.4}{app:models_catalog}{Models Catalog}
\appToCMain{C}{app:query_templates_families}{Query Templates and Five Query Families}
\appToCSub{C.1}{app:metric_design_overview}{Query-Centric Fidelity Overview}
\appToCSub{C.2}{app:workload_sources}{Workload Sources}
\appToCSub{C.3}{app:template_taxonomy_catalog}{44-Template Taxonomy Catalog}
\appToCSub{C.4}{app:query_families}{Five Query Families}
\appToCSub{C.4.1}{app:experiment_subgroup}{Subgroup}
\appToCSub{C.4.2}{app:experiment_conditional}{Conditional}
\appToCSub{C.4.3}{app:tail_rarity_diagnostics}{Tail / Rarity}
\appToCSub{C.4.4}{app:experiment_missingness}{Missingness}
\appToCSub{C.4.5}{app:experiment_cardinality_range}{Cardinality / Range}
\appToCMain{D}{app:query_generation}{Query Grounding}
\appToCSub{D.1}{app:query_generation_overview}{Pipeline Overview}
\appToCSub{D.2}{app:query_template_construction}{Stage 1: Template Construction}
\appToCSub{D.3}{app:query_grounding}{Stage 2: Dataset Grounding and SQL Realization}
\appToCSub{D.4}{app:agent_api_usage}{Agent API Usage by Dataset}
\appToCMain{E}{app:query_scoring}{Query Scoring Standard}
\appToCSub{E.1}{app:scoring_applicability}{Reporting and Applicability Rules}
\appToCSub{E.2}{app:classical_distance_fidelity}{Classical Distance-Based Scoring Rules}
\appToCMain{F}{app:overall_evaluation_analysis}{Overall Evaluation Analysis}
\appToCSub{F.1}{app:overall_benchmark_summary}{Overall Benchmark Summary}
\appToCSub{F.2}{app:overall_query_score_heatmap}{Overall Query Score Heatmap}
\appToCMain{G}{app:extended_experiments}{Extended Experiments}
\appToCSub{G.1}{app:additional_experiments}{Stability Across Query-Workload Rebuilds}
\appToCSub{G.2}{app:structural_query_coverage}{Structural Coverage of the Full Generated Workload}
\appToCMain{H}{app:discussion_future}{Discussion and Future Works}
\endgroup

\vfill
\noindent\rule{\textwidth}{0.4pt}

\clearpage
\section{Related Works}
\label{app:related_works}

This appendix expands the three strands of related work introduced in
Section~\ref{sec:related_works}: tabular generative models, their evaluation
benchmarks, and database query benchmarks. These strands distinguish how
synthetic tables are generated, which properties are evaluated, and how
analytical queries can provide reusable evaluation tasks.

\subsection{Tabular Generative Models}
\label{app:related_tabular_models}

Tabular synthesis models records whose columns may have different types
and semantics. Preserving plausible values in each column is one part of
the problem; maintaining dependencies between columns and support for
less frequent values is also relevant to downstream use. Different
generative approaches encode these relationships through different model
structures. Their architectural differences motivate evaluation across
multiple analytical properties rather than reliance on one summary score.

\textbf{Bayesian and tree-based approaches.}
DataSynthesizer uses a Bayesian-network representation to model
dependencies between attributes~\citep{ping2017datasynthesizer}.
Adversarial Random Forests (ARF) instead use tree-based partitions for
density estimation and synthetic-data generation~\citep{watson2023arf}.
These approaches provide alternatives to neural synthesis and are useful
comparators when assessing whether different representations preserve
grouped aggregates and conditional structure.

\textbf{Adversarial and variational models.}
CTGAN and TVAE adapt adversarial and variational learning to mixed-type
tabular data~\citep{xu2019ctgan}. CTGAN uses conditional generation,
whereas TVAE learns a latent-variable representation. Both are established
baselines for tabular synthesis. Our evaluation concerns the analytical
answers supported by their outputs, rather than their training objectives.

\textbf{Diffusion, transformer-based, and flow-matching models.}
TabDDPM, TabSyn, and TabDiff apply diffusion-based generation to tabular
data, with different representations of mixed-type variables
\citep{kotelnikov2023tabddpm,zhang2024tabsyn,shi2025tabdiff}.
ForestDiffusion uses tree-based components in diffusion and flow-based
generative modeling~\citep{jolicoeurmartineau2023forestdiffusion}.
REaLTabFormer uses transformer-based generation, while TabPFGen explores
generation using pretrained tabular foundation models
\citep{solatorio2023realtabformer,ma2024tabpfgen}.
TabbyFlow provides a flow-matching approach to tabular synthesis
\citep{guzmancordero2025tabbyflow}.
These mechanisms need not preserve the same analytical properties to the
same degree. \BenchmarkName{} therefore evaluates their generated tables
through a common set of dataset-grounded analytical queries.

\subsection{Tabular Generative Model Benchmarks}
\label{app:related_synthetic_benchmarks}

\textbf{Evaluation frameworks and standardized comparisons.}
Synthcity and SynthEval provide reusable tools for evaluating synthetic
data through collections of metrics
\citep{qian2023synthcity,lautrup2024syntheval}.
The data-centric evaluation of Hansen et al. and SDGym support comparisons
of tabular synthesis methods across datasets
\citep{hansen2023reimagining,sdgym2026}.
Such resources make distributional fidelity, predictive utility, privacy,
or computational cost easier to assess systematically. These dimensions
answer different questions: distributional metrics compare statistical
properties, predictive utility measures usefulness for specified learning
tasks, and privacy evaluation concerns information exposure. None of
these concepts should be treated as interchangeable with the preservation
of a particular analytical query answer.

\textbf{Specialized evaluation settings.}
SDNist and the associated Diverse Community Data resource support
evaluation in public data-release scenarios
\citep{sen2023diverse,task2023sdnist}.
SyntheRela addresses multi-table synthesis, while SynthRO focuses on
health-oriented synthetic-data validation
\citep{hudovernik2024benchmarking,santangelo2025synthro}.
Sidorenko et al. consider multidimensional evaluation of mixed and
contextual data, and Tao et al. benchmark differentially private
tabular-data generation
\citep{sidorenko2025multidim,tao2021dpbenchmark}.
TabArena is an adjacent benchmark for tabular predictive learning rather
than an evaluation of synthetic-table generators
\citep{erickson2025tabarena}. The interpretation of each comparison must
therefore account for its target setting and evaluation object.

\textbf{Structural fidelity and TabStruct.}
TabStruct explicitly adds structural fidelity to conventional evaluation
dimensions and studies 13 tabular generative models across 29 datasets
\citep{jiang2026tabstruct}. When a ground-truth structural causal model
is available, it evaluates the preservation of implied conditional
independence statements. For datasets without this information, its
\emph{global utility} metric treats each variable as a prediction target,
predicts it from the remaining variables, and aggregates normalized
predictive utility across targets. TabStruct reports empirical alignment
between global utility and its conditional-independence-based structural
scores. This is a broader structural assessment than predictive utility
for a single designated target.

Our evaluation addresses a complementary object: the answers returned by
dataset-grounded analytical SQL queries. In \BenchmarkName{}, a query
family identifies analytical properties such as subgroup summaries, tail
coverage, or missingness patterns. It is not a causal graph, and its
conditional queries should not be equated with causal conditional
independence tests. Likewise, a high query-centric fidelity score does
not establish causal-structure recovery. The distinction concerns what
is directly measured: prediction across variables and causal structural
consistency in TabStruct, versus analytical answer preservation in
\BenchmarkName{}.

\begingroup
\small
\setlength{\tabcolsep}{4pt}
\renewcommand{\arraystretch}{1.12}
\begin{longtable}{@{}p{0.35\textwidth}p{\dimexpr0.65\textwidth-2\tabcolsep\relax}@{}}
\caption{\textbf{Evaluation settings and objects of related resources.}
Entries summarize the role of each resource, rather than asserting the
absence of capabilities not listed here.}
\label{tab:appendix_benchmark_resource_settings}\\
\toprule
Resource & Evaluation setting and object \\
\midrule
\endfirsthead
\multicolumn{2}{l}{\textit{Table~\thetable{} continued.}}\\
\toprule
Resource & Evaluation setting and object \\
\midrule
\endhead
\midrule
\multicolumn{2}{r}{\textit{Continued on next page.}}\\
\endfoot
\bottomrule
\endlastfoot
Synthcity~\citep{qian2023synthcity} & Reusable synthetic-data framework with multiple evaluation metrics. \\
SynthEval~\citep{lautrup2024syntheval} & Synthetic-data evaluation through a configurable metric suite. \\
Data-centric benchmark~\citep{hansen2023reimagining} & Comparative evaluation of synthetic tabular data quality. \\
SDGym~\citep{sdgym2026} & Benchmarking infrastructure for tabular synthesis methods. \\
SDNist~\citep{sen2023diverse,task2023sdnist} & Evaluation associated with public data-release scenarios. \\
SyntheRela~\citep{hudovernik2024benchmarking} & Multi-table synthetic data and relational fidelity. \\
SynthRO~\citep{santangelo2025synthro} & Health-oriented synthetic-data validation. \\
Sidorenko et al.~\citep{sidorenko2025multidim} & Multidimensional evaluation of mixed and contextual data. \\
DP benchmark~\citep{tao2021dpbenchmark} & Differentially private tabular synthesis and privacy--utility evaluation. \\
TabArena~\citep{erickson2025tabarena} & Tabular predictive learning, rather than synthetic-data generation. \\
TabStruct~\citep{jiang2026tabstruct} & Structural fidelity through conditional independence and global utility, alongside conventional dimensions. \\
\BenchmarkName{} & Analytical SQL answer preservation across five query families, alongside distance-based metrics and cost. \\
\end{longtable}
\endgroup

\subsection{Database Query Benchmarks}
\label{app:related_query_benchmarks}

\textbf{Parameterized decision-support queries.}
TPC-H and TPC-DS encode analytical workloads through schemas,
parameterized queries, and controlled execution procedures
\citep{poess2000new,nambiar2006making,poess2007workload,tpchspec,tpcdsspec}.
Their central role is to evaluate database systems under specified
decision-support workloads. For our purposes, their reusable query
patterns provide public provenance for analytical operations, including
filtering, grouping, aggregation, and ordering.

\textbf{Public analytical workloads.}
ClickBench provides a web-analytics dataset and SQL queries for analytical
database comparisons~\citep{clickbench2022}.
H2O db-benchmark compares data-processing systems through operations such
as grouping and joins~\citep{h2odbbenchmark}.
RTABench targets real-time application analytics
\citep{rtabench2025}. These resources offer additional examples of
recurring analytical operations. Their original workloads may include
relational or system-specific features beyond the single-table setting
evaluated in the current \BenchmarkName{} release.

\textbf{From system workloads to synthetic-data evaluation.}
\BenchmarkName{} extracts reusable analytical templates from public SQL
sources and grounds applicable templates to each dataset. It does not
reproduce the complete database benchmarks or their execution protocols.
The evaluation asks whether synthetic data preserves the answers induced
by these grounded queries. Source attribution and construction details
are provided in Appendix~\ref{app:query_template_construction};
grounding and output-specific scoring are described in
Appendices~\ref{app:query_grounding} and~\ref{app:query_scoring}.

\subsection{Distance-Based vs. Query-Centric Fidelity}
\label{app:related_fidelity_gap}

This appendix gives additional context for the motivating ClickBench
experiment in Section~\ref{sec:mismatch-two-types}. The goal of this
experiment is not to replace the full \BenchmarkName{} evaluation, but to
test in a realistic analytical workload whether distance-based fidelity and
query-centric fidelity expose different behavior. It also checks whether
the queries generated by our template-grounding pipeline resemble real
analytical queries in terms of their score distribution.

\textbf{Experimental purpose.}
We use ClickBench~\citep{clickbench2022} as a real-world analytical
workload because it contains public queries written for analytical database
evaluation. For each synthetic table, we compare two views of fidelity. The
first view is the classical distance-based view, which summarizes global
statistical similarity between the real and synthetic tables. The second
view is the query-centric view, which asks whether the same analytical SQL
queries return similar answers on the real and synthetic tables. The
scoring rules for both views follow the definitions in
Appendix~\ref{app:query_scoring}, with the distance-based metrics detailed
in Appendix~\ref{app:classical_distance_fidelity}.

\textbf{Official queries and generated queries.}
We compare the official ClickBench queries with QClick2.4, our generated
ClickBench-style query workload produced by the same query-construction
pipeline used in \BenchmarkName{}. Because the two workloads contain
different numbers of scorable queries, we do not compare them by raw query
index. Instead, we sort queries in each workload from high to low by their
mean query score across the evaluated synthetic generators and align the
two curves by ranked percentile. This lets us compare the score profile of
real analytical queries with the score profile of our generated queries
under the same synthetic-data evaluation setting.

\textbf{Interpretation.}
This experiment serves two roles. First, it illustrates that distance-based
fidelity and query-centric fidelity can diverge in a realistic workload: a
synthetic table may look close under global distance metrics while still
failing to preserve analytical query answers. Second, it provides a sanity
check for our generated workload. If QClick2.4 produced only trivial or
unrepresentative queries, its ranked score curve would behave very
differently from the official ClickBench workload. Instead, comparing the
two ranked curves helps assess whether our generated queries occupy a
similar analytical difficulty regime while still providing broader
coverage.

\clearpage
\section{Synthetic Data Experiment}
\label{app:experiment}

\subsection{Benchmark Scope}
\label{app:evaluation_scope}
The released benchmark evaluates \NumModels{} synthetic tabular generators over \NumDatasets{} datasets. The dataset and model rosters are kept visible in one place below.

\begin{figure*}[!b]
\centering
\includegraphics[width=\textwidth]{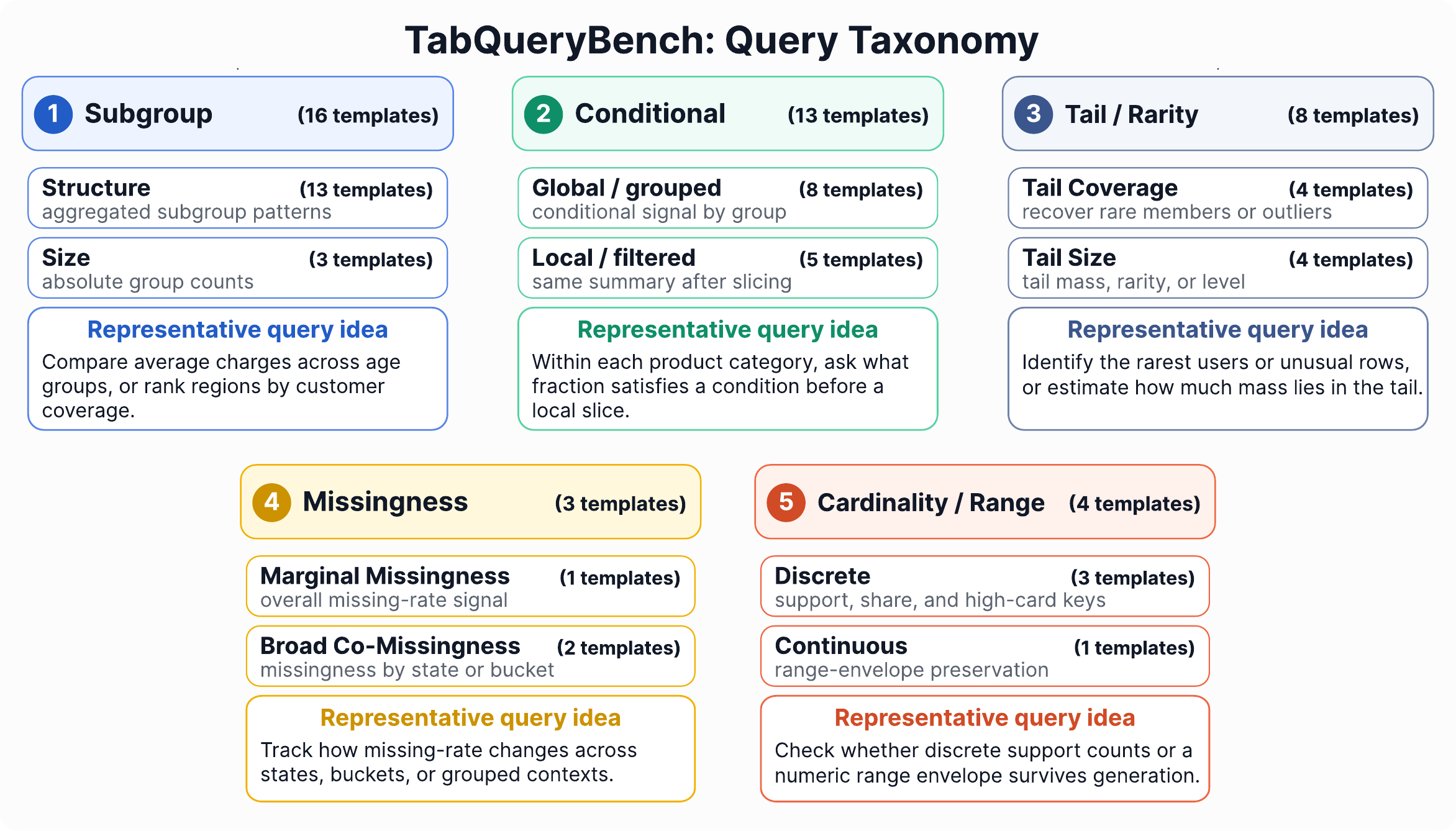}
\caption{The Query Template Taxonomy.
}
\label{fig:stage1-template-taxonomy}
\end{figure*}

\begin{table}[t]
\centering
\captionsetup{skip=3pt}
\begingroup
\definecolor{BindingBlue}{HTML}{1397B8}
\definecolor{SkeletonBind}{HTML}{D46B08}
\definecolor{FamilySubgroup}{HTML}{225CC7}
\definecolor{FamilyConditional}{HTML}{0F8F67}
\definecolor{FamilyTail}{HTML}{3B568C}
\definecolor{FamilyMissing}{HTML}{B17A00}
\definecolor{FamilyCardinality}{HTML}{D14B26}
\definecolor{ExampleFrame}{HTML}{D8DEE8}
\definecolor{ExampleFill}{HTML}{FBFCFE}
\definecolor{MutedText}{HTML}{5C6470}

\newcommand{\groundedtoken}[1]{\textcolor{BindingBlue}{#1}}
\newcommand{\bindmap}[1]{\textcolor{SkeletonBind}{[\texttt{#1}]}}
\newcommand{\fieldlabel}[2]{\textcolor{#1}{\bfseries\scriptsize #2}\par\vspace{0.03em}}
\newcommand{\questiontext}[1]{{\footnotesize #1\par}}
\newcommand{\groundedquestiontext}[1]{{\footnotesize\itshape #1\par}}
\newcommand{\sqltext}[1]{%
  {\ttfamily\scriptsize
  \setlength{\parskip}{0pt}%
  #1\par}%
}
\newcommand{\familycard}[4]{%
  \begingroup
  \setlength{\fboxsep}{3.6pt}%
  \setlength{\fboxrule}{0.45pt}%
  \fcolorbox{#1}{#1!7}{%
    \begin{minipage}[t]{\dimexpr\linewidth-2\fboxsep-2\fboxrule\relax}
      \vspace{0pt}
      {\color{#1}\bfseries #2}\par
      \vspace{0.12em}
      {\scriptsize\textcolor{MutedText}{#3}}\par
      {\scriptsize\textcolor{MutedText}{#4}}\par
      \vspace{0pt}
    \end{minipage}%
  }%
  \endgroup
}
\newcommand{\examplecard}[4]{%
  \begingroup
  \setlength{\fboxsep}{4pt}%
  \setlength{\fboxrule}{0.35pt}%
  \fcolorbox{ExampleFrame}{ExampleFill}{%
    \begin{minipage}[t]{\dimexpr\linewidth-2\fboxsep-2\fboxrule\relax}
      \fieldlabel{#1}{QUESTION}
      \questiontext{#2}
      \vspace{0.18em}
      \fieldlabel{#1}{GROUNDED QUESTION}
      \groundedquestiontext{#3}
      \vspace{0.18em}
      \fieldlabel{#1}{GROUNDED SQL}
      \sqltext{#4}
    \end{minipage}%
  }%
  \endgroup
}
\newcommand{\exampleentry}[7]{%
  \begin{tabularx}{\linewidth}{@{}>{\raggedright\arraybackslash}p{0.22\linewidth}@{\hspace{0.55em}}>{\raggedright\arraybackslash}X@{}}
    \familycard{#1}{#2}{#3}{#4}
    &
    \examplecard{#1}{#5}{#6}{#7}
  \end{tabularx}
}

\setlength{\tabcolsep}{0pt}
\renewcommand{\arraystretch}{1.0}

\exampleentry{FamilySubgroup}
{Subgroup}
{template S-2.1}
{dataset c2}
{Find subgroup counts and rank groups by size.}
{Which \groundedtoken{buying} groups are most frequent?}
{SELECT \groundedtoken{buying} \bindmap{<group\_field>},\par
COUNT(*) AS row\_count\par
FROM \groundedtoken{c2} \bindmap{<table>}\par
GROUP BY \groundedtoken{buying} \bindmap{<group\_field>}\par
ORDER BY row\_count DESC;}

\vspace{0.22em}
\exampleentry{FamilyConditional}
{Conditional}
{template C-L1}
{dataset c6}
{Count rows per $(x, y)$ cell inside a local slice.}
{Within rows where \groundedtoken{Subtopic} equals Linear Transformations, how many rows fall into each \groundedtoken{Student Country} and \groundedtoken{Question Level} combination?}
{SELECT \groundedtoken{"Student Country"} \bindmap{<group\_x>},\par
\groundedtoken{"Question Level"} \bindmap{<group\_y>},\par
COUNT(*) AS row\_count\par
FROM \groundedtoken{"c6"} \bindmap{<table>}\par
WHERE \groundedtoken{"Subtopic" = 'Linear Transformations'}\par
\bindmap{<slice\_predicate>}\par
GROUP BY \groundedtoken{"Student Country"} \bindmap{<group\_x>},\par
\groundedtoken{"Question Level"} \bindmap{<group\_y>}\par
ORDER BY row\_count DESC;}

\vspace{0.22em}
\exampleentry{FamilyTail}
{Tail / rarity}
{template T-2.2}
{dataset m8}
{Count rows by group inside an upper-tail slice.}
{Among the top 3\% of \groundedtoken{balance}, which \groundedtoken{previous} values occur most often?}
{SELECT \groundedtoken{previous} \bindmap{<group\_field>},\par
COUNT(*) AS row\_count\par
FROM \groundedtoken{m8} \bindmap{<table>}\par
WHERE \groundedtoken{balance > PERCENTILE(balance, 0.97)}\par
\bindmap{<tail\_predicate>}\par
GROUP BY \groundedtoken{previous} \bindmap{<group\_field>};}

\vspace{0.22em}
\exampleentry{FamilyMissing}
{Missingness}
{template M-2.1}
{dataset c5}
{Compute a target missing rate for each state.}
{For each \groundedtoken{odor} category, what fraction of rows has missing \groundedtoken{stalk-root}?}
{SELECT \groundedtoken{"odor"} \bindmap{<state\_col>},\par
AVG(CASE WHEN \groundedtoken{"stalk-root"} \bindmap{<target\_col>} IS NULL\par
THEN 1.0 ELSE 0.0 END) AS missing\_rate\par
FROM \groundedtoken{"c5"} \bindmap{<table>}\par
GROUP BY \groundedtoken{"odor"} \bindmap{<state\_col>}\par
ORDER BY missing\_rate DESC;}

\vspace{0.22em}
\exampleentry{FamilyCardinality}
{Cardinality / range}
{template K-1.1}
{dataset c18}
{Enumerate a high-cardinality attribute by support.}
{How many rows are associated with each wine \groundedtoken{title}?}
{SELECT \groundedtoken{title} \bindmap{<value\_col>} AS category\_value,\par
COUNT(*) AS support\par
FROM \groundedtoken{c18} \bindmap{<table>}\par
WHERE \groundedtoken{title} \bindmap{<value\_col>} IS NOT NULL\par
GROUP BY \groundedtoken{title} \bindmap{<value\_col>}\par
ORDER BY support DESC, category\_value;}
\endgroup

\caption{\textbf{One question and SQL example for each of the five
query families.} The \textcolor[HTML]{1397B8}{blue text} marks
dataset-specific grounding, and the \textcolor{orange!85!black}{orange
notes} show which template slot each grounded SQL fragment instantiates. Details in Table~\ref{tab:appendix_template_taxonomy_full_catalog}.}

\label{tab:five-family-grounded-sql}
\end{table}

\subsection{Experimental Settings Details}
\label{app:evaluation_settings_details}
This appendix records the implementation details behind the compact evaluation setup in Section~\ref{sec:evaluation_settings}. The main text keeps only the common evaluation contract; the items below make the data split, SQL grounding, execution infrastructure, and model-tuning policy auditable.

\textbf{Data split and synthetic output contract.}
For each dataset, the preparation pipeline creates disjoint training and test splits with proportions $80\%/20\%$ using a fixed split seed and stratification when the task is classificatory. Each generator observes only the training split and is asked to synthesize a table with the same schema and, by default, the same number of rows as that split. Query fidelity and distance-based fidelity both use the real training table as their reference. The validation and test splits are retained for model-development or future downstream evaluation and are not used in the current scoring pipeline.

\textbf{Generator suite.}
The benchmark covers \NumModels{} tabular generators spanning Bayesian-network, tree-based, adversarial, variational, diffusion, transformer-based, and flow-matching families. The evaluated generators are BayesNet, ARF, CTGAN, TVAE, TabDDPM, TabSyn, TabDiff, ForestDiffusion, RealTabFormer, TabPFGen, and TabbyFlow. All generators use the same dataset split and output-size requirement.

\textbf{SQL grounding and validation.}
For each dataset, \BenchmarkName{} constructs a grounded SQL workload from the fixed template library described in Section~\ref{sec:query_grounding}. The agent selects applicable templates and dataset columns. The profiler, binder, and validator are deterministic. The SQL realizer is the only LLM-assisted component: it calls the GLM-5.3 API inside a constrained template-to-SQL realization step, with a mean API cost of $0.78$ per dataset. The validator rejects invalid SQL, incompatible realizations, and queries that fail execution; failed realizations are retried under the same fixed template policy.

\textbf{Runtime and infrastructure.}
Each generator is wrapped behind a common interface for fitting, sampling, schema restoration, and output validation. We package each generator in its own Docker image so that incompatible dependencies remain isolated while the produced synthetic tables follow a shared output contract. Runtime is decomposed into training time and generation time for the quality--cost analysis. The audited runtime experiments use the same hardware platform with an RTX Pro 6000 GPU.

\textbf{Resolved model configurations.}
The benchmark runner uses model-specific resolved configurations rather than one common hyperparameter search space. Unless a run records an explicit override, ARF uses 30 trees, 10 adversarial iterations, minimum node size 5, \texttt{delta}=0, early stopping, and a fixed seed; BayesNet uses a Chow--Liu structure with mutual-information edge weights, up to 120K fitting rows and 25K structure-learning rows, with discretization bins and categorical-level caps reduced for wide or very large tables. CTGAN uses 50 epochs, embedding dimension 4, $(8,8)$ generator and discriminator hidden dimensions, batch size 2, \texttt{pac}=1, and a 50K training-row cap. TVAE uses 300 epochs, batch size 500, embedding dimension 128, and $(128,128)$ encoder/decoder dimensions. RealTabFormer uses 100 epochs in the reported default configuration.

For the diffusion and flow-based adapters, ForestDiffusion uses an XGBoost-backed model with 20 estimators, \texttt{n\_t}=10, \texttt{duplicate\_K}=5, maximum depth 4, and one worker in the default configuration. TabDDPM uses 1,000 diffusion timesteps, learning rate $10^{-3}$, zero dropout and weight decay, and size-dependent training steps, batch size, and hidden-layer width; for example, tables below 10K training rows use 10K steps and $(512,512)$ hidden layers, while larger tables use larger tiers. TabDiff and TabbyFlow use 500 training steps, learning rate $10^{-3}$, batch size 4,096, and a sampling batch size of 10,000. TabDiff uses 50 diffusion timesteps, while TabbyFlow uses the flow-matching configuration with a 100-epoch learning-rate warmup.

TabPFGen uses pretrained TabPFN components without model fitting, caps the fit set at 50K rows, uses 1,000 SGLD steps with step size 0.01 and noise scale 0.01, and generates in memory-bounded chunks. TabSyn trains a VAE followed by latent diffusion on the full training split; its defaults are 4,000 VAE epochs and 10,001 diffusion epochs, with batch sizes adjusted for table width, and 50 function-evaluation steps at sampling time. When a run supplies an explicit epoch override, the TabSyn diffusion epoch count is resolved as approximately $2.5$ times the VAE epoch count. These settings are recorded in the run artifacts when available; model-specific safety fallbacks, such as reduced sampling batches after GPU out-of-memory errors or learning-rate halving after numerical divergence, are treated as recovery behavior rather than additional tuning trials.

\textbf{Preset and override policy.}
The main benchmark uses the resolved model configurations above. The repository also contains a \texttt{fast} preset for separate workload stress tests; it changes selected settings, including TabDiff and TabbyFlow training steps and ForestDiffusion's noise levels, duplicate factor, thread count, and training-row cap. Those stress-test overrides are not used to redefine the main benchmark results. Command-line arguments and explicit environment variables take precedence over preset values, and failed or technically invalid runs remain visible in the benchmark bookkeeping rather than being silently removed.

\onecolumn

\subsection{Dataset Catalog}
\label{app:dataset_catalog}

\begingroup
\scriptsize
\setlength{\tabcolsep}{3.0pt}
\renewcommand{\arraystretch}{1.08}
\setlength{\LTleft}{0pt}
\setlength{\LTright}{0pt}
\begin{longtable}{@{}>{\raggedright\arraybackslash}p{\dimexpr0.046296\linewidth-0.462963\tabcolsep\relax}%
                  >{\raggedright\arraybackslash}p{\dimexpr0.046296\linewidth-0.462963\tabcolsep\relax}%
                  >{\raggedright\arraybackslash}p{\dimexpr0.262346\linewidth-2.623457\tabcolsep\relax}%
                  >{\raggedleft\arraybackslash}p{\dimexpr0.064815\linewidth-0.648148\tabcolsep\relax}%
                  >{\raggedleft\arraybackslash}p{\dimexpr0.040123\linewidth-0.401235\tabcolsep\relax}%
                  >{\raggedright\arraybackslash}p{\dimexpr0.540123\linewidth-5.401235\tabcolsep\relax}@{}}
\caption{49 dataset catalog.}
\label{tab:appendix_dataset_catalog_all}
\label{tab:appendix_dataset_catalog_c}
\label{tab:appendix_dataset_catalog_m}
\label{tab:appendix_dataset_catalog_n}\\
\toprule
Series & No. & Original dataset name & Rows & Cols & Original source URL \\
\midrule
\endfirsthead

\caption[]{49 dataset catalog (continued).}\\
\toprule
Series & No. & Original dataset name & Rows & Cols & Original source URL \\
\midrule
\endhead

\midrule
\multicolumn{6}{r}{\textit{Continued on next page}}\\
\endfoot

\bottomrule
\endlastfoot

C & c1 & Jungle Chess 2pcs Raw Endgame Complete & 44819 & 7 & \url{https://www.openml.org/d/41027} \\
C & c2 & Car Evaluation & 1728 & 7 & \url{https://archive.ics.uci.edu/dataset/19/car+evaluation} \\
C & c3 & Splice junction Gene Sequences & 3189 & 3 & \url{https://archive.ics.uci.edu/dataset/69/molecular+biology+splice+junction+gene+sequences} \\
C & c4 & Chess King Rook Vs King Pawn & 3196 & 37 & \url{https://archive.ics.uci.edu/dataset/22/chess+king+rook+vs+king+pawn} \\
C & c5 & Mushroom & 8416 & 23 & \url{https://archive.ics.uci.edu/dataset/73/mushroom} \\
C & c6 & Dataset For Assessing ML In Higher Education & 9546 & 8 & \url{https://archive.ics.uci.edu/dataset/1031/dataset+for+assessing+mathematics+learning+in+higher+education} \\
C & c7 & Nursery & 12960 & 9 & \url{https://www.openml.org/d/26} \\
C & c8 & Phishing Websites & 11055 & 31 & \url{https://archive.ics.uci.edu/dataset/327/phishing+websites} \\
C & c9 & Amazon Employee Access Challenge & 32769 & 10 & \url{https://www.kaggle.com/c/amazon-employee-access-challenge} \\
C & c10 & Poker Hand & 1025010 & 11 & \url{https://archive.ics.uci.edu/dataset/158/poker+hand} \\
C & c11 & Connect 4 & 67557 & 43 & \url{https://archive.ics.uci.edu/dataset/26/connect-4} \\
C & c12 & Internet-Advertisements & 3279 & 1559 & \url{https://www.openml.org/d/40978} \\
C & c13 & Us Census Data 1990 & 2458285 & 69 & \url{https://archive.ics.uci.edu/dataset/116/us+census+data+1990} \\
C & c14 & Cat In The Dat & 300000 & 25 & \url{https://www.kaggle.com/c/cat-in-the-dat} \\
C & c15 & Cat In The Dat Ii & 600000 & 25 & \url{https://www.kaggle.com/c/cat-in-the-dat-ii} \\
C & c16 & Fivethirtyeight Comic Characters Dataset & 6896 & 13 & \url{https://www.kaggle.com/datasets/fivethirtyeight/fivethirtyeight-comic-characters-dataset} \\
C & c17 & Netflix Shows & 8807 & 12 & \url{https://www.kaggle.com/datasets/shivamb/netflix-shows} \\
C & c18 & Wine Reviews & 129971 & 14 & \url{https://www.kaggle.com/datasets/zynicide/wine-reviews} \\
C & c19 & Trending YouTube Video Statistics and Comments & 40949 & 16 & \url{https://www.kaggle.com/datasets/datasnaek/youtube-new} \\
M & m1 & Remote Worker Productivity & 1500 & 30 & \url{https://huggingface.co/datasets/nprak26/remote-worker-productivity} \\
M & m2 & Car Insurance Claim Prediction & 58592 & 44 & \url{https://www.kaggle.com/datasets/ifteshanajnin/carinsuranceclaimprediction-classification} \\
M & m3 & Hotel Booking Demand & 119390 & 32 & \url{https://www.kaggle.com/datasets/jessemostipak/hotel-booking-demand} \\
M & m4 & Medical Insurance Charges & 2772 & 7 & \url{https://huggingface.co/datasets/rahulvyasm/medical_insurance_data} \\
M & m5 & Predict Students Dropout And Academic Success & 4424 & 37 & \url{https://archive.ics.uci.edu/dataset/697/predict+students+dropout+and+academic+success} \\
M & m6 & Online Shoppers Purchasing Intention Dataset & 12330 & 18 & \url{https://archive.ics.uci.edu/dataset/468/online+shoppers+purchasing+intention+dataset} \\
M & m7 & Stroke Prediction Dataset & 5110 & 12 & \url{https://www.kaggle.com/datasets/fedesoriano/stroke-prediction-dataset} \\
M & m8 & Bank Marketing & 45211 & 17 & \url{https://archive.ics.uci.edu/dataset/222/bank+marketing} \\
M & m9 & HR Analytics: Job Change of Data Scientists & 19158 & 14 & \url{https://www.kaggle.com/datasets/arashnic/hr-analytics-job-change-of-data-scientists} \\
M & m10 & Mobile Price Classification & 2000 & 21 & \url{https://www.kaggle.com/datasets/iabhishekofficial/mobile-price-classification} \\
M & m11 & Health Insurance Cross Sell Prediction & 381109 & 12 & \url{https://www.kaggle.com/datasets/anmolkumar/health-insurance-cross-sell-prediction} \\
N & n1 & Spambase & 4601 & 58 & \url{https://www.openml.org/d/44} \\
N & n2 & Airfoil Self-Noise & 1503 & 6 & \url{https://archive.ics.uci.edu/dataset/291/airfoil+self-noise} \\
N & n3 & Wine Quality & 4898 & 12 & \url{https://archive.ics.uci.edu/dataset/186/wine+quality} \\
N & n4 & Communities And Crime & 1994 & 128 & \url{https://archive.ics.uci.edu/dataset/183/communities+and+crime} \\
N & n5 & Superconductivity & 21263 & 82 & \url{https://www.openml.org/d/44964} \\
N & n6 & BEED: Bangalore EEG Epilepsy Dataset & 8000 & 17 & \url{https://archive.ics.uci.edu/dataset/1134/beed:+bangalore+eeg+epilepsy+dataset} \\
N & n7 & Anuran Calls Mfccs & 7195 & 26 & \url{https://archive.ics.uci.edu/dataset/406/anuran+calls+mfccs} \\
N & n8 & Secom & 1567 & 593 & \url{https://archive.ics.uci.edu/dataset/179/secom} \\
N & n9 & Pen Based Recognition Of Handwritten Digits & 10992 & 17 & \url{https://archive.ics.uci.edu/dataset/81/pen-based+recognition+of+handwritten+digits} \\
N & n10 & Dry Bean Dataset & 13611 & 17 & \url{https://archive.ics.uci.edu/dataset/602/dry+bean+dataset} \\
N & n11 & Magic Gamma Telescope & 19019 & 11 & \url{https://archive.ics.uci.edu/dataset/159/magic+gamma+telescope} \\
N & n12 & Skin Segmentation & 245057 & 4 & \url{https://archive.ics.uci.edu/dataset/229/skin+segmentation} \\
N & n13 & Combined Cycle Power Plant & 9568 & 5 & \url{https://archive.ics.uci.edu/dataset/294/combined+cycle+power+plant} \\
N & n14 & First-Order Theorem Proving & 2000 & 52 & \url{https://www.openml.org/d/44663} \\
N & n15 & COIL2000 & 9822 & 86 & \url{https://www.openml.org/d/298} \\
N & n16 & Credit Card Fraud Detection & 284807 & 31 & \url{https://www.kaggle.com/datasets/mlg-ulb/creditcardfraud} \\
N & n17 & Statlog Shuttle & 14500 & 10 & \url{https://archive.ics.uci.edu/dataset/148/statlog+shuttle} \\
N & n18 & APS Failure at Scania Trucks & 76000 & 171 & \url{https://archive.ics.uci.edu/dataset/421/aps+failure+at+scania+trucks} \\
N & n19 & Fashion-MNIST & 70000 & 785 & \url{https://www.openml.org/d/40996} \\
\end{longtable}
\par
\endgroup

\subsection{Models Catalog}
\label{app:models_catalog}

\begin{center}
\small
\setlength{\tabcolsep}{0pt}
\renewcommand{\arraystretch}{1.12}
\newcommand{\ModelColor}[2]{\textcolor{#1}{\rule{0.9em}{0.9em}}\,{\scriptsize\texttt{#2}}}

\begin{tabularx}{\linewidth}{@{}
>{\raggedright\arraybackslash}p{0.95in}
@{\hspace{3pt}}
X
@{\hspace{6pt}}
>{\raggedright\arraybackslash}p{0.90in}
@{\hspace{10pt}}
>{\raggedright\arraybackslash}p{0.95in}
@{}}

\toprule
\textbf{Model family} & \textbf{Model} & \textbf{Scope} & \textbf{Color} \\
\midrule
\rowcolor{RuleGray!35}
\multicolumn{4}{@{}l@{}}{\textbf{Transformer-based}} \\
 & REaLTabFormer~\citep{solatorio2023realtabformer} & Single + relational & \ModelColor{modelrealtabformer}{\#6F4C9B} \\
\addlinespace[2pt]
\rowcolor{RuleGray!35}
\multicolumn{4}{@{}l@{}}{\textbf{Diffusion-based}} \\
 & TabDDPM~\citep{kotelnikov2023tabddpm} & Single table & \ModelColor{modeltabddpm}{\#004488} \\
 & TabSyn~\citep{zhang2024tabsyn} & Single table & \ModelColor{modeltabsyn}{\#4477AA} \\
 & ForestDiffusion~\citep{jolicoeurmartineau2023forestdiffusion} & Single table & \ModelColor{modelforestdiffusion}{\#66CCEE} \\
 & TabDiff~\citep{shi2025tabdiff} & Single table & \ModelColor{modeltabdiff}{\#009988} \\
\addlinespace[2pt]
\rowcolor{RuleGray!35}
\multicolumn{4}{@{}l@{}}{\textbf{Adversarial / energy-based}} \\
 & CTGAN~\citep{xu2019ctgan} & Single table & \ModelColor{modelctgan}{\#CC3311} \\
 & ARF~\citep{watson2023arf} & Single table & \ModelColor{modelarf}{\#EE7733} \\
 & TabPFGen~\citep{ma2024tabpfgen} & Single table & \ModelColor{modeltabpfgen}{\#DDAA33} \\
\addlinespace[2pt]
\rowcolor{RuleGray!35}
\multicolumn{4}{@{}l@{}}{\textbf{VAE / latent / flow-based}} \\
 & TVAE~\citep{xu2019ctgan} & Single table & \ModelColor{modeltvae}{\#882255} \\
 & TabbyFlow~\citep{guzmancordero2025tabbyflow} & Single table & \ModelColor{modeltabbyflow}{\#CC6677} \\
\addlinespace[2pt]
\rowcolor{RuleGray!35}
\multicolumn{4}{@{}l@{}}{\textbf{Probabilistic / synthetic database}} \\
 & BayesNet (DataSynthesizer)~\citep{ping2017datasynthesizer} & Single table & \ModelColor{modelbayesnet}{\#666666} \\
\bottomrule
\end{tabularx}
\captionof{table}{Models currently instantiated in the benchmark roster, grouped by shared modeling backbone.}
\label{tab:model_taxonomy}
\end{center}

\clearpage
\section{Query Templates and Five Query Families}
\label{app:query_templates_families}
\label{app:metric_designs}
\label{app:evaluation_designs}

\subsection{Query-Centric Fidelity Overview}
\label{app:metric_design_overview}
\BenchmarkName{} treats query-centric fidelity as a workload-facing measure: a synthetic table is useful for analytical workflows only when it preserves the answers to the types of queries users are likely to run. The benchmark therefore reports family-level signals instead of relying on a single aggregate resemblance score.

\subsection{Workload Sources}
\label{app:workload_sources}

The current appendix-facing template inventory surfaces 44 templates in
total. Table~\ref{tab:appendix_direct_template_sources} lists the
public benchmark, repository, documentation, and paper sources that
directly materialize into one or more released templates.

\begingroup

\defcitealias{tpchspec}{TPC, 1999}
\defcitealias{tpcdsspec}{TPC, 2017}
\defcitealias{cwidaTPCDSQualification}{CWI, n.d.}
\defcitealias{altinityTPCDS}{Altinity, n.d.}
\defcitealias{bigqueryApproxAgg}{Google Cloud, n.d.}
\defcitealias{clickhouseAggregateDocs}{ClickHouse, n.d.}
\defcitealias{snowflake_percentile_cont}{Snowflake, 2026}
\defcitealias{snowflake_width_bucket}{Snowflake, 2026}
\defcitealias{trino_aggregate_functions}{Trino, 2026}
\defcitealias{druidSQLFunctions}{Apache Druid, n.d.}
\begin{center}
\small
\setlength{\tabcolsep}{3.8pt}
\renewcommand{\arraystretch}{1.07}
\begin{tabularx}{\linewidth}{@{}X>{\raggedright\arraybackslash}p{0.86in}>{\centering\arraybackslash}p{0.52in}@{}}
\toprule
 \textbf{Direct source} & \textbf{Type} & \textbf{Templates} \\
\midrule
\rowcolor{RuleGray!35}
\multicolumn{3}{@{}l@{}}{\textbf{Benchmarks and benchmark specifications}} \\
ClickBench~\citep{clickbench2022} & Benchmark & 8 \\
H2O db-benchmark~\citep{h2odbbenchmark} & Benchmark & 9 \\
TPC-H / TPC-DS benchmark specs~\citep{tpchspec};~\citep{tpcdsspec} & Benchmark specs & 1 \\
TPC-DS benchmark family~\citep{tpcdsspec} & Benchmark spec & 3 \\
TPC-H qgen~\citep{tpchspec} & Benchmark & 4 \\
\addlinespace[2pt]
\rowcolor{RuleGray!35}
\multicolumn{3}{@{}l@{}}{\textbf{Repositories and benchmark-adjacent sources}} \\
TPC-DS qualification repo~\citep{cwidaTPCDSQualification} & TPC-DS repository & 2 \\
TPC-DS Altinity repo~\citep{altinityTPCDS} & TPC-DS repository & 2 \\
RTABench \texttt{order\_events} slice~\citep{rtabench2025} & Benchmark + GitHub & 2 \\
\addlinespace[2pt]
\rowcolor{RuleGray!35}
\multicolumn{3}{@{}l@{}}{\textbf{Documentation sources}} \\
BigQuery approximate aggregate docs~\citep{bigqueryApproxAgg} & Documentation & 2 \\
ClickHouse aggregate docs~\citep{clickhouseAggregateDocs} & Documentation & 1 \\
Snowflake \texttt{PERCENTILE\_CONT} docs~\citep{snowflake_percentile_cont} & Documentation & 2 \\
Snowflake \texttt{WIDTH\_BUCKET} docs~\citep{snowflake_width_bucket} & Documentation & 1 \\
Trino aggregate docs~\citep{trino_aggregate_functions} & Documentation & 3 \\
Apache Druid SQL-function docs~\citep{druidSQLFunctions} & Documentation & 1 \\
\addlinespace[2pt]
\rowcolor{RuleGray!35}
\multicolumn{3}{@{}l@{}}{\textbf{Paper source}} \\
Preserving Missing Data Distribution in Synthetic Data~\citep{wang2023missingdist} & Paper & 3 \\
\bottomrule
\end{tabularx}
\captionof{table}{Sources that directly materialize into one or more templates in the released library.}
\label{tab:appendix_direct_template_sources}
\end{center}
\endgroup

\subsection{44-Template Taxonomy Catalog}
\label{app:template_taxonomy_catalog}

The compact taxonomy summary in Appendix~\ref{app:query_families}
collapses related templates into family-level rows. The longtable below
expands that same appendix-facing taxonomy back to the template level
while keeping all 44 released templates in one consistent row format.

\begingroup
\tiny
\setlength{\tabcolsep}{1.6pt}
\renewcommand{\arraystretch}{1.14}
\setlength{\LTleft}{0pt}
\setlength{\LTright}{0pt}
\begin{longtable}{@{}>{\raggedright\arraybackslash}p{\dimexpr0.133956\linewidth-1.339564\tabcolsep\relax}%
                  >{\centering\arraybackslash}p{\dimexpr0.059190\linewidth-0.591900\tabcolsep\relax}%
                  >{\raggedright\arraybackslash}p{\dimexpr0.236760\linewidth-2.367601\tabcolsep\relax}%
                  >{\raggedright\arraybackslash}p{\dimexpr0.099688\linewidth-0.996885\tabcolsep\relax}%
                  >{\raggedright\arraybackslash}p{\dimexpr0.146417\linewidth-1.464174\tabcolsep\relax}%
                  >{\raggedright\arraybackslash}p{\dimexpr0.323988\linewidth-3.239875\tabcolsep\relax}@{}}
\caption{\textbf{44-template taxonomy catalog.}}
\label{tab:appendix_template_taxonomy_full_catalog}\\
\toprule
Template ID & No. & Canonical template & Family & Subfamily & Upstream source \\
\midrule
\endfirsthead

\caption[]{\textbf{44-template taxonomy catalog (continued).}}\\
\toprule
Template ID & No. & Canonical template & Family & Subfamily & Upstream source \\
\midrule
\endhead

\midrule
\multicolumn{6}{r}{\textit{Continued on next page}}\\
\endfoot

\bottomrule
\endlastfoot

\nolinkurl{grp_cond_rate} & C-G1 & Condition Rate by Group & Conditional & Global Structure & TPC-H, TPC-DS \\
\nolinkurl{grp_ratio_2cond} & C-G2 & Two-Condition Ratio by Group & Conditional & Global Structure & TPC-DS \\
\nolinkurl{ds_within_grp_share} & C-G3 & Item Share Within Each Group & Conditional & Global Structure & TPC-DS qualification \\
\nolinkurl{win_part_avg} & C-G4 & Partition Average by Group & Conditional & Global Structure & H2O db-benchmark \\
\nolinkurl{2d_tgt_rate} & C-G5 & 2D Target-Rate Surface & Conditional & Global Structure & TPC-DS \\
\nolinkurl{binned_num_grp_avg} & C-G6 & Binned Group Average & Conditional & Global Structure & Snowflake WIDTH\_BUCKET docs \\
\nolinkurl{grp_disp_rank} & C-G7 & Group Dispersion Rank & Conditional & Global Structure & Trino aggregate docs \\
\nolinkurl{ds_base_gated_rank} & C-G8 & Baseline-Gated Extreme Ranking & Conditional & Global Structure & TPC-DS Altinity \\
\nolinkurl{flt_2d_grp_count} & C-L1 & Filtered 2D Count Surface & Conditional & Local Slice & TPC-DS \\
\nolinkurl{med_flt_num} & C-L2 & Filtered Median Slice & Conditional & Local Slice & Snowflake PERCENTILE\_CONT docs \\
\nolinkurl{cond_grp_quants} & C-L3 & Filtered Group Quantiles & Conditional & Local Slice & ClickHouse aggregate docs \\
\nolinkurl{rta_time_bucket_cnt} & C-L4 & Filtered Time-Bucket Count & Conditional & Local Slice & RTABench order\_events \\
\nolinkurl{rta_time_bucket_mavg} & C-L5 & Time-Bucket Moving Average by Group & Conditional & Local Slice & RTABench order\_events \\
\midrule
\nolinkurl{cb_grp_distinct_topk} & S-1.1 & Distinct-Coverage Ranking by Group & Subgroup & Structure & ClickBench \\
\nolinkurl{cb_flt_distinct_topk} & S-1.2 & Filtered Distinct-Coverage Ranking by Group & Subgroup & Structure & ClickBench \\
\nolinkurl{h2o_grp_sum} & S-1.3 & Total-Measure Ranking by Group & Subgroup & Structure & H2O db-benchmark \\
\nolinkurl{grp_avg_num} & S-1.4 & Mean-Measure Comparison by Group & Subgroup & Structure & H2O db-benchmark \\
\nolinkurl{guarded_grp_avg} & S-1.5 & Support-Gated Mean Comparison by Group & Subgroup & Structure & H2O db-benchmark \\
\nolinkurl{ds_topk_grp_sum} & S-1.6 & Filtered Total-Measure Ranking by Group & Subgroup & Structure & TPC-DS qualification \\
\nolinkurl{h2o_2d_grp_sum} & S-1.7 & Two-Dimensional Total-Measure Ranking & Subgroup & Structure & H2O db-benchmark \\
\nolinkurl{2d_grp_avg} & S-1.8 & Two-Dimensional Mean Comparison & Subgroup & Structure & H2O db-benchmark \\
\nolinkurl{cb_grp_summary_topk} & S-1.9 & Support-and-Mean Ranking by Group & Subgroup & Structure & ClickBench \\
\nolinkurl{tpch_2d_summary} & S-1.10 & Filtered Two-Dimensional Measure Summary & Subgroup & Structure & TPC-H qgen \\
\nolinkurl{h2o_2d_robust} & S-1.11 & Two-Dimensional Robust Measure Summary & Subgroup & Structure & H2O db-benchmark \\
\nolinkurl{tpch_max_agg_win} & S-1.12 & Top Aggregate Winner by Group & Subgroup & Structure & TPC-H qgen \\
\nolinkurl{wtd_topk_sum} & S-1.13 & Support-Gated Weighted Total Ranking by Group & Subgroup & Structure & BigQuery approx docs \\
\nolinkurl{cb_grp_count} & S-2.1 & Group Count Distribution & Subgroup & Size & ClickBench \\
\nolinkurl{cb_flt_grp_count} & S-2.2 & Filtered Group Count Ranking & Subgroup & Size & ClickBench \\
\nolinkurl{cb_2d_topk_count} & S-2.3 & Two-Dimensional Group Count Ranking & Subgroup & Size & ClickBench \\
\midrule
\nolinkurl{quant_tail_slice} & T-1.1 & Top-Quantile Tail Members & Tail / Rarity & Tail Coverage & Snowflake PERCENTILE\_CONT docs \\
\nolinkurl{global_z_outliers} & T-1.2 & Global Z-Score Outliers & Tail / Rarity & Tail Coverage & Trino aggregate docs \\
\nolinkurl{h2o_topn_in_grp} & T-1.3 & Within-Group Top-N Extremes & Tail / Rarity & Tail Coverage & H2O db-benchmark \\
\nolinkurl{ds_subgrp_base_outlier} & T-1.4 & Subgroup-Relative Outliers & Tail / Rarity & Tail Coverage & TPC-DS Altinity \\
\nolinkurl{tpch_rel_total_thr} & T-2.1 & Groups Above a Total-Share Threshold & Tail / Rarity & Tail Size & TPC-H qgen \\
\nolinkurl{tpch_thr_grp_rank} & T-2.2 & Above-Threshold Group Ranking & Tail / Rarity & Tail Size & TPC-H qgen \\
\nolinkurl{grp_pct_point} & T-2.3 & Group-wise Percentile Level & Tail / Rarity & Tail Size & BigQuery approx docs \\
\nolinkurl{thr_rarity_cdf} & T-2.4 & Threshold Exceedance Rarity & Tail / Rarity & Tail Size & Apache Druid SQL docs \\
\midrule
\nolinkurl{miss_rate_marg} & M-1.1 & Marginal Missing Rate & Missingness & Marginal Missingness & Preserving Missing Data Distribution in Synthetic Data \\
\nolinkurl{miss_rate_disc_state} & M-2.1 & Missing Rate by Discrete State & Missingness & Broad Co-Missingness & Preserving Missing Data Distribution in Synthetic Data \\
\nolinkurl{miss_rate_cont_bucket} & M-2.2 & Missing Rate by Continuous Bucket & Missingness & Broad Co-Missingness & Preserving Missing Data Distribution in Synthetic Data \\
\midrule
\nolinkurl{card_supp_rank_prof} & K-1.1 & Support Rank Profile & Cardinality / Range & Discrete & ClickBench \\
\nolinkurl{card_dist_share_prof} & K-1.2 & Distinct Share Profile & Cardinality / Range & Discrete & ClickBench \\
\nolinkurl{card_cont_range_env} & K-1.3 & Continuous Range Envelope Profile & Cardinality / Range & Continuous & Trino aggregate docs \\
\nolinkurl{card_hi_card_resp} & K-2.1 & High-Cardinality Response Stability & Cardinality / Range & Discrete & H2O db-benchmark \\
\end{longtable}
\endgroup

\subsection{Five Query Families}
\label{app:query_families}

\subsubsection{Subgroup}
\label{app:experiment_subgroup}

\textbf{Goal.}
Evaluate whether synthetic data preserves subgroup-level analytical
behavior rather than only global marginals.

\textbf{Template-side branches.}
\begin{enumerate}
    \item \textbf{Structure (13 templates).} Grouped summaries that
    keep the subgroup object fixed but vary the summary signal,
    including totals, means, distinct coverage, robust summaries, and
    winner-style aggregate views.
    \item \textbf{Size (3 templates).} Absolute subgroup count or
    support views over one-dimensional and two-dimensional groupings.
\end{enumerate}

Table~\ref{tab:appendix_subgroup_templates} lists every Subgroup
template and explains the analytical behavior it probes and the reason
for its branch assignment.
\begingroup
\scriptsize
\setlength{\tabcolsep}{2.4pt}
\renewcommand{\arraystretch}{1.08}
\setlength{\LTleft}{0pt}
\setlength{\LTright}{0pt}
\begin{longtable}{@{}>{\raggedright\arraybackslash}p{\dimexpr0.090343\linewidth-0.542056\tabcolsep\relax}%
                  >{\raggedright\arraybackslash}p{\dimexpr0.218069\linewidth-1.308411\tabcolsep\relax}%
                  >{\raggedright\arraybackslash}p{\dimexpr0.345794\linewidth-2.074766\tabcolsep\relax}%
                  >{\raggedright\arraybackslash}p{\dimexpr0.345794\linewidth-2.074766\tabcolsep\relax}@{}}
\caption{\textbf{Subgroup template inventory.} Each row states the
analytical behavior tested by the template and why it belongs to the
Structure or Size branch.}
\label{tab:appendix_subgroup_templates}\\
\toprule
\textbf{ID} & \textbf{Canonical template} & \textbf{What it probes} &
\textbf{Why this branch} \\
\midrule
\endfirsthead

\caption[]{\textbf{Subgroup template inventory (continued).}}\\
\toprule
\textbf{ID} & \textbf{Canonical template} & \textbf{What it probes} &
\textbf{Why this branch} \\
\midrule
\endhead

\midrule
\multicolumn{4}{r}{\textit{Continued on next page}}\\
\endfoot

\bottomrule
\endlastfoot

\texttt{S-1.1} &
Distinct-Coverage Ranking by Group &
Ranks groups by the number of distinct values or entities they cover,
testing whether synthetic data preserves relative subgroup breadth. &
\emph{Structure}: the target is the ordering and profile of subgroup
coverage, rather than the raw number of rows in each group. \\

\texttt{S-1.2} &
Filtered Distinct-Coverage Ranking by Group &
Repeats distinct-coverage ranking after applying a valid filter, testing
whether subgroup breadth remains stable in a restricted analytical view. &
\emph{Structure}: the query compares group-level coverage behavior under
a scoped view; the filter changes context but not the subgroup object. \\

\texttt{S-1.3} &
Total-Measure Ranking by Group &
Ranks groups by a total such as a sum, exposing whether the same
subgroups dominate an aggregate measure. &
\emph{Structure}: it evaluates the relative aggregate pattern across
groups, including which groups appear at the top of the ranking. \\

\texttt{S-1.4} &
Mean-Measure Comparison by Group &
Compares the average of a numeric measure across groups and tests
whether group-level central tendency is preserved. &
\emph{Structure}: the score concerns the pattern of group means, not
group population size alone. \\

\texttt{S-1.5} &
Support-Gated Mean Comparison by Group &
Compares group means while excluding groups with insufficient support,
testing robust subgroup comparisons under a minimum-support rule. &
\emph{Structure}: the support gate controls which groups are eligible,
while the measured object remains the cross-group mean pattern. \\

\texttt{S-1.6} &
Filtered Total-Measure Ranking by Group &
Ranks group-level totals inside a filtered view, testing whether
aggregate dominance is preserved after restricting the rows considered. &
\emph{Structure}: it probes the ordering of subgroup aggregate behavior;
the filter is a query scope rather than a separate conditional slice
diagnostic. \\

\texttt{S-1.7} &
Two-Dimensional Total-Measure Ranking &
Ranks pairs of groups by a total measure, testing whether joint
subgroup structure survives beyond one-dimensional grouping. &
\emph{Structure}: the central object is the two-dimensional aggregate
surface and its ranking pattern. \\

\texttt{S-1.8} &
Two-Dimensional Mean Comparison &
Compares mean measures across combinations of two grouping attributes,
testing preservation of joint subgroup-level averages. &
\emph{Structure}: it evaluates the arrangement of group-pair means,
rather than the marginal support count of the pairs. \\

\texttt{S-1.9} &
Support-and-Mean Ranking by Group &
Ranks groups using a summary that combines subgroup support and a
numeric measure, testing whether a multi-signal group ordering is kept. &
\emph{Structure}: the output is a group-level ranking driven by the
summary profile, not a standalone count distribution. \\

\texttt{S-1.10} &
Filtered Two-Dimensional Measure Summary &
Summarizes a two-dimensional aggregate surface after a filter, testing
whether joint subgroup relationships remain usable in a scoped query. &
\emph{Structure}: it preserves the shape of a filtered group-pair
summary; the primary object is still the aggregate structure. \\

\texttt{S-1.11} &
Two-Dimensional Robust Measure Summary &
Compares robust summaries across group pairs to reduce sensitivity to
extreme rows while testing joint subgroup behavior. &
\emph{Structure}: robustness changes the summary operator, while the
evaluated object remains the two-dimensional subgroup pattern. \\

\texttt{S-1.12} &
Top Aggregate Winner by Group &
Identifies the group that wins under an aggregate comparison, testing
whether the identity of the leading subgroup is preserved. &
\emph{Structure}: the query probes a winner identity and ordering
relationship rather than the amount of mass in each group. \\

\texttt{S-1.13} &
Support-Gated Weighted Total Ranking by Group &
Ranks groups by a weighted total after enforcing a support condition,
testing whether weighted subgroup dominance is preserved reliably. &
\emph{Structure}: the target is the weighted ranking among eligible
groups; support affects validity but is not the outcome being measured. \\

\texttt{S-2.1} &
Group Count Distribution &
Measures the number of rows assigned to each group and the resulting
distribution of subgroup sizes. &
\emph{Size}: the query directly targets how much mass each subgroup
contains, without asking which numeric summary it produces. \\

\texttt{S-2.2} &
Filtered Group Count Ranking &
Ranks groups by row count inside a filtered view, testing whether
relative subgroup sizes remain stable after restricting the data. &
\emph{Size}: the measured quantity is subgroup population, while the
filter only defines the rows included in the count. \\

\texttt{S-2.3} &
Two-Dimensional Group Count Ranking &
Ranks group pairs by their row counts, testing whether joint support
mass is preserved across a two-dimensional partition. &
\emph{Size}: the query focuses on the amount of support in each group
pair, rather than on a value-based aggregate over those rows. \\

\end{longtable}
\endgroup

\subsubsection{Conditional}
\label{app:experiment_conditional}

\textbf{Goal.}
Evaluate whether conditional behavior is preserved both at full grouped
scope and after narrowing attention to local slices.

\textbf{Template-side branches.}
\begin{enumerate}
    \item \textbf{Global structure (8 templates).} Grouped
    conditional summaries such as condition rates, grouped ratios,
    grouped averages, grouped dispersion, and other full-scope
    conditional views.
    \item \textbf{Local slices (5 templates).} The same
    grouped conditional scaffold after adding a predicate-defined slice
    or local support gate.
\end{enumerate}

Table~\ref{tab:appendix_conditional_templates} lists every Conditional
template and explains the analytical behavior it probes and the reason
for its branch assignment.
\begingroup
\scriptsize
\setlength{\tabcolsep}{2.4pt}
\renewcommand{\arraystretch}{1.08}
\setlength{\LTleft}{0pt}
\setlength{\LTright}{0pt}
\begin{longtable}{@{}>{\raggedright\arraybackslash}p{\dimexpr0.090343\linewidth-0.542056\tabcolsep\relax}%
                  >{\raggedright\arraybackslash}p{\dimexpr0.218069\linewidth-1.308411\tabcolsep\relax}%
                  >{\raggedright\arraybackslash}p{\dimexpr0.345794\linewidth-2.074766\tabcolsep\relax}%
                  >{\raggedright\arraybackslash}p{\dimexpr0.345794\linewidth-2.074766\tabcolsep\relax}@{}}
\caption{\textbf{Conditional template inventory.} Each row states the
analytical behavior tested by the template and why it belongs to the
Global Structure or Local Slice branch.}
\label{tab:appendix_conditional_templates}\\
\toprule
\textbf{ID} & \textbf{Canonical template} & \textbf{What it probes} &
\textbf{Why this branch} \\
\midrule
\endfirsthead
\caption[]{\textbf{Conditional template inventory (continued).}}\\
\toprule
\textbf{ID} & \textbf{Canonical template} & \textbf{What it probes} &
\textbf{Why this branch} \\
\midrule
\endhead
\midrule
\multicolumn{4}{r}{\textit{Continued on next page}}\\
\endfoot
\bottomrule
\endlastfoot

\texttt{C-G1} & Condition Rate by Group &
Measures how often a target condition holds within each group, testing
whether conditional incidence varies across groups in the same way. &
\emph{Global Structure}: it evaluates a full-table grouped conditional
relationship without a predicate-defined local slice. \\

\texttt{C-G2} & Two-Condition Ratio by Group &
Compares the rates of two conditions within each group, testing whether
relative conditional behavior is preserved across groups. &
\emph{Global Structure}: the query describes a grouped ratio over the
full available scope rather than a filtered subpopulation. \\

\texttt{C-G3} & Item Share Within Each Group &
Measures the share contributed by an item or category inside each group,
testing within-group composition. &
\emph{Global Structure}: the outcome is a group-level composition pattern
over the full table. \\

\texttt{C-G4} & Partition Average by Group &
Computes a partition-aware average for each group, testing whether
grouped central tendency remains aligned with the partition structure. &
\emph{Global Structure}: it probes a grouped conditional summary without
introducing an explicit local filter. \\

\texttt{C-G5} & 2D Target-Rate Surface &
Builds a target-rate surface over pairs of grouping attributes, testing
whether conditional behavior is preserved jointly across two dimensions. &
\emph{Global Structure}: the target is a full-scope two-dimensional
conditional surface, not a local slice. \\

\texttt{C-G6} & Binned Group Average &
Bins a numeric attribute and compares group averages across the bins,
testing conditional trends over a discretized numeric axis. &
\emph{Global Structure}: binning organizes the full-scope conditional
relationship rather than restricting rows to one selected slice. \\

\texttt{C-G7} & Group Dispersion Rank &
Ranks groups by a dispersion statistic, testing whether conditional
variability and group ordering are preserved. &
\emph{Global Structure}: it evaluates a grouped variability pattern at
global scope. \\

\texttt{C-G8} & Baseline-Gated Extreme Ranking &
Ranks groups by an extreme conditional value after applying a baseline
eligibility rule, testing structured extremes across groups. &
\emph{Global Structure}: the gate controls valid groups, while the
measured object remains the global grouped conditional ranking. \\

\texttt{C-L1} & Filtered 2D Count Surface &
Counts group pairs inside a predicate-defined slice, testing whether
joint conditional support survives local filtering. &
\emph{Local Slice}: the defining operation is a \texttt{WHERE}-based
restriction before evaluating the two-dimensional grouped result. \\

\texttt{C-L2} & Filtered Median Slice &
Computes a median-style numeric summary after restricting rows to a
local subpopulation. &
\emph{Local Slice}: the query explicitly measures behavior inside a
predicate-defined slice rather than across the full table. \\

\texttt{C-L3} & Filtered Group Quantiles &
Measures grouped quantiles within a filtered view, testing local
distributional structure beyond averages or counts. &
\emph{Local Slice}: the quantile summary is evaluated only after the
local predicate has isolated a subpopulation. \\

\texttt{C-L4} & Filtered Time-Bucket Count &
Counts rows in time buckets after filtering, testing whether local
temporal support and grouped activity remain usable. &
\emph{Local Slice}: both the temporal aggregation and its answer are
conditioned on a predicate-defined local view. \\

\texttt{C-L5} & Time-Bucket Moving Average by Group &
Computes a moving average over time buckets for each group, testing
whether temporal trends and group trajectories are preserved. &
\emph{Local Slice}: it belongs to the filtered temporal workload branch
and evaluates a scoped group-by-time pattern. \\

\end{longtable}
\endgroup

For the paired local-vs-global diagnostic in Figure~\ref{fig:local-vs-defiltered-scatter}, we group filtered conditional queries by their SQL template rather than by dataset or model. Each local query first uses a \texttt{WHERE} predicate to isolate a subpopulation, then evaluates an aggregation inside that local slice. Its global counterpart is constructed by deleting only that \texttt{WHERE} predicate while keeping the remaining \texttt{SELECT}, \texttt{GROUP BY}, and \texttt{ORDER BY} structure unchanged. The five local-slice template parts are:

\begin{itemize}[leftmargin=1.6em,itemsep=0.1em,topsep=0.2em]
    \item \textbf{Part~\ding{172}: Filtered two-dimensional group count} ($228$ paired query instances). Counts the joint support of two grouped columns inside a predicate-defined local slice.
    \item \textbf{Part~\ding{173}: Filtered median slice} ($29$ paired query instances). Computes a median-style numeric summary after filtering to a local subpopulation.
    \item \textbf{Part~\ding{174}: Filtered group quantile} ($97$ paired query instances). Measures whether grouped quantile structure is preserved inside a filtered conditional view.
    \item \textbf{Part~\ding{175}: Filtered time-bucket count} ($52$ paired query instances). Evaluates time-bucketed counts under a local filter, matching dashboard-style filtered temporal summaries.
    \item \textbf{Part~\ding{176}: Time-bucket moving average by group} ($11$ paired query instances). Computes a moving average over time-bucketed subgroup counts inside a predicate-defined local slice.
\end{itemize}

\begin{figure*}[!htbp]
\centering
\begin{subfigure}[t]{0.32\textwidth}
  \centering
  \resizebox{\linewidth}{!}{\centering
\begingroup
\definecolor{tplL1}{HTML}{1F77B4}
\definecolor{tplL2}{HTML}{FF7F0E}
\definecolor{tplL3}{HTML}{2CA02C}
\definecolor{tplL4}{HTML}{D62728}
\definecolor{tplL5}{HTML}{9467BD}
\begin{tikzpicture}
\begin{axis}[
width=0.98\linewidth,
height=0.82\linewidth,
xmin=0.0, xmax=1.0,
ymin=0.0, ymax=1.0,
xlabel={Global counterpart score},
ylabel={Local-slice score},
tick label style={font=\normalsize},
label style={font=\normalsize},
xtick={0,0.2,0.4,0.6,0.8,1.0},
ytick={0,0.2,0.4,0.6,0.8,1.0},
xmajorgrids=false, ymajorgrids=false,
grid=none,
axis line style={draw=black!70},
tick style={draw=black!70},
clip=false,
legend columns=1,
legend cell align=left,
legend style={draw=none, fill=none, font=\normalsize, at={(0.05,0.95)}, anchor=north west, row sep=0.04cm},
]
\addplot[black, dashed, line width=1.0pt, forget plot] coordinates {(0,0) (1,1)};
\addplot+[only marks, mark=*, mark size=0.95pt, draw=tplL1, fill=tplL1, fill opacity=0.32, draw opacity=0.45] coordinates { (1.000000,1.000000) (1.000000,1.000000) (1.000000,1.000000) (1.000000,1.000000) (1.000000,1.000000) (1.000000,1.000000) (0.400000,0.400000) (1.000000,1.000000) (1.000000,1.000000) (1.000000,1.000000) (1.000000,1.000000) (1.000000,1.000000) (1.000000,1.000000) (1.000000,1.000000) (1.000000,1.000000) (1.000000,1.000000) (0.000000,0.000000) (1.000000,1.000000) (1.000000,1.000000) (1.000000,1.000000) (1.000000,1.000000) (0.056592,0.048502) (0.001968,0.003377) (0.000196,0.000000) (0.000000,0.000000) (0.059032,0.037397) (0.000000,0.000000) (0.000000,0.000000) (0.000000,0.000000) (0.005772,0.002547) (0.000000,0.000000) (0.019376,0.022343) (1.000000,1.000000) (1.000000,1.000000) (1.000000,1.000000) (1.000000,1.000000) (1.000000,1.000000) (1.000000,1.000000) (1.000000,1.000000) (1.000000,1.000000) (0.500000,0.500000) (1.000000,1.000000) (1.000000,1.000000) (0.617647,0.500000) (0.600000,0.470588) (0.512195,0.390244) (0.500000,0.380952) (0.954545,0.888889) (1.000000,0.941176) (0.875000,0.666667) (0.525000,0.410256) (1.000000,1.000000) (0.677419,0.571429) (0.600000,0.457143) (1.000000,0.866667) (0.937500,0.812500) (0.937500,0.866667) (0.937500,0.812500) (1.000000,1.000000) (1.000000,0.923077) (0.933333,0.928571) (0.937500,0.812500) (1.000000,1.000000) (0.937500,1.000000) (1.000000,0.923077) (1.000000,1.000000) (1.000000,1.000000) (1.000000,1.000000) (0.000000,0.000000) (1.000000,1.000000) (1.000000,1.000000) (1.000000,1.000000) (1.000000,1.000000) (1.000000,1.000000) (0.000000,0.000000) (1.000000,1.000000) (1.000000,1.000000) (1.000000,1.000000) (1.000000,1.000000) (1.000000,1.000000) (1.000000,1.000000) (1.000000,1.000000) (1.000000,0.833333) (1.000000,1.000000) (1.000000,1.000000) (1.000000,1.000000) (0.666667,0.500000) (0.000000,0.000000) (0.001875,0.001871) (0.000207,0.000122) (0.000000,0.000000) (0.887403,0.861965) (0.000000,0.000000) (0.000000,0.000000) (0.000000,0.000000) (0.000000,0.000000) (0.000000,0.000000) (0.000069,0.000044) (1.000000,1.000000) (1.000000,1.000000) (1.000000,1.000000) (1.000000,1.000000) (1.000000,1.000000) (1.000000,1.000000) (1.000000,1.000000) (1.000000,1.000000) (1.000000,1.000000) (1.000000,1.000000) (1.000000,1.000000) (1.000000,1.000000) (1.000000,1.000000) (1.000000,1.000000) (1.000000,1.000000) (1.000000,1.000000) (1.000000,1.000000) (0.055556,0.000000) (1.000000,1.000000) (1.000000,1.000000) (1.000000,1.000000) (0.277778,0.166667) (0.685083,0.680000) (0.631250,0.620155) (0.604762,0.556757) (0.323024,0.250000) (0.855172,0.817391) (0.046358,0.040650) (0.418719,0.387640) (0.393782,0.353293) (0.239726,0.279661) (0.864060,0.829787) (0.722939,0.717770) (0.617936,0.596479) (0.439920,0.429293) (0.035785,0.026713) (0.000994,0.001161) (0.000000,0.000000) (0.719273,0.718584) (0.845703,0.789352) (0.797998,0.785106) (0.348907,0.324042) (1.000000,0.970588) (0.970588,0.911765) (0.823529,0.848485) (0.058824,0.060606) (1.000000,0.970588) (1.000000,0.970588) (0.058824,0.060606) (1.000000,0.970588) (1.000000,1.000000) (0.970588,0.941176) (0.676471,0.666667) (1.000000,1.000000) (0.793103,0.807692) (0.766667,0.777778) (0.000000,0.000000) (0.956522,1.000000) (0.956522,0.809524) (0.724138,0.760000) (0.766667,0.800000) (0.956522,0.952381) (0.846154,1.000000) (0.521739,0.380952) (1.000000,1.000000) (1.000000,1.000000) (1.000000,1.000000) (1.000000,1.000000) (1.000000,1.000000) (1.000000,1.000000) (0.687500,0.593750) (1.000000,1.000000) (1.000000,1.000000) (1.000000,1.000000) (1.000000,1.000000) (1.000000,1.000000) (1.000000,1.000000) (1.000000,1.000000) (0.250000,0.250000) (1.000000,1.000000) (0.250000,0.250000) (0.250000,0.250000) (0.250000,0.250000) (1.000000,1.000000) (0.143713,0.172662) (1.000000,1.000000) (0.003585,0.003881) (0.000000,0.000000) (0.000000,0.000000) (0.000238,0.000000) (0.588060,0.579661) (0.029955,0.029208) (0.004839,0.005435) (0.036439,0.049470) (0.000000,0.000000) (0.071315,0.097540) (0.000000,0.000000) (0.762712,0.741379) (1.000000,0.976744) (0.740000,0.714286) (0.846154,0.840000) (0.888889,0.930233) (0.891304,0.866667) (0.020000,0.021739) (0.833333,0.811321) (1.000000,1.000000) (0.000000,0.000000) (0.734694,0.744681) (0.000000,0.000000) (0.802187,0.280320) (0.722391,0.154321) (0.000000,0.000000) (0.906103,0.490667) (0.000000,0.000000) (0.000000,0.000000) (0.000000,0.000000) (0.000000,0.000000) (0.000000,0.000000) (0.781962,0.293624) (0.000000,1.000000) (0.755102,1.000000) (0.043210,1.000000) (0.000000,1.000000) (0.712500,1.000000) (0.000000,1.000000) (0.000000,1.000000) (0.000000,1.000000) (0.000000,1.000000) (0.000000,1.000000) (0.618557,1.000000) };
\end{axis}
\end{tikzpicture}
\endgroup}
  \caption{Part~\ding{172}: 2D count}
\end{subfigure}\hfill
\begin{subfigure}[t]{0.32\textwidth}
  \centering
  \resizebox{\linewidth}{!}{\centering
\begingroup
\definecolor{tplL1}{HTML}{1F77B4}
\definecolor{tplL2}{HTML}{FF7F0E}
\definecolor{tplL3}{HTML}{2CA02C}
\definecolor{tplL4}{HTML}{D62728}
\definecolor{tplL5}{HTML}{9467BD}
\begin{tikzpicture}
\begin{axis}[
width=0.98\linewidth,
height=0.82\linewidth,
xmin=0.0, xmax=1.0,
ymin=0.0, ymax=1.0,
xlabel={Global counterpart score},
ylabel={Local-slice score},
tick label style={font=\normalsize},
label style={font=\normalsize},
xtick={0,0.2,0.4,0.6,0.8,1.0},
ytick={0,0.2,0.4,0.6,0.8,1.0},
xmajorgrids=false, ymajorgrids=false,
grid=none,
axis line style={draw=black!70},
tick style={draw=black!70},
clip=false,
legend columns=1,
legend cell align=left,
legend style={draw=none, fill=none, font=\normalsize, at={(0.05,0.95)}, anchor=north west, row sep=0.04cm},
]
\addplot[black, dashed, line width=1.0pt, forget plot] coordinates {(0,0) (1,1)};
\addplot+[only marks, mark=*, mark size=0.95pt, draw=tplL2, fill=tplL2, fill opacity=0.32, draw opacity=0.45] coordinates { (0.000000,0.000000) (0.000000,0.000000) (0.000000,0.000000) (0.000000,0.000000) (0.000000,0.000000) (0.000000,0.000000) (0.000000,0.000000) (0.000000,0.000000) (0.000000,0.000000) (0.000000,0.000000) (0.000000,0.000000) (0.000000,0.000000) (0.000000,0.000000) (0.000000,0.000000) (0.000000,0.000000) (0.000000,0.000000) (0.000000,0.000000) (0.000000,0.000000) (0.000000,0.000000) (0.000000,0.000000) (0.000000,0.000000) (0.000000,0.000000) (0.000000,0.000000) (0.000000,0.000000) (0.000000,0.000000) (0.000000,0.000000) (0.000000,0.000000) (0.000000,0.000000) (0.000000,0.000000) };
\end{axis}
\end{tikzpicture}
\endgroup}
  \caption{Part~\ding{173}: median slice}
\end{subfigure}\hfill
\begin{subfigure}[t]{0.32\textwidth}
  \centering
  \resizebox{\linewidth}{!}{\centering
\begingroup
\definecolor{tplL1}{HTML}{1F77B4}
\definecolor{tplL2}{HTML}{FF7F0E}
\definecolor{tplL3}{HTML}{2CA02C}
\definecolor{tplL4}{HTML}{D62728}
\definecolor{tplL5}{HTML}{9467BD}
\begin{tikzpicture}
\begin{axis}[
width=0.98\linewidth,
height=0.82\linewidth,
xmin=0.0, xmax=1.0,
ymin=0.0, ymax=1.0,
xlabel={Global counterpart score},
ylabel={Local-slice score},
tick label style={font=\normalsize},
label style={font=\normalsize},
xtick={0,0.2,0.4,0.6,0.8,1.0},
ytick={0,0.2,0.4,0.6,0.8,1.0},
xmajorgrids=false, ymajorgrids=false,
grid=none,
axis line style={draw=black!70},
tick style={draw=black!70},
clip=false,
legend columns=1,
legend cell align=left,
legend style={draw=none, fill=none, font=\normalsize, at={(0.05,0.95)}, anchor=north west, row sep=0.04cm},
]
\addplot[black, dashed, line width=1.0pt, forget plot] coordinates {(0,0) (1,1)};
\addplot+[only marks, mark=*, mark size=0.95pt, draw=tplL3, fill=tplL3, fill opacity=0.32, draw opacity=0.45] coordinates { (1.000000,1.000000) (1.000000,1.000000) (1.000000,1.000000) (1.000000,1.000000) (1.000000,1.000000) (1.000000,1.000000) (1.000000,1.000000) (1.000000,1.000000) (1.000000,1.000000) (1.000000,1.000000) (1.000000,1.000000) (1.000000,1.000000) (1.000000,1.000000) (1.000000,1.000000) (1.000000,1.000000) (1.000000,1.000000) (1.000000,1.000000) (0.000000,0.000000) (0.000000,0.000000) (1.000000,1.000000) (0.000000,0.000000) (1.000000,0.916667) (1.000000,1.000000) (1.000000,1.000000) (1.000000,1.000000) (0.714286,0.714286) (1.000000,1.000000) (0.714286,0.714286) (0.714286,0.714286) (0.714286,0.714286) (0.833333,0.833333) (0.714286,0.714286) (1.000000,1.000000) (1.000000,1.000000) (1.000000,1.000000) (1.000000,1.000000) (1.000000,1.000000) (1.000000,1.000000) (1.000000,1.000000) (1.000000,1.000000) (1.000000,1.000000) (1.000000,1.000000) (1.000000,1.000000) (1.000000,1.000000) (1.000000,1.000000) (1.000000,1.000000) (1.000000,1.000000) (0.250000,0.250000) (1.000000,1.000000) (0.250000,0.250000) (0.250000,0.250000) (0.250000,0.250000) (1.000000,1.000000) (0.088889,0.400000) (1.000000,1.000000) (0.000000,0.000000) (0.733990,0.391144) (0.226837,0.000000) (0.000000,0.000000) (0.800595,0.808333) (0.000000,0.000000) (0.000000,0.000000) (0.000000,0.000000) (0.000000,0.000000) (0.000000,0.000000) (0.778711,0.758333) (0.333333,1.000000) (0.666667,1.000000) (0.200000,0.000000) (0.666667,0.500000) (0.333333,1.000000) (0.333333,0.000000) (0.666667,1.000000) (0.000000,0.000000) (0.666667,1.000000) (0.333333,0.000000) (0.000000,0.000000) (0.000103,0.000074) (0.000000,0.000000) (0.000000,0.000000) (0.762476,0.704814) (0.000257,0.000155) (0.000000,0.000000) (0.000051,0.000000) (0.000103,0.000164) (0.000051,0.000074) (0.000000,0.000000) (0.007686,0.027692) (0.000618,0.001439) (0.000000,0.000000) (0.000000,0.000000) (0.097561,0.090909) (0.000000,0.000000) (0.022005,0.000000) (0.000000,0.000000) (0.023041,0.036765) (0.000000,0.000000) };
\end{axis}
\end{tikzpicture}
\endgroup}
  \caption{Part~\ding{174}: group quantile}
\end{subfigure}

\vspace{0.6em}
\begin{subfigure}[t]{0.32\textwidth}
  \centering
  \resizebox{\linewidth}{!}{\centering
\begingroup
\definecolor{tplL1}{HTML}{1F77B4}
\definecolor{tplL2}{HTML}{FF7F0E}
\definecolor{tplL3}{HTML}{2CA02C}
\definecolor{tplL4}{HTML}{D62728}
\definecolor{tplL5}{HTML}{9467BD}
\begin{tikzpicture}
\begin{axis}[
width=0.98\linewidth,
height=0.82\linewidth,
xmin=0.0, xmax=1.0,
ymin=0.0, ymax=1.0,
xlabel={Global counterpart score},
ylabel={Local-slice score},
tick label style={font=\normalsize},
label style={font=\normalsize},
xtick={0,0.2,0.4,0.6,0.8,1.0},
ytick={0,0.2,0.4,0.6,0.8,1.0},
xmajorgrids=false, ymajorgrids=false,
grid=none,
axis line style={draw=black!70},
tick style={draw=black!70},
clip=false,
legend columns=1,
legend cell align=left,
legend style={draw=none, fill=none, font=\normalsize, at={(0.05,0.95)}, anchor=north west, row sep=0.04cm},
]
\addplot[black, dashed, line width=1.0pt, forget plot] coordinates {(0,0) (1,1)};
\addplot+[only marks, mark=*, mark size=0.95pt, draw=tplL4, fill=tplL4, fill opacity=0.32, draw opacity=0.45] coordinates { (1.000000,0.991525) (0.991525,1.000000) (1.000000,0.991525) (0.000000,0.000000) (0.177966,0.119658) (0.000000,0.000000) (0.000000,0.000000) (0.000000,0.000000) (0.991525,0.991453) (0.000000,0.000000) (0.983051,0.914530) (0.904762,1.000000) (0.078125,0.454545) (0.825397,0.500000) (0.000000,0.000000) (0.047619,0.000000) (0.000000,0.000000) (0.000000,0.000000) (0.857143,0.500000) (0.000000,0.000000) (0.412698,0.900000) (1.000000,1.000000) (1.000000,1.000000) (1.000000,1.000000) (0.000000,0.000000) (1.000000,1.000000) (0.000000,0.000000) (1.000000,1.000000) (0.000000,0.000000) (1.000000,1.000000) (0.000000,0.000000) (1.000000,1.000000) (1.000000,1.000000) (1.000000,1.000000) (1.000000,1.000000) (0.000000,0.000000) (1.000000,1.000000) (0.000000,0.000000) (0.000000,0.000000) (1.000000,1.000000) (0.000000,0.000000) (1.000000,1.000000) (0.000000,0.000000) (0.759494,0.652778) (0.708861,0.671053) (0.000000,0.000000) (0.647059,0.524590) (0.000000,0.000000) (0.000000,0.000000) (0.000000,0.000000) (0.000000,0.000000) (0.746667,0.765625) };
\end{axis}
\end{tikzpicture}
\endgroup}
  \caption{Part~\ding{175}: time-bucket count}
\end{subfigure}\hspace{0.08\textwidth}
\begin{subfigure}[t]{0.32\textwidth}
  \centering
  \resizebox{\linewidth}{!}{\centering
\begingroup
\definecolor{tplL1}{HTML}{1F77B4}
\definecolor{tplL2}{HTML}{FF7F0E}
\definecolor{tplL3}{HTML}{2CA02C}
\definecolor{tplL4}{HTML}{D62728}
\definecolor{tplL5}{HTML}{9467BD}
\begin{tikzpicture}
\begin{axis}[
width=0.98\linewidth,
height=0.82\linewidth,
xmin=0.0, xmax=1.0,
ymin=0.0, ymax=1.0,
xlabel={Global counterpart score},
ylabel={Local-slice score},
tick label style={font=\normalsize},
label style={font=\normalsize},
xtick={0,0.2,0.4,0.6,0.8,1.0},
ytick={0,0.2,0.4,0.6,0.8,1.0},
xmajorgrids=false, ymajorgrids=false,
grid=none,
axis line style={draw=black!70},
tick style={draw=black!70},
clip=false,
legend columns=1,
legend cell align=left,
legend style={draw=none, fill=none, font=\normalsize, at={(0.05,0.95)}, anchor=north west, row sep=0.04cm},
]
\addplot[black, dashed, line width=1.0pt, forget plot] coordinates {(0,0) (1,1)};
\addplot+[only marks, mark=*, mark size=0.95pt, draw=tplL5, fill=tplL5, fill opacity=0.32, draw opacity=0.45] coordinates { (0.952991,0.882096) (0.952586,0.892377) (0.919149,0.672566) (0.000000,0.000000) (0.116379,0.062780) (0.000000,0.000000) (0.000000,0.000000) (0.000000,0.000000) (0.982759,0.888889) (0.000000,0.000000) (0.892241,0.855895) };
\end{axis}
\end{tikzpicture}
\endgroup}
  \caption{Part~\ding{176}: moving average}
\end{subfigure}
\captionsetup{skip=2pt}
\caption{\textbf{Local-slice conditional query parts.} Each panel uses the same paired construction as Figure~\ref{fig:local-vs-defiltered-scatter}: the x-axis scores the query after deleting the local \texttt{WHERE} predicate, and the y-axis scores the original local-slice query. The diagonal marks equal fidelity between the global counterpart and the filtered local slice.}
\label{fig:appendix-local-slice-query-parts}
\end{figure*}

\subsubsection{Tail / Rarity}
\label{app:tail_rarity_diagnostics}

\textbf{Goal.}
Evaluate whether synthetic data preserves low-support and rare-region
behavior instead of washing rare cases into the distributional head.

\textbf{Template-side branches.}
\begin{enumerate}
    \item \textbf{Tail Coverage (4 templates).} Queries that identify
    rare members, outliers, or extreme cases and test whether the same
    tail-support region remains recoverable after synthesis.
    \item \textbf{Tail Size (4 templates).} Queries that measure how
    much mass, level, or threshold exceedance is assigned to a rare
    region and test whether synthetic data preserves the scale of that
    tail event.
\end{enumerate}

Table~\ref{tab:appendix_tail_templates} lists every Tail / Rarity
template and explains whether it probes rare-region identity or the mass
assigned to the rare region.
\begingroup
\scriptsize
\setlength{\tabcolsep}{2.4pt}
\renewcommand{\arraystretch}{1.08}
\setlength{\LTleft}{0pt}
\setlength{\LTright}{0pt}
\begin{longtable}{@{}>{\raggedright\arraybackslash}p{\dimexpr0.090343\linewidth-0.542056\tabcolsep\relax}%
                  >{\raggedright\arraybackslash}p{\dimexpr0.218069\linewidth-1.308411\tabcolsep\relax}%
                  >{\raggedright\arraybackslash}p{\dimexpr0.345794\linewidth-2.074766\tabcolsep\relax}%
                  >{\raggedright\arraybackslash}p{\dimexpr0.345794\linewidth-2.074766\tabcolsep\relax}@{}}
\caption{\textbf{Tail / Rarity template inventory.} Each row states
the rare-region behavior tested by the template and why it belongs to
the Tail Coverage or Tail Size branch.}
\label{tab:appendix_tail_templates}\\
\toprule
\textbf{ID} & \textbf{Canonical template} & \textbf{What it probes} &
\textbf{Why this branch} \\
\midrule
\endfirsthead
\caption[]{\textbf{Tail / Rarity template inventory (continued).}}\\
\toprule
\textbf{ID} & \textbf{Canonical template} & \textbf{What it probes} &
\textbf{Why this branch} \\
\midrule
\endhead
\midrule
\multicolumn{4}{r}{\textit{Continued on next page}}\\
\endfoot
\bottomrule
\endlastfoot

\texttt{T-1.1} & Top-Quantile Tail Members &
Identifies members in an extreme quantile region and tests whether the
same rare values remain recoverable after synthesis. &
\emph{Tail Coverage}: the primary object is the identity of the rare
members or support region, not its aggregate mass. \\

\texttt{T-1.2} & Global Z-Score Outliers &
Finds observations with unusually large standardized deviations from the
global center, testing preservation of globally rare extremes. &
\emph{Tail Coverage}: the query asks whether the rare outlier support is
still present and identifiable. \\

\texttt{T-1.3} & Within-Group Top-N Extremes &
Finds the top extreme observations separately within each group,
testing subgroup-relative rare support. &
\emph{Tail Coverage}: it evaluates which rare members occur in each
group, rather than only how much total mass is extreme. \\

\texttt{T-1.4} & Subgroup-Relative Outliers &
Detects observations that are extreme relative to their subgroup,
testing whether conditional rare cases survive synthesis. &
\emph{Tail Coverage}: the target is the identity of subgroup-relative
outliers and their recoverability. \\

\texttt{T-2.1} & Groups Above a Total-Share Threshold &
Counts or identifies groups whose total exceeds a specified share of
the overall total, testing the amount of mass in a rare high-contribution
region. &
\emph{Tail Size}: the query emphasizes how much aggregate mass crosses
the threshold rather than the exact identity of individual outliers. \\

\texttt{T-2.2} & Above-Threshold Group Ranking &
Ranks groups whose aggregate values exceed a threshold, testing the
scale and ordering of threshold-defined tail responses. &
\emph{Tail Size}: the main object is the magnitude of threshold-exceeding
group behavior and its concentration. \\

\texttt{T-2.3} & Group-wise Percentile Level &
Reports percentile levels within groups, testing whether the scale of
upper or lower tail responses is preserved. &
\emph{Tail Size}: percentile values characterize the level or mass of a
tail event rather than enumerating rare identities. \\

\texttt{T-2.4} & Threshold Exceedance Rarity &
Measures how rarely a value exceeds a specified threshold, testing the
amount of probability mass assigned to an extreme event. &
\emph{Tail Size}: it directly measures rare-event frequency or mass
relative to the threshold. \\

\end{longtable}
\endgroup

\textbf{Threshold-stress diagnostic.}
This appendix also gives the detailed definition and measurement protocol
for the tail and rarity diagnostic used in the main-text threshold-stress
analysis. Unless otherwise stated, the paper uses $\tau=3\%$ as the
default tail threshold; the threshold-stress figures additionally sweep
$\tau\in\{10\%,8\%,6\%,4\%,3\%,2\%,1\%,0.5\%,0.1\%\}$. The diagnostic is
complementary to the grounded Tail Coverage and Tail Size SQL templates
described in the template taxonomy.

\textbf{Column-level tail region.}
Let $R$ denote the real training split, let $S$ denote the synthetic
table, and let $\tau$ be a tail-mass threshold. The real tail region
$T_R^{(\tau)}$ is defined separately for each column. For a continuous
column, the empirical CDF of $R$ determines the lower and/or upper
quantile interval(s) whose total probability mass is $\tau$. For a
categorical column, categories are ordered by increasing frequency in
$R$, and the least frequent categories are accumulated until their
cumulative frequency reaches $\tau$. Thus, the categorical tail is a set
of rare states, whereas the continuous tail is a rare interval or union
of intervals.

\textbf{Tail size.}
Tail size measures whether the synthetic table places the appropriate
amount of mass in the real-defined tail. We first fix
$T_R^{(\tau)}$ using the real data, then count the real and synthetic
mass assigned to this same region. Let
\[
  m_R^{(\tau)}
  =
  \frac{\#\{x\in R:x\in T_R^{(\tau)}\}}{|R|},
  \qquad
  m_S^{(\tau)}
  =
  \frac{\#\{x\in S:x\in T_R^{(\tau)}\}}{|S|}.
\]
For categorical columns, the implementation uses the bounded relative-error
score
\[
  \operatorname{Size}^{(\tau)}
  =
  \operatorname{clip}\!\left(
    1-\frac{|m_S^{(\tau)}-m_R^{(\tau)}|}{m_R^{(\tau)}},
    0,1
\right),
\]
when $m_R^{(\tau)}>0$; when $m_R^{(\tau)}=0$, the score is $1$ if
$m_S^{(\tau)}=0$ and $0$ otherwise.
For continuous columns, the real lower and upper tail cutoffs define two
target tail masses. If $m_{S,-}^{(\tau)}$ and $m_{S,+}^{(\tau)}$ are the
synthetic masses below and above those real cutoffs, respectively, the
implementation uses
\[
  \operatorname{Size}^{(\tau)}
  =
  \frac{1}{2}
  \left[
    \min\!\left(\frac{m_{S,-}^{(\tau)}}{\tau},1\right)
    +
    \min\!\left(\frac{m_{S,+}^{(\tau)}}{\tau},1\right)
  \right].
\]
Thus the reported tail-size score is explicitly bounded in $[0,1]$.
It measures preservation of tail mass and does not require the synthetic
table to identify exactly the same rare states or interval endpoints.

\textbf{Tail coverage.}
Tail coverage measures whether the synthetic table recovers the identity
of the rare region. We independently construct the corresponding
synthetic tail $T_S^{(\tau)}$ using the same threshold rule and compute
its overlap with the real tail:
\[
  \operatorname{Coverage}^{(\tau)}
  =
  \frac{\mu\!\left(T_R^{(\tau)}\cap T_S^{(\tau)}\right)}
       {\mu\!\left(T_R^{(\tau)}\right)}.
\]
For categorical columns, $\mu$ is set cardinality, so coverage is the
fraction of real tail categories recovered by the synthetic tail. For
continuous columns, $\mu$ is interval length (or the corresponding
one-dimensional measure), so coverage is the fraction of the real tail
range covered by the synthetic tail range.

\textbf{Tail overall and threshold stress.}
The column-level tail overall score is the equally weighted combination
\[
  \operatorname{TailOverall}^{(\tau)}
  =
  \frac{1}{2}
  \left(
    \operatorname{Size}^{(\tau)}
    +
    \operatorname{Coverage}^{(\tau)}
  \right).
\]
Because both component scores are bounded in $[0,1]$, the resulting
TailOverall score is also bounded in $[0,1]$.
The threshold-stress analysis evaluates these quantities over
$\tau\in\{10\%,8\%,6\%,4\%,3\%,2\%,1\%,0.5\%,0.1\%\}$. The resulting
curves report absolute mean tail overall scores at each threshold, not
scores normalized to the $10\%$ baseline. Tightening $\tau$ therefore
tests whether a model continues to preserve increasingly rare support,
while the separate size and coverage curves distinguish loss of tail
mass from loss of rare-region identity.

\subsubsection{Missingness}
\label{app:experiment_missingness}

\textbf{Goal.}
Treat native missingness as an analyzable signal rather than as noise to
be removed before fidelity assessment.

\textbf{Template-side branches.}
\begin{enumerate}
    \item \textbf{Marginal Missingness (1 template).} Leaves the
    missing-rate signal unconditional and measures only the overall
    missing-rate profile.
    \item \textbf{Broad Co-Missingness (2 templates).} Reuses the same
    missingness target but conditions it on either a discrete state or a
    continuous bucket, producing broad state-conditioned missingness
    profiles.
\end{enumerate}

Table~\ref{tab:appendix_missingness_templates} lists every Missingness
template and explains whether it measures marginal missingness or
state-conditioned co-missingness.
\begingroup
\scriptsize
\setlength{\tabcolsep}{2.4pt}
\renewcommand{\arraystretch}{1.08}
\setlength{\LTleft}{0pt}
\setlength{\LTright}{0pt}
\begin{longtable}{@{}>{\raggedright\arraybackslash}p{\dimexpr0.090343\linewidth-0.542056\tabcolsep\relax}%
                  >{\raggedright\arraybackslash}p{\dimexpr0.218069\linewidth-1.308411\tabcolsep\relax}%
                  >{\raggedright\arraybackslash}p{\dimexpr0.345794\linewidth-2.074766\tabcolsep\relax}%
                  >{\raggedright\arraybackslash}p{\dimexpr0.345794\linewidth-2.074766\tabcolsep\relax}@{}}
\caption{\textbf{Missingness template inventory.} Each row states the
missingness behavior tested by the template and why it belongs to the
Marginal Missingness or Broad Co-Missingness branch.}
\label{tab:appendix_missingness_templates}\\
\toprule
\textbf{ID} & \textbf{Canonical template} & \textbf{What it probes} &
\textbf{Why this branch} \\
\midrule
\endfirsthead
\caption[]{\textbf{Missingness template inventory (continued).}}\\
\toprule
\textbf{ID} & \textbf{Canonical template} & \textbf{What it probes} &
\textbf{Why this branch} \\
\midrule
\endhead
\midrule
\multicolumn{4}{r}{\textit{Continued on next page}}\\
\endfoot
\bottomrule
\endlastfoot

\texttt{M-1.1} & Marginal Missing Rate &
Measures the overall fraction of missing values in a target column,
testing whether the global missingness level is preserved. &
\emph{Marginal Missingness}: it has no conditioning attribute and
therefore probes only the unconditional missing-rate profile. \\

\texttt{M-2.1} & Missing Rate by Discrete State &
Measures the missing rate of a target column separately for each
discrete state, testing structured co-missingness across categories. &
\emph{Broad Co-Missingness}: missingness is conditioned on a categorical
state, exposing dependence between the target and another attribute. \\

\texttt{M-2.2} & Missing Rate by Continuous Bucket &
Measures the missing rate of a target column across buckets of a
continuous conditioning attribute. &
\emph{Broad Co-Missingness}: bucketing creates a structured conditional
profile without requiring exact continuous-value matching. \\

\end{longtable}
\endgroup

\subsubsection{Cardinality / Range}
\label{app:experiment_cardinality_range}

\textbf{Goal.}
Evaluate whether observed support richness and numerical coverage
envelopes remain analytically usable after synthesis.

\textbf{Template-side branches.}
\begin{enumerate}
    \item \textbf{Discrete (3 templates).} Support/rank/concentration
    views over observed discrete values, including support ranking,
    support-share profiling, and high-cardinality response summaries.
    \item \textbf{Continuous (1 template).} Min/max and range-envelope
    views over numerical coverage.
\end{enumerate}

Table~\ref{tab:appendix_cardinality_templates} lists every
Cardinality / Range template and explains whether it probes discrete
support richness or continuous numerical coverage.
\begingroup
\scriptsize
\setlength{\tabcolsep}{2.4pt}
\renewcommand{\arraystretch}{1.08}
\setlength{\LTleft}{0pt}
\setlength{\LTright}{0pt}
\begin{longtable}{@{}>{\raggedright\arraybackslash}p{\dimexpr0.090343\linewidth-0.542056\tabcolsep\relax}%
                  >{\raggedright\arraybackslash}p{\dimexpr0.218069\linewidth-1.308411\tabcolsep\relax}%
                  >{\raggedright\arraybackslash}p{\dimexpr0.345794\linewidth-2.074766\tabcolsep\relax}%
                  >{\raggedright\arraybackslash}p{\dimexpr0.345794\linewidth-2.074766\tabcolsep\relax}@{}}
\caption{\textbf{Cardinality / Range template inventory.} Each row
states the support or range behavior tested by the template and why it
belongs to the Discrete or Continuous branch.}
\label{tab:appendix_cardinality_templates}\\
\toprule
\textbf{ID} & \textbf{Canonical template} & \textbf{What it probes} &
\textbf{Why this branch} \\
\midrule
\endfirsthead
\caption[]{\textbf{Cardinality / Range template inventory (continued).}}\\
\toprule
\textbf{ID} & \textbf{Canonical template} & \textbf{What it probes} &
\textbf{Why this branch} \\
\midrule
\endhead
\midrule
\multicolumn{4}{r}{\textit{Continued on next page}}\\
\endfoot
\bottomrule
\endlastfoot

\texttt{K-1.1} & Support Rank Profile &
Ranks observed discrete values or categories by support, testing
whether the ordering of common and rare states is retained. &
\emph{Discrete}: the target is the identity and rank of discrete support
values, not a continuous interval. \\

\texttt{K-1.2} & Distinct Share Profile &
Measures how support mass is distributed across distinct values,
testing concentration and share structure within a discrete domain. &
\emph{Discrete}: it describes shares over enumerated categories or
values and their support concentration. \\

\texttt{K-1.3} & Continuous Range Envelope Profile &
Measures the lower and upper numerical envelope of a continuous column,
testing whether its usable range remains covered. &
\emph{Continuous}: the analytical object is an interval or numerical
range rather than a finite set of category identities. \\

\texttt{K-2.1} & High-Cardinality Response Stability &
Tests whether responses remain stable when the query operates over a
large discrete support with many distinct values. &
\emph{Discrete}: the challenge comes from preserving a rich finite
support and its response behavior, even when the cardinality is high. \\

\end{longtable}
\endgroup

\clearpage
\section{Query Grounding}
\label{app:query_generation}

\subsection{Pipeline Overview}
\label{app:query_generation_overview}
The query-generation pipeline has two stages. It first extracts reusable analytical templates from public SQL workloads and assigns them to five query families---subgroup, conditional, tail/rarity, missingness, and cardinality/range. It then grounds applicable templates to each dataset schema, realizes executable SQL under fixed constraints, and records the resulting semantic answer contract. The source inventory and taxonomy are detailed in Appendix~\ref{app:workload_sources} and Appendix~\ref{app:template_taxonomy_catalog}; the corresponding answer-scoring rules are given in Appendix~\ref{app:query_scoring}.

\subsection{Stage 1: Template Construction}
\label{app:query_template_construction}

\textbf{Source acquisition and provenance.}
We construct the query-template library from public analytical workloads,
including TPC-H, TPC-DS, ClickBench, H2O db-benchmark, and RTABench.
Each source query is retained with its workload identifier, source-query
identifier, and bibliographic provenance. The purpose of this step is to
preserve reusable analytical intent rather than to reproduce the complete
execution protocol of any individual database benchmark.

\textbf{Canonical template construction.}
We abstract each source query into a single-table SQL skeleton whose
dataset-dependent elements are represented by typed roles and bounded
parameters. Typical roles include grouping columns, measure columns,
predicate columns, condition columns, entity columns, and time columns.
The skeleton preserves the analytical structure, while column names,
observed values, thresholds, and other dataset-specific choices remain
unbound until the grounding stage.

\textbf{Template records and family assignment.}
Each template record stores a stable identifier, analytical intent, SQL
skeleton, required and optional roles, binding constraints, query-family
assignment, source provenance, portability metadata, and release status.
Templates are assigned to the five query families described in
Appendix~\ref{app:query_families}; the family label describes the primary
analytical object of the template and does not change its source provenance.
The concrete source-to-template mappings are listed in
Appendix~\ref{app:workload_sources}.

\textbf{Screening and materialization policy.}
The construction process screens templates for duplicate identifiers,
missing provenance, invalid family mappings, unsupported SQL features, and
violations of the single-table benchmark contract. Templates that satisfy
the portability and execution requirements enter the core library. Other
well-specified templates may be retained as extensions or prior-only
material, but are not silently included in the released workload.

\textbf{Versioned construction and auditability.}
Template construction is deterministic and versioned. Each library build
records the hashes of its input catalogs and source-query bank, materialized
template counts, skipped mappings, duplicate identifiers, and workload
coverage. This makes every released grounded workload traceable to a
specific source library and provides an audit trail for later library
updates.

\subsection{Stage 2: Dataset Grounding and SQL Realization}
\label{app:query_grounding}

\textbf{Dataset profiling.}
For each input dataset, we first construct a typed profile containing the
schema, row count, column types, cardinalities, missingness indicators,
observed categorical values, and candidate numeric thresholds. This profile
defines the admissible search space for grounding and prevents the
realization step from inventing schema elements or predicate values.

\begin{figure*}[t]
\centering
\includegraphics[width=\textwidth]{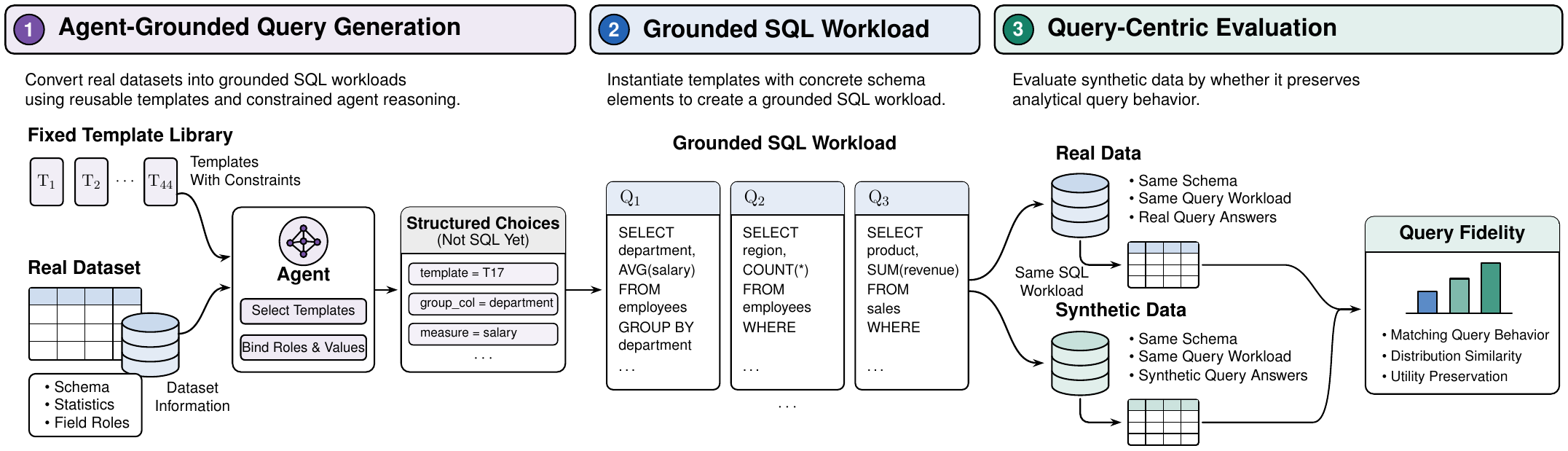}
\captionsetup{skip=2pt}
\caption{Agent query grounding pipeline.}
\label{fig:stage2-grounding-pipeline}
\end{figure*}

\textbf{Applicability screening and binding.}
For every ready template, the grounding procedure checks whether all
required roles have compatible candidates and whether structural constraints,
such as distinct grouping columns or ordered temporal roles, can be
satisfied. Templates with unavailable roles, incompatible types, or
unrealizable predicates are skipped with an explicit reason. For applicable
templates, the agent selects dataset columns and the configured binding procedure selects values from the
admissible candidate set; the template skeleton and its required constraints
remain fixed.

\textbf{Deterministic validation and realization.}
Bindings are validated against the template contract before SQL generation.
The validator checks required and unknown placeholders, candidate-column
membership, allowed operators and values, observed categorical values,
bounded numeric parameters, distinct-role constraints, and temporal
ordering. A deterministic renderer then instantiates the SQL skeleton with
the validated bindings. Thus, dataset-specific choices are separated from
the reusable analytical structure.

\textbf{Execution filtering and retry policy.}
Each realized query is executed in an isolated SQLite environment under a
fixed timeout. We retain only read-only, executable queries that return at
least one row and are not entirely NULL. Invalid bindings, invalid SQL,
timeouts, execution failures, empty results, and NULL-only results are
rejected and may be retried under the configured grounding budget. Accepted
queries are kept in deterministic template order.

\textbf{Released artifacts and audit records.}
The generated bundle contains the accepted SQL, a machine-readable query
registry, a manifest, and the SQLite materialization used for execution
checks. Each accepted query records its template identifier, source
provenance, template hash, concrete role bindings, semantic result contract,
output columns, and preview rows. Skipped templates retain explicit failure
reasons, while the manifest records input and template-library hashes,
runtime metadata, timings, and usage counters. These records allow a
released workload to be audited and regenerated without exposing raw model
responses or server-local paths.

\begin{algorithm}[!h]
\caption{Template Grounding and Constrained SQL Realization}
\label{alg:realize}
\scriptsize
\begin{algorithmic}[1]
\Require dataset schema and profile; template pool with \texttt{can\_vary} and \texttt{must\_fix}; retry budget $B$
\Ensure accepted grounded problems and accepted SQL realizations
\ForAll{templates that are bindable on the dataset}
  \State construct admissible dataset-specific bindings
  \ForAll{admissible bindings}
    \For{$b = 1$ to $B$}
      \State realize a grounded problem and its main/check SQL pair
      \State validate type compatibility, must-fix preservation, and executability
      \If{both SQL queries pass validation}
        \State accept the grounded problem and stop retrying
        \State \textbf{break}
      \EndIf
    \EndFor
  \EndFor
\EndFor
\end{algorithmic}
\end{algorithm}

Algorithm~\ref{alg:realize} summarizes the Stage 2 grounding and
constrained SQL realization loop used after a template has been selected
for a dataset.

\subsection{Agent API Usage by Dataset}
\label{app:agent_api_usage}

This appendix records dataset-level API usage.

This table group reports the recorded token usage of the GLM V11 workload builders in the
repository release manifests and per-dataset grounding inventories. Costs are computed per
million tokens using the provider rates shown below: GLM-5.3-Flash uses
$\$0.15$, $\$0.03$, and $\$0.50$; and GLM-5.3 uses $\$1.40$, $\$0.26$, and
$\$4.40$, respectively. These are standard API rates, not the separate Z.AI Coding Plan
quota accounting. The repository contains three released runs for each GLM model; the
GPT-5.4 grounding usage is reported in the table immediately above.\footnote{Pricing page
checked on 2026-09-22: \url{https://docs.z.ai/guides/overview/pricing}.}

\begingroup
\fontsize{6.6}{7.5}\selectfont
\setlength{\tabcolsep}{2.5pt}
\renewcommand{\arraystretch}{1.08}
\begin{center}
\captionof{table}{\textbf{Total recorded V11 grounding usage and estimated API cost.} Each row is one released run; repeated suffixes denote independent reruns. Dataset-level training rows and column counts are reported in the three tables below.}
\label{tab:appendix_v11_token_summary}
\begin{tabular}{@{}lrrrrrrr@{}}
\toprule
Run & Model & Queries & Input tok. & Cached in. & Output tok. & Total tok. & Cost (USD) \\
\midrule
v11.2.1-2\_glm-5.3-flash & GLM-5.3 Flash & 2,729 & 25,976,126 & 2,331,072 & 672,327 & 26,648,453 & 3.95 \\
v11.2.1-3\_glm-5.3-flash & GLM-5.3 Flash & 2,734 & 25,972,319 & 3,159,360 & 695,009 & 26,667,328 & 3.86 \\
v11.2.1\_glm-5.3-flash & GLM-5.3 Flash & 2,734 & 25,986,950 & 581,696 & 696,387 & 26,683,337 & 4.18 \\
v11.2.2-2\_glm-5.3 & GLM-5.3 & 2,761 & 25,957,914 & 1,272,256 & 642,610 & 26,600,524 & 37.72 \\
v11.2.2-3\_glm-5.3 & GLM-5.3 & 2,760 & 25,947,520 & 3,328,704 & 700,987 & 26,648,507 & 35.62 \\
v11.2.2\_glm-5.3 & GLM-5.3 & 2,763 & 25,917,017 & 697,088 & 628,253 & 26,545,270 & 38.25 \\
\midrule
\multicolumn{2}{@{}l}{\textbf{All recorded GLM runs}} & 16,481 & 155,757,846 & 11,370,176 & 4,035,573 & 159,793,419 & 123.58 \\
\bottomrule
\end{tabular}
\end{center}

\begin{longtable}{@{}lrrrrrrrr@{}}
\caption{\textbf{Dataset-level recorded usage for GLM-5.3 Flash.}}\\
\label{tab:appendix_v11_glm_53_flash_dataset_usage}\\
\toprule
Dataset & Train rows & Cols & Queries & Input tok. & Cached in. & Output tok. & Total tok. & Cost (USD) \\
\midrule
\endfirsthead
\multicolumn{9}{l}{\textit{Table~\thetable{} continued.}}\\
\toprule
Dataset & Train rows & Cols & Queries & Input tok. & Cached in. & Output tok. & Total tok. & Cost (USD) \\
\midrule
\endhead
\midrule
\multicolumn{9}{r}{\textit{Continued on next page.}}\\
\endfoot
\bottomrule
\endlastfoot
c1 & 35,855 & 7 & 78 & 196,279 & 41,344 & 35,143 & 231,422 & 0.04 \\
c2 & 1,382 & 7 & 74 & 111,577 & 14,272 & 28,515 & 140,092 & 0.03 \\
c3 & 2,551 & 3 & 53 & 99,837 & 14,656 & 48,144 & 147,981 & 0.04 \\
c4 & 2,557 & 37 & 102 & 350,215 & 71,616 & 28,398 & 378,613 & 0.06 \\
c5 & 6,732 & 23 & 188 & 357,025 & 27,392 & 30,634 & 387,659 & 0.07 \\
c6 & 7,636 & 8 & 115 & 200,077 & 33,664 & 42,892 & 242,969 & 0.05 \\
c7 & 10,368 & 9 & 84 & 128,403 & 9,280 & 27,709 & 156,112 & 0.03 \\
c8 & 8,844 & 31 & 103 & 446,067 & 40,704 & 37,367 & 483,434 & 0.08 \\
c9 & 26,215 & 10 & 158 & 294,305 & 41,280 & 45,961 & 340,266 & 0.06 \\
c10 & 820,008 & 11 & 102 & 279,888 & 12,928 & 48,916 & 328,804 & 0.06 \\
c11 & 54,045 & 43 & 107 & 533,205 & 10,304 & 32,496 & 565,701 & 0.09 \\
c12 & 2,623 & 1,559 & 134 & 16,275,641 & 1,811,264 & 54,088 & 16,329,729 & 2.25 \\
c13 & 1,966,628 & 69 & 118 & 1,385,199 & 194,304 & 44,800 & 1,429,999 & 0.21 \\
c14 & 240,000 & 25 & 144 & 527,476 & 58,816 & 47,094 & 574,570 & 0.10 \\
c15 & 480,000 & 25 & 253 & 534,431 & 59,136 & 42,172 & 576,603 & 0.09 \\
c16 & 5,516 & 13 & 255 & 290,332 & 34,752 & 43,887 & 334,219 & 0.06 \\
c17 & 7,045 & 12 & 231 & 270,733 & 6,912 & 37,710 & 308,443 & 0.06 \\
c18 & 103,976 & 14 & 251 & 379,469 & 6,848 & 36,275 & 415,744 & 0.07 \\
c19 & 32,759 & 16 & 217 & 651,774 & 35,904 & 42,952 & 694,726 & 0.11 \\
m1 & 1,200 & 30 & 171 & 607,711 & 38,400 & 41,891 & 649,602 & 0.11 \\
m2 & 46,873 & 44 & 152 & 747,716 & 116,544 & 47,448 & 795,164 & 0.12 \\
m3 & 95,512 & 32 & 242 & 680,027 & 59,968 & 48,583 & 728,610 & 0.12 \\
m4 & 2,217 & 7 & 107 & 173,539 & 6,976 & 34,245 & 207,784 & 0.04 \\
m5 & 3,539 & 37 & 179 & 763,702 & 47,424 & 41,089 & 804,791 & 0.13 \\
m6 & 9,864 & 18 & 168 & 392,160 & 75,584 & 39,145 & 431,305 & 0.07 \\
m7 & 4,088 & 12 & 227 & 245,685 & 18,368 & 36,079 & 281,764 & 0.05 \\
m8 & 36,168 & 17 & 249 & 327,447 & 38,336 & 39,174 & 366,621 & 0.06 \\
m9 & 15,326 & 14 & 245 & 318,108 & 45,376 & 40,045 & 358,153 & 0.06 \\
m10 & 1,600 & 21 & 164 & 443,730 & 6,656 & 49,438 & 493,168 & 0.09 \\
m11 & 304,887 & 12 & 160 & 279,180 & 31,040 & 34,648 & 313,828 & 0.06 \\
n1 & 3,680 & 58 & 171 & 1,155,956 & 68,672 & 53,153 & 1,209,109 & 0.19 \\
n2 & 1,202 & 6 & 111 & 184,436 & 21,824 & 32,630 & 217,066 & 0.04 \\
n3 & 3,918 & 12 & 173 & 329,433 & 15,360 & 36,948 & 366,381 & 0.07 \\
n4 & 1,595 & 128 & 283 & 3,109,939 & 328,640 & 41,941 & 3,151,880 & 0.45 \\
n5 & 17,010 & 82 & 177 & 2,415,926 & 83,584 & 42,230 & 2,458,156 & 0.37 \\
n6 & 6,400 & 17 & 176 & 429,287 & 61,184 & 41,786 & 471,073 & 0.08 \\
n7 & 5,756 & 26 & 180 & 839,130 & 29,248 & 44,448 & 883,578 & 0.14 \\
n8 & 1,253 & 593 & 284 & 13,525,624 & 384,320 & 72,680 & 13,598,304 & 2.02 \\
n9 & 8,794 & 17 & 176 & 428,065 & 39,360 & 43,966 & 472,031 & 0.08 \\
n10 & 10,888 & 17 & 180 & 563,229 & 7,232 & 42,177 & 605,406 & 0.10 \\
n11 & 15,215 & 11 & 168 & 325,431 & 26,176 & 44,603 & 370,034 & 0.07 \\
n12 & 196,045 & 4 & 95 & 142,027 & 6,976 & 37,189 & 179,216 & 0.04 \\
n13 & 7,654 & 5 & 111 & 174,840 & 14,720 & 39,776 & 214,616 & 0.04 \\
n14 & 1,600 & 52 & 175 & 1,326,726 & 89,728 & 50,912 & 1,377,638 & 0.21 \\
n15 & 7,857 & 86 & 115 & 1,805,712 & 153,984 & 39,153 & 1,844,865 & 0.27 \\
n16 & 227,845 & 31 & 173 & 971,557 & 7,360 & 47,047 & 1,018,604 & 0.17 \\
n17 & 11,600 & 10 & 155 & 267,126 & 19,520 & 46,601 & 313,727 & 0.06 \\
n18 & 60,800 & 171 & 287 & 4,189,781 & 238,528 & 50,006 & 4,239,787 & 0.62 \\
n19 & 56,000 & 785 & 176 & 17,460,232 & 1,465,664 & 49,539 & 17,509,771 & 2.47 \\
\end{longtable}
\par\smallskip
\begin{longtable}{@{}lrrrrrrrr@{}}
\caption{\textbf{Dataset-level recorded usage for GLM-5.3.}}\\
\label{tab:appendix_v11_glm_53_dataset_usage}\\
\toprule
Dataset & Train rows & Cols & Queries & Input tok. & Cached in. & Output tok. & Total tok. & Cost (USD) \\
\midrule
\endfirsthead
\multicolumn{9}{l}{\textit{Table~\thetable{} continued.}}\\
\toprule
Dataset & Train rows & Cols & Queries & Input tok. & Cached in. & Output tok. & Total tok. & Cost (USD) \\
\midrule
\endhead
\midrule
\multicolumn{9}{r}{\textit{Continued on next page.}}\\
\endfoot
\bottomrule
\endlastfoot
c1 & 35,855 & 7 & 78 & 196,283 & 52,544 & 46,535 & 242,818 & 0.42 \\
c2 & 1,382 & 7 & 76 & 118,037 & 25,536 & 29,787 & 147,824 & 0.27 \\
c3 & 2,551 & 3 & 54 & 102,366 & 12,544 & 48,260 & 150,626 & 0.34 \\
c4 & 2,557 & 37 & 102 & 350,190 & 46,336 & 24,222 & 374,412 & 0.54 \\
c5 & 6,732 & 23 & 186 & 335,115 & 93,952 & 21,420 & 356,535 & 0.46 \\
c6 & 7,636 & 8 & 120 & 205,088 & 36,288 & 39,724 & 244,812 & 0.42 \\
c7 & 10,368 & 9 & 88 & 144,437 & 25,344 & 30,933 & 175,370 & 0.31 \\
c8 & 8,844 & 31 & 104 & 446,636 & 51,648 & 32,411 & 479,047 & 0.71 \\
c9 & 26,215 & 10 & 157 & 286,835 & 49,024 & 33,692 & 320,527 & 0.49 \\
c10 & 820,008 & 11 & 102 & 279,286 & 48,384 & 33,323 & 312,609 & 0.48 \\
c11 & 54,045 & 43 & 106 & 518,286 & 101,760 & 28,802 & 547,088 & 0.74 \\
c12 & 2,623 & 1,559 & 134 & 16,227,063 & 1,390,528 & 48,828 & 16,275,891 & 21.35 \\
c13 & 1,966,628 & 69 & 120 & 1,386,089 & 179,776 & 36,799 & 1,422,888 & 1.90 \\
c14 & 240,000 & 25 & 146 & 527,723 & 20,928 & 36,429 & 564,152 & 0.88 \\
c15 & 480,000 & 25 & 255 & 533,377 & 34,560 & 38,334 & 571,711 & 0.88 \\
c16 & 5,516 & 13 & 255 & 289,615 & 58,304 & 27,786 & 317,401 & 0.46 \\
c17 & 7,045 & 12 & 231 & 270,920 & 19,008 & 42,857 & 313,777 & 0.55 \\
c18 & 103,976 & 14 & 252 & 379,466 & 65,152 & 49,557 & 429,023 & 0.68 \\
c19 & 32,759 & 16 & 219 & 651,207 & 18,944 & 39,613 & 690,820 & 1.06 \\
m1 & 1,200 & 30 & 171 & 607,398 & 66,496 & 57,781 & 665,179 & 1.03 \\
m2 & 46,873 & 44 & 155 & 746,126 & 77,312 & 46,164 & 792,290 & 1.16 \\
m3 & 95,512 & 32 & 249 & 679,295 & 89,920 & 36,594 & 715,889 & 1.01 \\
m4 & 2,217 & 7 & 108 & 177,667 & 32,896 & 36,114 & 213,781 & 0.37 \\
m5 & 3,539 & 37 & 180 & 763,990 & 59,648 & 32,997 & 796,987 & 1.15 \\
m6 & 9,864 & 18 & 168 & 392,311 & 47,552 & 41,299 & 433,610 & 0.68 \\
m7 & 4,088 & 12 & 233 & 244,448 & 29,056 & 49,876 & 294,324 & 0.53 \\
m8 & 36,168 & 17 & 249 & 326,717 & 41,792 & 39,118 & 365,835 & 0.58 \\
m9 & 15,326 & 14 & 248 & 316,952 & 41,088 & 43,408 & 360,360 & 0.59 \\
m10 & 1,600 & 21 & 168 & 454,575 & 42,048 & 35,289 & 489,864 & 0.74 \\
m11 & 304,887 & 12 & 162 & 279,459 & 7,488 & 37,016 & 316,475 & 0.55 \\
n1 & 3,680 & 58 & 179 & 1,155,936 & 69,056 & 38,547 & 1,194,483 & 1.71 \\
n2 & 1,202 & 6 & 114 & 193,787 & 11,968 & 44,510 & 238,297 & 0.45 \\
n3 & 3,918 & 12 & 176 & 329,330 & 17,728 & 41,813 & 371,143 & 0.62 \\
n4 & 1,595 & 128 & 287 & 3,101,626 & 188,224 & 39,977 & 3,141,603 & 4.30 \\
n5 & 17,010 & 82 & 177 & 2,417,208 & 215,680 & 38,353 & 2,455,561 & 3.31 \\
n6 & 6,400 & 17 & 179 & 428,796 & 50,432 & 39,779 & 468,575 & 0.72 \\
n7 & 5,756 & 26 & 180 & 838,625 & 84,736 & 37,900 & 876,525 & 1.24 \\
n8 & 1,253 & 593 & 286 & 13,489,251 & 396,608 & 62,033 & 13,551,284 & 18.71 \\
n9 & 8,794 & 17 & 180 & 427,235 & 28,992 & 40,828 & 468,063 & 0.74 \\
n10 & 10,888 & 17 & 178 & 563,041 & 46,784 & 47,332 & 610,373 & 0.94 \\
n11 & 15,215 & 11 & 168 & 325,158 & 18,240 & 47,565 & 372,723 & 0.64 \\
n12 & 196,045 & 4 & 94 & 141,539 & 23,616 & 42,079 & 183,618 & 0.36 \\
n13 & 7,654 & 5 & 111 & 174,613 & 25,664 & 43,229 & 217,842 & 0.41 \\
n14 & 1,600 & 52 & 179 & 1,325,251 & 125,568 & 52,230 & 1,377,481 & 1.94 \\
n15 & 7,857 & 86 & 118 & 1,809,425 & 165,376 & 44,168 & 1,853,593 & 2.54 \\
n16 & 227,845 & 31 & 178 & 971,465 & 70,528 & 44,433 & 1,015,898 & 1.48 \\
n17 & 11,600 & 10 & 156 & 265,683 & 25,280 & 35,926 & 301,609 & 0.50 \\
n18 & 60,800 & 171 & 288 & 4,180,327 & 361,856 & 39,773 & 4,220,100 & 5.61 \\
n19 & 56,000 & 785 & 180 & 17,447,198 & 505,856 & 46,407 & 17,493,605 & 24.05 \\
\end{longtable}
\par\smallskip
\endgroup
\clearpage

\clearpage
\section{Query Scoring Standard}
\label{app:query_scoring}
\label{app:scoring_standard}
This appendix documents the scoring protocol used by \BenchmarkName{} to compare
query answers on real and synthetic tables. The protocol is query-centric: each
valid query answer is routed to one semantic answer type, scored by one primary
metric, and represented as a similarity in $[0,1]$, where larger is better.
Auxiliary diagnostics, such as key overlap or raw distance values, are reported
for interpretation but are not blended into the primary query score.

\subsection{Reporting and Applicability Rules}
\label{app:scoring_applicability}

\textbf{Scorable query objects.}
The scoring protocol applies to semantic answers rather than raw table rows:
scalar aggregates, grouped aggregates, grouped counts, bounded rates or shares,
and ranked selections whose identity columns are visible in the output. It does
not score \texttt{SELECT *}, raw column projections without aggregation, metadata
queries, write statements, multi-table queries, or non-deterministic queries.
Top-$k$ queries are scorable only when the ranked identity and ordering are
deterministic.

\textbf{Single-primary rule.}
Each valid query has exactly one primary semantic scorer. The query-level score is
the similarity produced by that scorer:
\[
S_q = S_{\mathrm{semantic}}(q) \in [0,1].
\]
The benchmark does not average several semantic metrics inside a single query.
The scorer is selected from template metadata and the semantic answer type, not
from the observed numeric range of the answer. For example, a numeric aggregate
that happens to lie in $[0,1]$ is not treated as a rate unless the query semantics
declare it to be a rate, share, probability, or proportion.

\textbf{Validity and failures.}
A query is invalid, rather than assigned a semantic zero, when execution fails,
columns are missing or unexpected, values have incompatible types, a value is
non-finite, a count is negative or null, a rate falls outside its declared scale,
ranking identities are duplicated, or no scorer can be resolved. Invalid queries
are excluded from means but reported through valid-query counts and invalid-reason
counts, separating failures from valid but poor semantic agreement.

\textbf{Key alignment.}
For keyed answers, scoring uses the union of real and synthetic keys,
$K=K_R\cup K_S$, not only their intersection. A key present on both sides receives
the normal per-key similarity; a key present on only one side receives similarity
$0$ for numeric and rate answers, or probability mass $0$ on the missing side for
count distributions. Duplicate keys are summed only for count/support
distributions; duplicate numeric, rate, or ranking keys are invalid unless the
query contract gives an explicit resolution rule.

\textbf{Nulls and non-finite values.}
For numeric and rate-valued cells, SQL \texttt{NULL} is treated as a semantic
missing value: \texttt{NULL} versus \texttt{NULL} scores $1$, while
\texttt{NULL} versus a finite value scores $0$. NaN and $\pm\infty$ are invalid
inputs, not missing values. Count outputs must be finite, numeric, and
non-negative.

\subsection{Query-Centric Semantic Scores}
\label{app:semantic_query_scoring}

\textbf{Numeric magnitude.}
Scalar numeric aggregates and keyed numeric aggregates are scored by normalized
sMAPE similarity, following the established sMAPE literature
\citep{makridakis1993accuracy,chen2017bounded}. For finite real values $r$ and $s$,
\[
S_{\mathrm{num}}(r,s)=1-\frac{|r-s|}{|r|+|s|}.
\]
If both values are zero, the similarity is $1$; if exactly one value is zero, the
similarity is $0$. Signed values are allowed. For keyed answers, the final score
is the macro-average of per-key similarities over $K$.

\textbf{Count and support distributions.}
Grouped counts and support profiles are first normalized into distributions over
the union key space:
\[
P_R(k)=\frac{c_R(k)}{\sum_j c_R(j)}, \qquad
P_S(k)=\frac{c_S(k)}{\sum_j c_S(j)}.
\]
The primary score is the complement of total variation distance, a standard
probability metric \citep{gibbs2002choosing},
\[
S_{\mathrm{count}} =
1-\frac{1}{2}\sum_{k\in K}|P_R(k)-P_S(k)|.
\]
Thus this scorer evaluates whether support mass is allocated to the same keys,
not whether the absolute total row count is identical. Total mass is retained as a
diagnostic only.

\textbf{Rates, shares, and proportions.}
Rates, probabilities, shares, proportions, and missingness rates have a fixed
bounded scale. For values in $[0,1]$, the score is one minus mean absolute error
\citep{willmott2005mae}:
\[
S_{\mathrm{rate}} =
1-\frac{1}{|K|}\sum_{k\in K}|r_k-s_k|.
\]
For scalar rates, $K$ contains one implicit key. Percent-scale outputs are
converted only when the query metadata explicitly declares that scale.

\textbf{Top-$k$ rankings.}
Ranking queries compare ordered identities, not attached magnitudes. The primary
score is finite rank-biased overlap, introduced for comparing indefinite rankings,
with persistence $p=0.9$ \citep{webber2010rbo}:
\[
S_{\mathrm{rank}}=(1-p)\sum_{d=1}^{D}p^{d-1}A_d+p^D A_D,
\]
where $A_d$ is the overlap fraction between the real and synthetic prefixes of
length $d$, and $D$ is the requested ranking depth or the observed comparison
depth. Identical rankings score $1$ and disjoint non-empty rankings score $0$.
Queries with ambiguous tie-breaking are not used for official ranking scores
unless the SQL or template defines a deterministic order.

\subsection{Aggregation Across Queries}
\label{app:score_aggregation}

Query scores are first computed at the valid query level. Family-, dataset-, and
model-level summaries then average the corresponding valid query scores in the
reported group. Invalid queries are excluded from the numerical mean but are
counted separately, so a high score is interpreted together with its valid-query
coverage. When scorer-type summaries are reported, the benchmark uses macro means
within each semantic scorer type; when an overall query score is reported, it is
the mean over valid query scores in the stated reporting group.

\subsection{Classical Distance-Based Fidelity}
\label{app:classical_distance_fidelity}

The classical distance-based baseline compares the real training split against
the synthetic output at the table-distribution level. The held-out test split is
reserved for future downstream-task studies and is not used in this block.

\textbf{Categorical distributions.}
For categorical columns, the benchmark reports Jensen--Shannon distance and total
variation distance between aligned value distributions. Jensen--Shannon distance
is computed as the square root of Jensen--Shannon divergence with logarithm base
$2$ \citep{lin1991divergence}. Total variation distance is
\[
D_{\mathrm{TV}}(P,Q)=\frac{1}{2}\sum_x |P(x)-Q(x)|.
\]
We use the standard probability-metric interpretation of total variation
\citep{gibbs2002choosing}.

\textbf{Numerical distributions.}
For numerical columns, the benchmark reports Kolmogorov--Smirnov distance and a
normalized Wasserstein distance. Kolmogorov--Smirnov distance is the maximum gap
between empirical cumulative distribution functions over the combined value grid
\citep{massey1951kolmogorov}.
The normalized Wasserstein distance is computed from mean absolute quantile
differences on a fixed quantile grid, normalized by the real-column range with a
standard-deviation fallback, and capped to $[0,1]$. Wasserstein distance is used
in its modern optimal-transport sense \citep{peyre2019computational}.

\textbf{Distance-derived fidelity.}
For paper-facing comparisons, each lower-is-better distance is converted to
higher-is-better form as $1-D$. The overall classical fidelity score is the mean
of available Jensen--Shannon, Kolmogorov--Smirnov, total variation, and
normalized Wasserstein components.

\clearpage
\section{Overall Evaluation Analysis}
\label{app:overall_evaluation_analysis}
\label{app:extended_analysis}

\subsection{Overall Benchmark Summary}
\label{app:overall_benchmark_summary}

\begin{table}[!t]
\centering
{\definecolor{FirstPlace}{HTML}{1397B8}%
\definecolor{SecondPlace}{HTML}{7B45E5}%
\definecolor{ThirdPlace}{HTML}{F28E2B}%
\definecolor{OverallTint}{HTML}{F8F1DA}%
\definecolor{RuleGray}{HTML}{C8CDD3}%
\arrayrulecolor{RuleGray}%
\setlength{\tabcolsep}{4.0pt}%
\renewcommand{\arraystretch}{1.12}%
\normalsize
\resizebox{\textwidth}{!}{%
\begin{tabular}{@{}l c c c c c c c c@{}}
\toprule
\multirow{2}{*}{\textbf{Generator}} & \multicolumn{6}{c}{\textbf{Query Fidelity}} & \multicolumn{2}{c}{\textbf{Cost}} \\
\cmidrule(lr){2-7}\cmidrule(lr){8-9}
 & \cellcolor{OverallTint} Overall $\uparrow$ & Subgroup $\uparrow$ & Conditional $\uparrow$ & Tail / Rarity $\uparrow$ & Missingness $\uparrow$ & Cardinality / Range $\uparrow$ & Train min. $\downarrow$ & Gen. s/1k $\downarrow$ \\
\midrule
REAL & \cellcolor{OverallTint} 1.00$_{\pm 0.00}$ & 1.00$_{\pm 0.00}$ & 1.00$_{\pm 0.00}$ & 1.00$_{\pm 0.00}$ & 1.00$_{\pm 0.00}$ & 1.00$_{\pm 0.00}$ & \textemdash & \textemdash \\
RealTabFormer & \cellcolor{OverallTint} {\color{FirstPlace}\textbf{0.76$_{\pm 0.23}$}} & {\color{FirstPlace}\textbf{0.88$_{\pm 0.17}$}} & {\color{FirstPlace}\textbf{0.75$_{\pm 0.26}$}} & {\color{FirstPlace}\textbf{0.83$_{\pm 0.20}$}} & {\color{FirstPlace}\textbf{0.81$_{\pm 0.26}$}} & {\color{FirstPlace}\textbf{0.74$_{\pm 0.26}$}} & 54.93 & 10.45 \\
ARF & \cellcolor{OverallTint} {\color{SecondPlace}\textbf{0.64$_{\pm 0.23}$}} & {\color{SecondPlace}\textbf{0.85$_{\pm 0.22}$}} & {\color{SecondPlace}\textbf{0.67$_{\pm 0.29}$}} & {\color{SecondPlace}\textbf{0.77$_{\pm 0.21}$}} & {\color{ThirdPlace}\textbf{0.75$_{\pm 0.19}$}} & {\color{SecondPlace}\textbf{0.60$_{\pm 0.27}$}} & 9.22 & 0.84 \\
BayesNet & \cellcolor{OverallTint} {\color{ThirdPlace}\textbf{0.60$_{\pm 0.24}$}} & {\color{ThirdPlace}\textbf{0.80$_{\pm 0.22}$}} & {\color{ThirdPlace}\textbf{0.64$_{\pm 0.27}$}} & {\color{ThirdPlace}\textbf{0.69$_{\pm 0.25}$}} & 0.69$_{\pm 0.20}$ & 0.56$_{\pm 0.28}$ & 1.10 & 1.92 \\
TabDiff & \cellcolor{OverallTint} 0.59$_{\pm 0.20}$ & 0.72$_{\pm 0.27}$ & 0.51$_{\pm 0.29}$ & 0.69$_{\pm 0.24}$ & 0.55$_{\pm 0.26}$ & {\color{ThirdPlace}\textbf{0.58$_{\pm 0.22}$}} & 7.56 & 1.60 \\
TabSyn & \cellcolor{OverallTint} 0.57$_{\pm 0.24}$ & 0.73$_{\pm 0.32}$ & 0.52$_{\pm 0.28}$ & 0.67$_{\pm 0.27}$ & 0.55$_{\pm 0.26}$ & 0.55$_{\pm 0.26}$ & 23.07 & 0.71 \\
TabbyFlow & \cellcolor{OverallTint} 0.57$_{\pm 0.23}$ & 0.66$_{\pm 0.32}$ & 0.49$_{\pm 0.32}$ & 0.67$_{\pm 0.25}$ & 0.57$_{\pm 0.25}$ & 0.54$_{\pm 0.26}$ & 6.67 & 3.19 \\
CTGAN & \cellcolor{OverallTint} 0.55$_{\pm 0.23}$ & 0.66$_{\pm 0.28}$ & 0.52$_{\pm 0.26}$ & 0.62$_{\pm 0.23}$ & {\color{SecondPlace}\textbf{0.75$_{\pm 0.15}$}} & 0.51$_{\pm 0.28}$ & 9.08 & 0.94 \\
TVAE & \cellcolor{OverallTint} 0.54$_{\pm 0.24}$ & 0.65$_{\pm 0.27}$ & 0.49$_{\pm 0.27}$ & 0.60$_{\pm 0.26}$ & 0.65$_{\pm 0.18}$ & 0.50$_{\pm 0.27}$ & 4.12 & 1.89 \\
TabPFGen & \cellcolor{OverallTint} 0.53$_{\pm 0.24}$ & 0.71$_{\pm 0.29}$ & 0.53$_{\pm 0.29}$ & 0.69$_{\pm 0.23}$ & 0.71$_{\pm 0.21}$ & 0.49$_{\pm 0.28}$ & 0.00 & 45.29 \\
TabDDPM & \cellcolor{OverallTint} 0.46$_{\pm 0.27}$ & 0.52$_{\pm 0.36}$ & 0.37$_{\pm 0.31}$ & 0.57$_{\pm 0.31}$ & 0.56$_{\pm 0.34}$ & 0.45$_{\pm 0.28}$ & 2.80 & 6.22 \\
ForestDiffusion & \cellcolor{OverallTint} 0.44$_{\pm 0.23}$ & 0.56$_{\pm 0.31}$ & 0.41$_{\pm 0.29}$ & 0.60$_{\pm 0.28}$ & 0.58$_{\pm 0.22}$ & 0.39$_{\pm 0.24}$ & 7.25 & 0.72 \\
\bottomrule
\end{tabular}%
}
}

\caption{\textbf{Overall benchmark summary across 11 synthetic tabular generative models averaged over 49 datasets.} $\uparrow$ means higher-is-better and $\downarrow$ means lower-is-better. \texttt{REAL} is the self-comparison reference row, and top generative models are highlighted as {\color{FirstPlace}\textbf{First}}, {\color{SecondPlace}\textbf{Second}}, and {\color{ThirdPlace}\textbf{Third}}. The last two columns report median training time in minutes and median generation time in seconds per 1k generated rows.}
\label{tab:benchmark_overall_real_main}
\end{table}

\subsection{Overall Query Score Heatmap}
\label{app:overall_query_score_heatmap}
\begin{center}
\begin{minipage}{\linewidth}
\centering
\scriptsize
\setlength{\tabcolsep}{2.0pt}
\renewcommand{\arraystretch}{0.92}
\adjustbox{max width=\linewidth,max totalheight=0.73\textheight}{
\begin{tabular}{lccccccccccc}
\toprule
Dataset & ARF & Bayes & CTGAN & F-Diff & RTF & T-Flow & T-DDPM & T-Diff & TPF & T-Syn & TVAE \\
\midrule
c1 & \cellcolor[rgb]{0.170,0.538,0.288}{\color{white}0.970} & \cellcolor[rgb]{0.150,0.531,0.289}{\color{white}0.982} & \cellcolor[rgb]{0.381,0.613,0.278}{\color{white}0.844} & \cellcolor[rgb]{0.418,0.626,0.276}{\color{white}0.823} & \cellcolor[rgb]{0.128,0.523,0.290}{\color{white}0.995} & \cellcolor[rgb]{0.263,0.571,0.283}{\color{white}0.915} & \cellcolor[rgb]{0.498,0.655,0.272}{\color{black}0.775} & \cellcolor[rgb]{0.930,0.930,0.930}{\color{black}TF} & \cellcolor[rgb]{0.273,0.575,0.283}{\color{white}0.909} & \cellcolor[rgb]{0.246,0.565,0.284}{\color{white}0.925} & \cellcolor[rgb]{0.317,0.590,0.281}{\color{white}0.883} \\
c2 & \cellcolor[rgb]{0.235,0.561,0.285}{\color{white}0.932} & \cellcolor[rgb]{0.164,0.536,0.288}{\color{white}0.974} & \cellcolor[rgb]{0.344,0.600,0.279}{\color{white}0.867} & \cellcolor[rgb]{0.200,0.548,0.286}{\color{white}0.953} & \cellcolor[rgb]{0.206,0.551,0.286}{\color{white}0.949} & \cellcolor[rgb]{0.331,0.595,0.280}{\color{white}0.874} & \cellcolor[rgb]{0.785,0.279,0.189}{\color{white}0.063} & \cellcolor[rgb]{0.335,0.597,0.280}{\color{white}0.872} & \cellcolor[rgb]{0.500,0.656,0.272}{\color{black}0.774} & \cellcolor[rgb]{0.253,0.568,0.284}{\color{white}0.921} & \cellcolor[rgb]{0.206,0.551,0.286}{\color{white}0.949} \\
c3 & \cellcolor[rgb]{0.946,0.815,0.251}{\color{black}0.508} & \cellcolor[rgb]{0.943,0.768,0.244}{\color{black}0.458} & \cellcolor[rgb]{0.937,0.748,0.242}{\color{black}0.442} & \cellcolor[rgb]{0.858,0.504,0.214}{\color{black}0.245} & \cellcolor[rgb]{0.878,0.565,0.221}{\color{black}0.294} & \cellcolor[rgb]{0.866,0.530,0.217}{\color{black}0.266} & \cellcolor[rgb]{0.867,0.533,0.218}{\color{black}0.268} & \cellcolor[rgb]{0.868,0.534,0.218}{\color{black}0.269} & \cellcolor[rgb]{0.908,0.660,0.232}{\color{black}0.371} & \cellcolor[rgb]{0.877,0.563,0.221}{\color{black}0.292} & \cellcolor[rgb]{0.879,0.568,0.222}{\color{black}0.297} \\
c4 & \cellcolor[rgb]{0.151,0.531,0.289}{\color{white}0.982} & \cellcolor[rgb]{0.226,0.558,0.285}{\color{white}0.937} & \cellcolor[rgb]{0.323,0.592,0.280}{\color{white}0.879} & \cellcolor[rgb]{0.497,0.655,0.272}{\color{black}0.776} & \cellcolor[rgb]{0.141,0.528,0.289}{\color{white}0.987} & \cellcolor[rgb]{0.193,0.546,0.287}{\color{white}0.956} & \cellcolor[rgb]{0.158,0.534,0.288}{\color{white}0.977} & \cellcolor[rgb]{0.170,0.538,0.288}{\color{white}0.970} & \cellcolor[rgb]{0.356,0.604,0.279}{\color{white}0.860} & \cellcolor[rgb]{0.245,0.565,0.284}{\color{white}0.926} & \cellcolor[rgb]{0.177,0.540,0.287}{\color{white}0.966} \\
c5 & \cellcolor[rgb]{0.151,0.531,0.289}{\color{white}0.982} & \cellcolor[rgb]{0.233,0.560,0.285}{\color{white}0.933} & \cellcolor[rgb]{0.252,0.567,0.284}{\color{white}0.921} & \cellcolor[rgb]{0.554,0.675,0.269}{\color{black}0.742} & \cellcolor[rgb]{0.143,0.528,0.289}{\color{white}0.986} & \cellcolor[rgb]{0.227,0.558,0.285}{\color{white}0.936} & \cellcolor[rgb]{0.224,0.557,0.285}{\color{white}0.938} & \cellcolor[rgb]{0.247,0.565,0.284}{\color{white}0.925} & \cellcolor[rgb]{0.211,0.553,0.286}{\color{white}0.946} & \cellcolor[rgb]{0.240,0.563,0.284}{\color{white}0.928} & \cellcolor[rgb]{0.260,0.570,0.283}{\color{white}0.917} \\
c6 & \cellcolor[rgb]{0.650,0.709,0.265}{\color{black}0.685} & \cellcolor[rgb]{0.663,0.714,0.264}{\color{black}0.677} & \cellcolor[rgb]{0.889,0.795,0.253}{\color{black}0.542} & \cellcolor[rgb]{0.939,0.755,0.243}{\color{black}0.448} & \cellcolor[rgb]{0.308,0.587,0.281}{\color{white}0.888} & \cellcolor[rgb]{0.816,0.769,0.257}{\color{black}0.585} & \cellcolor[rgb]{0.927,0.719,0.239}{\color{black}0.419} & \cellcolor[rgb]{0.849,0.780,0.255}{\color{black}0.566} & \cellcolor[rgb]{0.638,0.705,0.265}{\color{black}0.691} & \cellcolor[rgb]{0.794,0.761,0.258}{\color{black}0.599} & \cellcolor[rgb]{0.841,0.777,0.256}{\color{black}0.571} \\
c7 & \cellcolor[rgb]{0.192,0.546,0.287}{\color{white}0.957} & \cellcolor[rgb]{0.169,0.537,0.288}{\color{white}0.971} & \cellcolor[rgb]{0.327,0.594,0.280}{\color{white}0.877} & \cellcolor[rgb]{0.781,0.264,0.187}{\color{white}0.052} & \cellcolor[rgb]{0.137,0.526,0.289}{\color{white}0.990} & \cellcolor[rgb]{0.239,0.563,0.284}{\color{white}0.929} & \cellcolor[rgb]{0.180,0.541,0.287}{\color{white}0.964} & \cellcolor[rgb]{0.250,0.567,0.284}{\color{white}0.922} & \cellcolor[rgb]{0.581,0.685,0.268}{\color{black}0.726} & \cellcolor[rgb]{0.781,0.265,0.187}{\color{white}0.052} & \cellcolor[rgb]{0.347,0.601,0.279}{\color{white}0.865} \\
c8 & \cellcolor[rgb]{0.215,0.554,0.285}{\color{white}0.943} & \cellcolor[rgb]{0.163,0.535,0.288}{\color{white}0.974} & \cellcolor[rgb]{0.318,0.591,0.281}{\color{white}0.882} & \cellcolor[rgb]{0.165,0.536,0.288}{\color{white}0.973} & \cellcolor[rgb]{0.129,0.523,0.290}{\color{white}0.995} & \cellcolor[rgb]{0.244,0.564,0.284}{\color{white}0.926} & \cellcolor[rgb]{0.745,0.743,0.260}{\color{black}0.628} & \cellcolor[rgb]{0.308,0.587,0.281}{\color{white}0.888} & \cellcolor[rgb]{0.131,0.524,0.289}{\color{white}0.994} & \cellcolor[rgb]{0.214,0.554,0.286}{\color{white}0.944} & \cellcolor[rgb]{0.339,0.598,0.280}{\color{white}0.869} \\
c9 & \cellcolor[rgb]{0.870,0.542,0.219}{\color{black}0.276} & \cellcolor[rgb]{0.834,0.429,0.206}{\color{black}0.185} & \cellcolor[rgb]{0.811,0.357,0.198}{\color{white}0.126} & \cellcolor[rgb]{0.800,0.323,0.194}{\color{white}0.099} & \cellcolor[rgb]{0.301,0.584,0.281}{\color{white}0.893} & \cellcolor[rgb]{0.946,0.777,0.245}{\color{black}0.465} & \cellcolor[rgb]{0.816,0.374,0.200}{\color{white}0.141} & \cellcolor[rgb]{0.944,0.769,0.244}{\color{black}0.459} & \cellcolor[rgb]{0.843,0.457,0.209}{\color{black}0.207} & \cellcolor[rgb]{0.934,0.738,0.241}{\color{black}0.434} & \cellcolor[rgb]{0.815,0.369,0.199}{\color{white}0.137} \\
c10 & \cellcolor[rgb]{0.149,0.530,0.289}{\color{white}0.983} & \cellcolor[rgb]{0.310,0.588,0.281}{\color{white}0.887} & \cellcolor[rgb]{0.295,0.582,0.282}{\color{white}0.896} & \cellcolor[rgb]{0.173,0.539,0.287}{\color{white}0.969} & \cellcolor[rgb]{0.120,0.520,0.290}{\color{white}1.000} & \cellcolor[rgb]{0.236,0.561,0.284}{\color{white}0.931} & \cellcolor[rgb]{0.182,0.542,0.287}{\color{white}0.963} & \cellcolor[rgb]{0.389,0.616,0.277}{\color{white}0.840} & \cellcolor[rgb]{0.432,0.632,0.275}{\color{white}0.814} & \cellcolor[rgb]{0.349,0.602,0.279}{\color{white}0.864} & \cellcolor[rgb]{0.694,0.725,0.263}{\color{black}0.658} \\
c11 & \cellcolor[rgb]{0.133,0.525,0.289}{\color{white}0.992} & \cellcolor[rgb]{0.142,0.528,0.289}{\color{white}0.987} & \cellcolor[rgb]{0.249,0.566,0.284}{\color{white}0.923} & \cellcolor[rgb]{0.290,0.581,0.282}{\color{white}0.899} & \cellcolor[rgb]{0.124,0.521,0.290}{\color{white}0.998} & \cellcolor[rgb]{0.150,0.531,0.289}{\color{white}0.982} & \cellcolor[rgb]{0.159,0.534,0.288}{\color{white}0.977} & \cellcolor[rgb]{0.159,0.534,0.288}{\color{white}0.977} & \cellcolor[rgb]{0.293,0.582,0.282}{\color{white}0.897} & \cellcolor[rgb]{0.139,0.527,0.289}{\color{white}0.989} & \cellcolor[rgb]{0.145,0.529,0.289}{\color{white}0.985} \\
c12 & \cellcolor[rgb]{0.429,0.630,0.275}{\color{white}0.816} & \cellcolor[rgb]{0.389,0.616,0.277}{\color{white}0.840} & \cellcolor[rgb]{0.494,0.653,0.272}{\color{black}0.778} & \cellcolor[rgb]{0.799,0.763,0.258}{\color{black}0.596} & \cellcolor[rgb]{0.234,0.561,0.285}{\color{white}0.932} & \cellcolor[rgb]{0.776,0.248,0.185}{\color{white}0.039} & \cellcolor[rgb]{0.930,0.930,0.930}{\color{black}TF} & \cellcolor[rgb]{0.930,0.930,0.930}{\color{black}TF} & \cellcolor[rgb]{0.672,0.717,0.264}{\color{black}0.671} & \cellcolor[rgb]{0.543,0.671,0.270}{\color{black}0.748} & \cellcolor[rgb]{0.470,0.645,0.273}{\color{white}0.792} \\
c13 & \cellcolor[rgb]{0.355,0.604,0.279}{\color{white}0.860} & \cellcolor[rgb]{0.661,0.713,0.264}{\color{black}0.678} & \cellcolor[rgb]{0.947,0.779,0.245}{\color{black}0.467} & \cellcolor[rgb]{0.786,0.758,0.258}{\color{black}0.603} & \cellcolor[rgb]{0.357,0.605,0.279}{\color{white}0.859} & \cellcolor[rgb]{0.572,0.681,0.268}{\color{black}0.731} & \cellcolor[rgb]{0.917,0.805,0.252}{\color{black}0.526} & \cellcolor[rgb]{0.706,0.729,0.262}{\color{black}0.651} & \cellcolor[rgb]{0.792,0.300,0.191}{\color{white}0.081} & \cellcolor[rgb]{0.453,0.639,0.274}{\color{white}0.802} & \cellcolor[rgb]{0.949,0.787,0.246}{\color{black}0.474} \\
c14 & \cellcolor[rgb]{0.280,0.577,0.282}{\color{white}0.905} & \cellcolor[rgb]{0.298,0.584,0.282}{\color{white}0.894} & \cellcolor[rgb]{0.609,0.694,0.267}{\color{black}0.709} & \cellcolor[rgb]{0.916,0.685,0.235}{\color{black}0.391} & \cellcolor[rgb]{0.360,0.606,0.279}{\color{white}0.857} & \cellcolor[rgb]{0.846,0.779,0.255}{\color{black}0.568} & \cellcolor[rgb]{0.822,0.393,0.202}{\color{white}0.156} & \cellcolor[rgb]{0.799,0.762,0.258}{\color{black}0.596} & \cellcolor[rgb]{0.319,0.591,0.281}{\color{white}0.881} & \cellcolor[rgb]{0.850,0.781,0.255}{\color{black}0.566} & \cellcolor[rgb]{0.920,0.695,0.236}{\color{black}0.399} \\
c15 & \cellcolor[rgb]{0.359,0.605,0.279}{\color{white}0.858} & \cellcolor[rgb]{0.501,0.656,0.272}{\color{black}0.773} & \cellcolor[rgb]{0.673,0.717,0.264}{\color{black}0.671} & \cellcolor[rgb]{0.886,0.794,0.254}{\color{black}0.544} & \cellcolor[rgb]{0.245,0.565,0.284}{\color{white}0.926} & \cellcolor[rgb]{0.709,0.730,0.262}{\color{black}0.649} & \cellcolor[rgb]{0.930,0.930,0.930}{\color{black}TF} & \cellcolor[rgb]{0.703,0.728,0.262}{\color{black}0.653} & \cellcolor[rgb]{0.461,0.642,0.274}{\color{white}0.797} & \cellcolor[rgb]{0.701,0.727,0.262}{\color{black}0.654} & \cellcolor[rgb]{0.917,0.805,0.252}{\color{black}0.526} \\
c16 & \cellcolor[rgb]{0.554,0.675,0.269}{\color{black}0.742} & \cellcolor[rgb]{0.854,0.782,0.255}{\color{black}0.563} & \cellcolor[rgb]{0.737,0.740,0.261}{\color{black}0.633} & \cellcolor[rgb]{0.912,0.672,0.233}{\color{black}0.381} & \cellcolor[rgb]{0.473,0.646,0.273}{\color{white}0.790} & \cellcolor[rgb]{0.872,0.548,0.219}{\color{black}0.280} & \cellcolor[rgb]{0.930,0.930,0.930}{\color{black}TF} & \cellcolor[rgb]{0.951,0.791,0.247}{\color{black}0.476} & \cellcolor[rgb]{0.930,0.930,0.930}{\color{black}TF} & \cellcolor[rgb]{0.948,0.781,0.246}{\color{black}0.469} & \cellcolor[rgb]{0.952,0.796,0.247}{\color{black}0.481} \\
c17 & \cellcolor[rgb]{0.678,0.719,0.263}{\color{black}0.668} & \cellcolor[rgb]{0.679,0.720,0.263}{\color{black}0.667} & \cellcolor[rgb]{0.691,0.724,0.263}{\color{black}0.660} & \cellcolor[rgb]{0.854,0.492,0.213}{\color{black}0.235} & \cellcolor[rgb]{0.906,0.654,0.231}{\color{black}0.366} & \cellcolor[rgb]{0.846,0.465,0.210}{\color{black}0.214} & \cellcolor[rgb]{0.799,0.322,0.194}{\color{white}0.098} & \cellcolor[rgb]{0.876,0.560,0.221}{\color{black}0.290} & \cellcolor[rgb]{0.768,0.751,0.259}{\color{black}0.614} & \cellcolor[rgb]{0.873,0.552,0.220}{\color{black}0.284} & \cellcolor[rgb]{0.942,0.814,0.251}{\color{black}0.511} \\
c18 & \cellcolor[rgb]{0.639,0.705,0.265}{\color{black}0.691} & \cellcolor[rgb]{0.947,0.815,0.251}{\color{black}0.508} & \cellcolor[rgb]{0.959,0.817,0.250}{\color{black}0.497} & \cellcolor[rgb]{0.864,0.523,0.216}{\color{black}0.261} & \cellcolor[rgb]{0.445,0.636,0.275}{\color{white}0.806} & \cellcolor[rgb]{0.876,0.560,0.221}{\color{black}0.290} & \cellcolor[rgb]{0.930,0.930,0.930}{\color{black}TF} & \cellcolor[rgb]{0.930,0.930,0.930}{\color{black}TF} & \cellcolor[rgb]{0.930,0.930,0.930}{\color{black}TF} & \cellcolor[rgb]{0.828,0.411,0.204}{\color{white}0.170} & \cellcolor[rgb]{0.943,0.766,0.244}{\color{black}0.456} \\
c19 & \cellcolor[rgb]{0.405,0.622,0.276}{\color{white}0.830} & \cellcolor[rgb]{0.895,0.619,0.227}{\color{black}0.338} & \cellcolor[rgb]{0.956,0.807,0.249}{\color{black}0.490} & \cellcolor[rgb]{0.830,0.416,0.204}{\color{white}0.174} & \cellcolor[rgb]{0.872,0.546,0.219}{\color{black}0.279} & \cellcolor[rgb]{0.825,0.401,0.203}{\color{white}0.162} & \cellcolor[rgb]{0.930,0.930,0.930}{\color{black}TF} & \cellcolor[rgb]{0.831,0.421,0.205}{\color{white}0.178} & \cellcolor[rgb]{0.731,0.738,0.261}{\color{black}0.637} & \cellcolor[rgb]{0.865,0.526,0.217}{\color{black}0.263} & \cellcolor[rgb]{0.924,0.709,0.237}{\color{black}0.410} \\

m1 & \cellcolor[rgb]{0.950,0.789,0.246}{\color{black}0.475} & \cellcolor[rgb]{0.668,0.716,0.264}{\color{black}0.674} & \cellcolor[rgb]{0.857,0.783,0.255}{\color{black}0.561} & \cellcolor[rgb]{0.946,0.777,0.245}{\color{black}0.465} & \cellcolor[rgb]{0.823,0.771,0.257}{\color{black}0.582} & \cellcolor[rgb]{0.812,0.767,0.257}{\color{black}0.588} & \cellcolor[rgb]{0.925,0.710,0.238}{\color{black}0.411} & \cellcolor[rgb]{0.769,0.752,0.259}{\color{black}0.614} & \cellcolor[rgb]{0.958,0.815,0.249}{\color{black}0.496} & \cellcolor[rgb]{0.830,0.774,0.256}{\color{black}0.577} & \cellcolor[rgb]{0.928,0.719,0.239}{\color{black}0.419} \\
m2 & \cellcolor[rgb]{0.537,0.669,0.270}{\color{black}0.752} & \cellcolor[rgb]{0.599,0.691,0.267}{\color{black}0.715} & \cellcolor[rgb]{0.692,0.724,0.263}{\color{black}0.660} & \cellcolor[rgb]{0.951,0.817,0.250}{\color{black}0.506} & \cellcolor[rgb]{0.821,0.770,0.257}{\color{black}0.583} & \cellcolor[rgb]{0.719,0.734,0.261}{\color{black}0.643} & \cellcolor[rgb]{0.944,0.770,0.244}{\color{black}0.460} & \cellcolor[rgb]{0.930,0.930,0.930}{\color{black}TF} & \cellcolor[rgb]{0.586,0.686,0.268}{\color{black}0.723} & \cellcolor[rgb]{0.622,0.699,0.266}{\color{black}0.701} & \cellcolor[rgb]{0.779,0.755,0.259}{\color{black}0.608} \\
m3 & \cellcolor[rgb]{0.556,0.676,0.269}{\color{black}0.740} & \cellcolor[rgb]{0.755,0.747,0.260}{\color{black}0.622} & \cellcolor[rgb]{0.916,0.804,0.252}{\color{black}0.526} & \cellcolor[rgb]{0.936,0.746,0.242}{\color{black}0.441} & \cellcolor[rgb]{0.184,0.543,0.287}{\color{white}0.962} & \cellcolor[rgb]{0.617,0.698,0.266}{\color{black}0.704} & \cellcolor[rgb]{0.930,0.930,0.930}{\color{black}TF} & \cellcolor[rgb]{0.577,0.683,0.268}{\color{black}0.728} & \cellcolor[rgb]{0.808,0.766,0.257}{\color{black}0.590} & \cellcolor[rgb]{0.486,0.651,0.273}{\color{white}0.782} & \cellcolor[rgb]{0.900,0.799,0.253}{\color{black}0.536} \\
m4 & \cellcolor[rgb]{0.679,0.720,0.263}{\color{black}0.667} & \cellcolor[rgb]{0.698,0.726,0.262}{\color{black}0.656} & \cellcolor[rgb]{0.753,0.746,0.260}{\color{black}0.623} & \cellcolor[rgb]{0.952,0.817,0.250}{\color{black}0.504} & \cellcolor[rgb]{0.367,0.608,0.278}{\color{white}0.853} & \cellcolor[rgb]{0.899,0.798,0.253}{\color{black}0.536} & \cellcolor[rgb]{0.855,0.496,0.213}{\color{black}0.239} & \cellcolor[rgb]{0.620,0.699,0.266}{\color{black}0.702} & \cellcolor[rgb]{0.730,0.738,0.261}{\color{black}0.637} & \cellcolor[rgb]{0.571,0.681,0.269}{\color{black}0.731} & \cellcolor[rgb]{0.805,0.765,0.257}{\color{black}0.592} \\
m5 & \cellcolor[rgb]{0.779,0.755,0.259}{\color{black}0.608} & \cellcolor[rgb]{0.957,0.809,0.249}{\color{black}0.491} & \cellcolor[rgb]{0.922,0.702,0.237}{\color{black}0.405} & \cellcolor[rgb]{0.929,0.725,0.239}{\color{black}0.424} & \cellcolor[rgb]{0.269,0.573,0.283}{\color{white}0.911} & \cellcolor[rgb]{0.562,0.678,0.269}{\color{black}0.737} & \cellcolor[rgb]{0.930,0.727,0.239}{\color{black}0.425} & \cellcolor[rgb]{0.678,0.719,0.263}{\color{black}0.668} & \cellcolor[rgb]{0.840,0.777,0.256}{\color{black}0.571} & \cellcolor[rgb]{0.597,0.690,0.267}{\color{black}0.716} & \cellcolor[rgb]{0.904,0.647,0.230}{\color{black}0.360} \\
m6 & \cellcolor[rgb]{0.322,0.592,0.280}{\color{white}0.880} & \cellcolor[rgb]{0.932,0.734,0.240}{\color{black}0.431} & \cellcolor[rgb]{0.891,0.606,0.226}{\color{black}0.327} & \cellcolor[rgb]{0.858,0.505,0.214}{\color{black}0.246} & \cellcolor[rgb]{0.512,0.660,0.271}{\color{black}0.767} & \cellcolor[rgb]{0.768,0.752,0.259}{\color{black}0.614} & \cellcolor[rgb]{0.933,0.735,0.240}{\color{black}0.431} & \cellcolor[rgb]{0.652,0.710,0.265}{\color{black}0.683} & \cellcolor[rgb]{0.945,0.772,0.245}{\color{black}0.461} & \cellcolor[rgb]{0.631,0.702,0.266}{\color{black}0.696} & \cellcolor[rgb]{0.895,0.618,0.227}{\color{black}0.337} \\
m7 & \cellcolor[rgb]{0.707,0.730,0.262}{\color{black}0.650} & \cellcolor[rgb]{0.642,0.707,0.265}{\color{black}0.689} & \cellcolor[rgb]{0.740,0.741,0.260}{\color{black}0.631} & \cellcolor[rgb]{0.708,0.730,0.262}{\color{black}0.650} & \cellcolor[rgb]{0.422,0.628,0.276}{\color{white}0.821} & \cellcolor[rgb]{0.623,0.700,0.266}{\color{black}0.701} & \cellcolor[rgb]{0.665,0.715,0.264}{\color{black}0.676} & \cellcolor[rgb]{0.611,0.695,0.267}{\color{black}0.708} & \cellcolor[rgb]{0.670,0.716,0.264}{\color{black}0.673} & \cellcolor[rgb]{0.607,0.694,0.267}{\color{black}0.710} & \cellcolor[rgb]{0.807,0.765,0.257}{\color{black}0.591} \\
m8 & \cellcolor[rgb]{0.665,0.715,0.264}{\color{black}0.675} & \cellcolor[rgb]{0.835,0.775,0.256}{\color{black}0.574} & \cellcolor[rgb]{0.833,0.775,0.256}{\color{black}0.576} & \cellcolor[rgb]{0.810,0.766,0.257}{\color{black}0.589} & \cellcolor[rgb]{0.200,0.549,0.286}{\color{white}0.952} & \cellcolor[rgb]{0.528,0.666,0.271}{\color{black}0.757} & \cellcolor[rgb]{0.882,0.578,0.223}{\color{black}0.305} & \cellcolor[rgb]{0.881,0.574,0.222}{\color{black}0.301} & \cellcolor[rgb]{0.765,0.750,0.259}{\color{black}0.616} & \cellcolor[rgb]{0.886,0.592,0.224}{\color{black}0.316} & \cellcolor[rgb]{0.838,0.776,0.256}{\color{black}0.573} \\
m9 & \cellcolor[rgb]{0.401,0.620,0.277}{\color{white}0.833} & \cellcolor[rgb]{0.413,0.625,0.276}{\color{white}0.826} & \cellcolor[rgb]{0.449,0.638,0.274}{\color{white}0.804} & \cellcolor[rgb]{0.831,0.774,0.256}{\color{black}0.577} & \cellcolor[rgb]{0.266,0.572,0.283}{\color{white}0.913} & \cellcolor[rgb]{0.813,0.767,0.257}{\color{black}0.588} & \cellcolor[rgb]{0.678,0.719,0.263}{\color{black}0.668} & \cellcolor[rgb]{0.703,0.728,0.262}{\color{black}0.653} & \cellcolor[rgb]{0.569,0.680,0.269}{\color{black}0.733} & \cellcolor[rgb]{0.514,0.661,0.271}{\color{black}0.765} & \cellcolor[rgb]{0.680,0.720,0.263}{\color{black}0.667} \\
m10 & \cellcolor[rgb]{0.875,0.790,0.254}{\color{black}0.551} & \cellcolor[rgb]{0.515,0.661,0.271}{\color{black}0.765} & \cellcolor[rgb]{0.704,0.729,0.262}{\color{black}0.652} & \cellcolor[rgb]{0.781,0.756,0.259}{\color{black}0.606} & \cellcolor[rgb]{0.446,0.636,0.274}{\color{white}0.806} & \cellcolor[rgb]{0.763,0.750,0.259}{\color{black}0.617} & \cellcolor[rgb]{0.897,0.625,0.228}{\color{black}0.342} & \cellcolor[rgb]{0.723,0.735,0.261}{\color{black}0.641} & \cellcolor[rgb]{0.795,0.761,0.258}{\color{black}0.598} & \cellcolor[rgb]{0.707,0.730,0.262}{\color{black}0.651} & \cellcolor[rgb]{0.643,0.707,0.265}{\color{black}0.689} \\
m11 & \cellcolor[rgb]{0.747,0.744,0.260}{\color{black}0.627} & \cellcolor[rgb]{0.955,0.818,0.250}{\color{black}0.503} & \cellcolor[rgb]{0.723,0.735,0.261}{\color{black}0.641} & \cellcolor[rgb]{0.958,0.815,0.249}{\color{black}0.496} & \cellcolor[rgb]{0.295,0.583,0.282}{\color{white}0.896} & \cellcolor[rgb]{0.645,0.708,0.265}{\color{black}0.687} & \cellcolor[rgb]{0.531,0.667,0.270}{\color{black}0.755} & \cellcolor[rgb]{0.448,0.637,0.274}{\color{white}0.805} & \cellcolor[rgb]{0.936,0.811,0.251}{\color{black}0.514} & \cellcolor[rgb]{0.610,0.695,0.267}{\color{black}0.708} & \cellcolor[rgb]{0.830,0.774,0.256}{\color{black}0.577} \\

n1 & \cellcolor[rgb]{0.929,0.725,0.239}{\color{black}0.424} & \cellcolor[rgb]{0.862,0.517,0.216}{\color{black}0.255} & \cellcolor[rgb]{0.798,0.319,0.193}{\color{white}0.096} & \cellcolor[rgb]{0.862,0.516,0.216}{\color{black}0.255} & \cellcolor[rgb]{0.941,0.760,0.243}{\color{black}0.452} & \cellcolor[rgb]{0.905,0.650,0.231}{\color{black}0.363} & \cellcolor[rgb]{0.933,0.737,0.241}{\color{black}0.433} & \cellcolor[rgb]{0.842,0.778,0.256}{\color{black}0.570} & \cellcolor[rgb]{0.804,0.336,0.195}{\color{white}0.110} & \cellcolor[rgb]{0.923,0.704,0.237}{\color{black}0.406} & \cellcolor[rgb]{0.796,0.312,0.193}{\color{white}0.091} \\
n2 & \cellcolor[rgb]{0.957,0.811,0.249}{\color{black}0.493} & \cellcolor[rgb]{0.910,0.664,0.232}{\color{black}0.374} & \cellcolor[rgb]{0.916,0.685,0.235}{\color{black}0.391} & \cellcolor[rgb]{0.893,0.612,0.226}{\color{black}0.332} & \cellcolor[rgb]{0.553,0.675,0.269}{\color{black}0.742} & \cellcolor[rgb]{0.927,0.717,0.238}{\color{black}0.417} & \cellcolor[rgb]{0.887,0.794,0.253}{\color{black}0.544} & \cellcolor[rgb]{0.929,0.724,0.239}{\color{black}0.423} & \cellcolor[rgb]{0.940,0.758,0.243}{\color{black}0.450} & \cellcolor[rgb]{0.951,0.817,0.250}{\color{black}0.505} & \cellcolor[rgb]{0.918,0.691,0.235}{\color{black}0.396} \\
n3 & \cellcolor[rgb]{0.875,0.555,0.220}{\color{black}0.287} & \cellcolor[rgb]{0.878,0.566,0.221}{\color{black}0.296} & \cellcolor[rgb]{0.822,0.391,0.202}{\color{white}0.154} & \cellcolor[rgb]{0.850,0.479,0.212}{\color{black}0.225} & \cellcolor[rgb]{0.402,0.621,0.277}{\color{white}0.832} & \cellcolor[rgb]{0.950,0.788,0.246}{\color{black}0.474} & \cellcolor[rgb]{0.863,0.520,0.216}{\color{black}0.258} & \cellcolor[rgb]{0.957,0.811,0.249}{\color{black}0.493} & \cellcolor[rgb]{0.855,0.494,0.213}{\color{black}0.237} & \cellcolor[rgb]{0.913,0.803,0.252}{\color{black}0.528} & \cellcolor[rgb]{0.823,0.394,0.202}{\color{white}0.156} \\
n4 & \cellcolor[rgb]{0.907,0.656,0.231}{\color{black}0.367} & \cellcolor[rgb]{0.900,0.634,0.229}{\color{black}0.350} & \cellcolor[rgb]{0.869,0.537,0.218}{\color{black}0.272} & \cellcolor[rgb]{0.892,0.608,0.226}{\color{black}0.329} & \cellcolor[rgb]{0.473,0.646,0.273}{\color{white}0.790} & \cellcolor[rgb]{0.916,0.684,0.235}{\color{black}0.391} & \cellcolor[rgb]{0.874,0.554,0.220}{\color{black}0.286} & \cellcolor[rgb]{0.932,0.732,0.240}{\color{black}0.429} & \cellcolor[rgb]{0.907,0.656,0.231}{\color{black}0.368} & \cellcolor[rgb]{0.938,0.750,0.242}{\color{black}0.444} & \cellcolor[rgb]{0.877,0.564,0.221}{\color{black}0.293} \\
n5 & \cellcolor[rgb]{0.943,0.767,0.244}{\color{black}0.457} & \cellcolor[rgb]{0.931,0.730,0.240}{\color{black}0.428} & \cellcolor[rgb]{0.906,0.653,0.231}{\color{black}0.366} & \cellcolor[rgb]{0.912,0.671,0.233}{\color{black}0.380} & \cellcolor[rgb]{0.959,0.820,0.250}{\color{black}0.500} & \cellcolor[rgb]{0.947,0.780,0.245}{\color{black}0.468} & \cellcolor[rgb]{0.819,0.383,0.201}{\color{white}0.148} & \cellcolor[rgb]{0.942,0.763,0.244}{\color{black}0.454} & \cellcolor[rgb]{0.927,0.719,0.239}{\color{black}0.419} & \cellcolor[rgb]{0.924,0.708,0.237}{\color{black}0.410} & \cellcolor[rgb]{0.912,0.672,0.233}{\color{black}0.381} \\
n6 & \cellcolor[rgb]{0.880,0.571,0.222}{\color{black}0.299} & \cellcolor[rgb]{0.740,0.741,0.260}{\color{black}0.631} & \cellcolor[rgb]{0.898,0.629,0.228}{\color{black}0.346} & \cellcolor[rgb]{0.864,0.522,0.216}{\color{black}0.260} & \cellcolor[rgb]{0.441,0.635,0.275}{\color{white}0.809} & \cellcolor[rgb]{0.863,0.785,0.255}{\color{black}0.558} & \cellcolor[rgb]{0.862,0.785,0.255}{\color{black}0.558} & \cellcolor[rgb]{0.896,0.797,0.253}{\color{black}0.538} & \cellcolor[rgb]{0.880,0.573,0.222}{\color{black}0.301} & \cellcolor[rgb]{0.861,0.785,0.255}{\color{black}0.559} & \cellcolor[rgb]{0.537,0.669,0.270}{\color{black}0.752} \\
n7 & \cellcolor[rgb]{0.935,0.811,0.251}{\color{black}0.515} & \cellcolor[rgb]{0.841,0.778,0.256}{\color{black}0.571} & \cellcolor[rgb]{0.925,0.713,0.238}{\color{black}0.414} & \cellcolor[rgb]{0.949,0.787,0.246}{\color{black}0.473} & \cellcolor[rgb]{0.900,0.634,0.229}{\color{black}0.350} & \cellcolor[rgb]{0.882,0.792,0.254}{\color{black}0.547} & \cellcolor[rgb]{0.832,0.423,0.205}{\color{white}0.180} & \cellcolor[rgb]{0.955,0.818,0.250}{\color{black}0.503} & \cellcolor[rgb]{0.953,0.799,0.248}{\color{black}0.483} & \cellcolor[rgb]{0.883,0.793,0.254}{\color{black}0.546} & \cellcolor[rgb]{0.947,0.781,0.246}{\color{black}0.469} \\
n8 & \cellcolor[rgb]{0.932,0.733,0.240}{\color{black}0.430} & \cellcolor[rgb]{0.947,0.780,0.245}{\color{black}0.467} & \cellcolor[rgb]{0.918,0.689,0.235}{\color{black}0.395} & \cellcolor[rgb]{0.911,0.668,0.233}{\color{black}0.378} & \cellcolor[rgb]{0.886,0.794,0.254}{\color{black}0.544} & \cellcolor[rgb]{0.890,0.603,0.226}{\color{black}0.325} & \cellcolor[rgb]{0.930,0.930,0.930}{\color{black}TF} & \cellcolor[rgb]{0.929,0.723,0.239}{\color{black}0.422} & \cellcolor[rgb]{0.875,0.555,0.220}{\color{black}0.286} & \cellcolor[rgb]{0.930,0.726,0.239}{\color{black}0.424} & \cellcolor[rgb]{0.902,0.642,0.230}{\color{black}0.356} \\
n9 & \cellcolor[rgb]{0.902,0.641,0.230}{\color{black}0.356} & \cellcolor[rgb]{0.324,0.593,0.280}{\color{white}0.879} & \cellcolor[rgb]{0.426,0.629,0.275}{\color{white}0.818} & \cellcolor[rgb]{0.897,0.625,0.228}{\color{black}0.343} & \cellcolor[rgb]{0.214,0.553,0.286}{\color{white}0.944} & \cellcolor[rgb]{0.616,0.697,0.266}{\color{black}0.705} & \cellcolor[rgb]{0.581,0.685,0.268}{\color{black}0.726} & \cellcolor[rgb]{0.609,0.694,0.267}{\color{black}0.709} & \cellcolor[rgb]{0.885,0.588,0.224}{\color{black}0.313} & \cellcolor[rgb]{0.600,0.691,0.267}{\color{black}0.714} & \cellcolor[rgb]{0.398,0.619,0.277}{\color{white}0.835} \\
n10 & \cellcolor[rgb]{0.936,0.745,0.242}{\color{black}0.440} & \cellcolor[rgb]{0.896,0.620,0.227}{\color{black}0.339} & \cellcolor[rgb]{0.937,0.748,0.242}{\color{black}0.442} & \cellcolor[rgb]{0.892,0.609,0.226}{\color{black}0.330} & \cellcolor[rgb]{0.890,0.603,0.225}{\color{black}0.325} & \cellcolor[rgb]{0.943,0.767,0.244}{\color{black}0.457} & \cellcolor[rgb]{0.873,0.551,0.220}{\color{black}0.283} & \cellcolor[rgb]{0.945,0.774,0.245}{\color{black}0.463} & \cellcolor[rgb]{0.942,0.763,0.244}{\color{black}0.454} & \cellcolor[rgb]{0.918,0.688,0.235}{\color{black}0.394} & \cellcolor[rgb]{0.936,0.745,0.242}{\color{black}0.439} \\
n11 & \cellcolor[rgb]{0.922,0.703,0.237}{\color{black}0.405} & \cellcolor[rgb]{0.931,0.730,0.240}{\color{black}0.427} & \cellcolor[rgb]{0.860,0.510,0.215}{\color{black}0.250} & \cellcolor[rgb]{0.915,0.680,0.234}{\color{black}0.387} & \cellcolor[rgb]{0.929,0.809,0.251}{\color{black}0.518} & \cellcolor[rgb]{0.916,0.685,0.235}{\color{black}0.391} & \cellcolor[rgb]{0.909,0.661,0.232}{\color{black}0.372} & \cellcolor[rgb]{0.922,0.701,0.237}{\color{black}0.404} & \cellcolor[rgb]{0.915,0.681,0.234}{\color{black}0.388} & \cellcolor[rgb]{0.925,0.713,0.238}{\color{black}0.413} & \cellcolor[rgb]{0.876,0.559,0.221}{\color{black}0.290} \\
n12 & \cellcolor[rgb]{0.414,0.625,0.276}{\color{white}0.825} & \cellcolor[rgb]{0.909,0.661,0.232}{\color{black}0.372} & \cellcolor[rgb]{0.382,0.614,0.278}{\color{white}0.844} & \cellcolor[rgb]{0.872,0.546,0.219}{\color{black}0.279} & \cellcolor[rgb]{0.194,0.546,0.286}{\color{white}0.956} & \cellcolor[rgb]{0.824,0.771,0.256}{\color{black}0.581} & \cellcolor[rgb]{0.746,0.744,0.260}{\color{black}0.627} & \cellcolor[rgb]{0.810,0.766,0.257}{\color{black}0.589} & \cellcolor[rgb]{0.895,0.618,0.227}{\color{black}0.337} & \cellcolor[rgb]{0.495,0.654,0.272}{\color{black}0.777} & \cellcolor[rgb]{0.353,0.603,0.279}{\color{white}0.861} \\
n13 & \cellcolor[rgb]{0.854,0.491,0.213}{\color{black}0.234} & \cellcolor[rgb]{0.871,0.543,0.219}{\color{black}0.276} & \cellcolor[rgb]{0.824,0.399,0.202}{\color{white}0.160} & \cellcolor[rgb]{0.819,0.383,0.201}{\color{white}0.148} & \cellcolor[rgb]{0.930,0.930,0.930}{\color{black}TF} & \cellcolor[rgb]{0.869,0.537,0.218}{\color{black}0.272} & \cellcolor[rgb]{0.792,0.299,0.191}{\color{white}0.079} & \cellcolor[rgb]{0.930,0.930,0.930}{\color{black}TF} & \cellcolor[rgb]{0.862,0.515,0.216}{\color{black}0.254} & \cellcolor[rgb]{0.841,0.451,0.208}{\color{black}0.202} & \cellcolor[rgb]{0.824,0.397,0.202}{\color{white}0.159} \\
n14 & \cellcolor[rgb]{0.901,0.638,0.229}{\color{black}0.353} & \cellcolor[rgb]{0.910,0.666,0.233}{\color{black}0.376} & \cellcolor[rgb]{0.867,0.533,0.218}{\color{black}0.268} & \cellcolor[rgb]{0.870,0.542,0.219}{\color{black}0.276} & \cellcolor[rgb]{0.913,0.673,0.233}{\color{black}0.381} & \cellcolor[rgb]{0.930,0.930,0.930}{\color{black}TF} & \cellcolor[rgb]{0.867,0.532,0.217}{\color{black}0.267} & \cellcolor[rgb]{0.931,0.731,0.240}{\color{black}0.428} & \cellcolor[rgb]{0.882,0.578,0.223}{\color{black}0.305} & \cellcolor[rgb]{0.831,0.419,0.205}{\color{white}0.177} & \cellcolor[rgb]{0.861,0.512,0.215}{\color{black}0.252} \\
n15 & \cellcolor[rgb]{0.633,0.703,0.266}{\color{black}0.694} & \cellcolor[rgb]{0.346,0.601,0.279}{\color{white}0.865} & \cellcolor[rgb]{0.914,0.678,0.234}{\color{black}0.385} & \cellcolor[rgb]{0.955,0.806,0.248}{\color{black}0.489} & \cellcolor[rgb]{0.361,0.606,0.279}{\color{white}0.857} & \cellcolor[rgb]{0.833,0.775,0.256}{\color{black}0.576} & \cellcolor[rgb]{0.910,0.666,0.233}{\color{black}0.376} & \cellcolor[rgb]{0.552,0.674,0.269}{\color{black}0.743} & \cellcolor[rgb]{0.860,0.784,0.255}{\color{black}0.559} & \cellcolor[rgb]{0.574,0.682,0.268}{\color{black}0.730} & \cellcolor[rgb]{0.643,0.707,0.265}{\color{black}0.689} \\
n16 & \cellcolor[rgb]{0.872,0.548,0.219}{\color{black}0.281} & \cellcolor[rgb]{0.849,0.477,0.211}{\color{black}0.224} & \cellcolor[rgb]{0.867,0.531,0.217}{\color{black}0.267} & \cellcolor[rgb]{0.829,0.415,0.204}{\color{white}0.174} & \cellcolor[rgb]{0.844,0.461,0.209}{\color{black}0.211} & \cellcolor[rgb]{0.870,0.540,0.218}{\color{black}0.274} & \cellcolor[rgb]{0.826,0.405,0.203}{\color{white}0.166} & \cellcolor[rgb]{0.866,0.530,0.217}{\color{black}0.266} & \cellcolor[rgb]{0.868,0.534,0.218}{\color{black}0.269} & \cellcolor[rgb]{0.867,0.532,0.218}{\color{black}0.268} & \cellcolor[rgb]{0.847,0.469,0.210}{\color{black}0.217} \\
n17 & \cellcolor[rgb]{0.930,0.728,0.240}{\color{black}0.426} & \cellcolor[rgb]{0.745,0.743,0.260}{\color{black}0.628} & \cellcolor[rgb]{0.837,0.776,0.256}{\color{black}0.573} & \cellcolor[rgb]{0.853,0.488,0.213}{\color{black}0.232} & \cellcolor[rgb]{0.704,0.728,0.262}{\color{black}0.653} & \cellcolor[rgb]{0.930,0.930,0.930}{\color{black}TF} & \cellcolor[rgb]{0.804,0.764,0.257}{\color{black}0.593} & \cellcolor[rgb]{0.765,0.750,0.259}{\color{black}0.616} & \cellcolor[rgb]{0.852,0.486,0.212}{\color{black}0.231} & \cellcolor[rgb]{0.958,0.813,0.249}{\color{black}0.494} & \cellcolor[rgb]{0.843,0.778,0.256}{\color{black}0.570} \\
n18 & \cellcolor[rgb]{0.918,0.689,0.235}{\color{black}0.394} & \cellcolor[rgb]{0.883,0.581,0.223}{\color{black}0.307} & \cellcolor[rgb]{0.921,0.699,0.236}{\color{black}0.402} & \cellcolor[rgb]{0.886,0.591,0.224}{\color{black}0.315} & \cellcolor[rgb]{0.126,0.522,0.290}{\color{white}0.996} & \cellcolor[rgb]{0.935,0.742,0.241}{\color{black}0.437} & \cellcolor[rgb]{0.930,0.930,0.930}{\color{black}TF} & \cellcolor[rgb]{0.935,0.741,0.241}{\color{black}0.436} & \cellcolor[rgb]{0.874,0.553,0.220}{\color{black}0.284} & \cellcolor[rgb]{0.928,0.721,0.239}{\color{black}0.420} & \cellcolor[rgb]{0.889,0.599,0.225}{\color{black}0.322} \\
n19 & \cellcolor[rgb]{0.934,0.740,0.241}{\color{black}0.435} & \cellcolor[rgb]{0.923,0.705,0.237}{\color{black}0.407} & \cellcolor[rgb]{0.807,0.765,0.257}{\color{black}0.591} & \cellcolor[rgb]{0.929,0.724,0.239}{\color{black}0.422} & \cellcolor[rgb]{0.607,0.694,0.267}{\color{black}0.710} & \cellcolor[rgb]{0.951,0.792,0.247}{\color{black}0.477} & \cellcolor[rgb]{0.917,0.688,0.235}{\color{black}0.394} & \cellcolor[rgb]{0.732,0.739,0.261}{\color{black}0.636} & \cellcolor[rgb]{0.891,0.605,0.226}{\color{black}0.326} & \cellcolor[rgb]{0.930,0.930,0.930}{\color{black}TF} & \cellcolor[rgb]{0.648,0.709,0.265}{\color{black}0.685} \\
\bottomrule
\end{tabular}
}

\captionof{table}{Dataset-by-generative-model overall query score heatmap for the v11.2.2\_glm-5.3 workload. TF indicates technical failure due to unsupported generation or prohibitively long runtime.}
\label{tab:overall-query-dataset-model-heatmap}
\end{minipage}
\end{center}

\clearpage
\section{Extended Experiments}
\label{app:extended_experiments}

\subsection{Stability Across Query-Workload Rebuilds}
\label{app:additional_experiments}
This appendix branch is reserved for additional analysis artifacts that do not fit into the main benchmark summary, including comparisons between benchmark-derived queries and \BenchmarkName{} queries.

\begin{table}[!h]
\centering
{\small
\setlength{\tabcolsep}{3.4pt}
\renewcommand{\arraystretch}{1.08}
\begin{adjustbox}{max width=\textwidth}
\begin{tabular}{@{}lcccccccccc@{}}
\toprule
Generator & Haiku-4.5 & GLM-5.3F-1 & GLM-5.3F-2 & GLM-5.3F-3 & GLM-5.3-1 & GLM-5.3-2 & GLM-5.3-3 & Mean & SD & Range \\
\midrule
RealTabFormer & 0.779 & 0.777 & 0.780 & 0.766 & 0.766 & 0.773 & 0.761 & 0.772 & 0.008 & 0.019 \\
ARF & 0.653 & 0.657 & 0.669 & 0.664 & 0.653 & 0.662 & 0.649 & 0.658 & 0.007 & 0.020 \\
BayesNet & 0.615 & 0.625 & 0.625 & 0.615 & 0.618 & 0.623 & 0.610 & 0.619 & 0.006 & 0.016 \\
TabDiff & 0.596 & 0.596 & 0.614 & 0.607 & 0.593 & 0.606 & 0.591 & 0.600 & 0.009 & 0.023 \\
TabSyn & 0.569 & 0.579 & 0.593 & 0.584 & 0.582 & 0.584 & 0.577 & 0.581 & 0.007 & 0.024 \\
TabbyFlow & 0.581 & 0.582 & 0.589 & 0.579 & 0.573 & 0.578 & 0.576 & 0.580 & 0.005 & 0.016 \\
CTGAN & 0.546 & 0.561 & 0.558 & 0.544 & 0.543 & 0.561 & 0.549 & 0.552 & 0.008 & 0.018 \\
TabPFGen & 0.544 & 0.549 & 0.539 & 0.531 & 0.542 & 0.544 & 0.525 & 0.539 & 0.008 & 0.024 \\
TVAE & 0.541 & 0.561 & 0.555 & 0.541 & 0.544 & 0.552 & 0.541 & 0.548 & 0.008 & 0.020 \\
TabDDPM & 0.438 & 0.467 & 0.477 & 0.478 & 0.467 & 0.475 & 0.469 & 0.467 & 0.014 & 0.041 \\
ForestDiffusion & 0.451 & 0.461 & 0.464 & 0.455 & 0.460 & 0.459 & 0.448 & 0.457 & 0.006 & 0.016 \\
\bottomrule
\end{tabular}
\end{adjustbox}
}

\caption{\textbf{Query-score stability across six AI workload builders.} Scores are averaged over the 49-dataset benchmark set for each generator, excluding \texttt{dataevolve}. Haiku-4.5 denotes \texttt{v11.1.4}; Flash-1/2/3 denote \texttt{v11.2.1}, \texttt{v11.2.1-2}, and \texttt{v11.2.1-3}; GLM-1/2/3 denote \texttt{v11.2.2}, \texttt{v11.2.2-2}, and \texttt{v11.2.2-3}.}
\label{tab:appendix_query_rank_stability_v11_ai_model_scores}
\end{table}

\subsection{Structural Coverage of the Full Generated Workload}
\label{app:structural_query_coverage}

\subsubsection{Purpose and comparison universes}
\label{app:tsed_protocol}

The query-grounding pipeline is designed to produce executable, schema-aware
queries, but executable queries can still cover only a narrow part of the
structural space found in public analytical workloads. We therefore perform a
separate workload-level audit. This audit evaluates the query workload itself,
rather than synthetic-table fidelity or agreement between query answers.

The source corpus contains $334$ queries from twelve public sources. Of these,
$320$ parse successfully and yield $286$ unique normalized Detail-Level shapes.
For the main result, we use the complete \texttt{v11.2.2\_glm-5.3} generated
workload: all $2{,}763$ valid generated SQL queries and its $44$ unique
normalized shapes, including queries without an explicit public-source
provenance mapping. This is the result reported as the full generated
workload.

For comparison with the earlier source-grounded protocol, we also retain a
provenance-matched subset. This subset contains $229$ generated queries and
$25$ unique shapes whose template provenance resolves to one of the public
source workloads \texttt{S01}--\texttt{S12}. The remaining $2{,}534$ valid
generated queries are not invalid or unparseable; they are simply outside that
provenance filter. The matched subset is therefore reported only as a
sensitivity analysis, not as the complete generated workload.

\begin{table}[!t]
\centering
\small
\renewcommand{\arraystretch}{1.12}
\setlength{\tabcolsep}{3.5pt}
\begin{tabular}{lrrrr}
\toprule
Comparison universe & Generated SQL & Generated shapes &
Source $\to$ Generated & Generated $\to$ Source \\
\midrule
Full generated workload & 2,763 & 44 & 0.478 & 0.702 \\
\bottomrule
\end{tabular}

\caption{Full-workload structural coverage.
The main result retains every valid generated SQL query from
\texttt{v11.2.2\_glm-5.3}. 
Scores are computed over unique normalized Detail-Level shapes.}
\label{tab:appendix_tsed_workload_summary}
\end{table}

\subsubsection{Detail-Level normalization and TSED construction}
\label{app:tsed_tree_construction}

Before computing TSED, each SQL query is converted to a schema-masked
Detail-Level skeleton. Dataset names, column names, and other
schema-specific identifiers are replaced by placeholders, while SQL
operators, clause roles, aggregate functions, nesting, and ordering structure
are retained. Identical skeleton strings are deduplicated before the
frequency-independent comparison, so each unique shape receives one equal
weight. The audit therefore compares $286$ source shapes against $44$
generated shapes, rather than comparing all $320\times 2{,}763$ SQL pairs.

Each skeleton is parsed into an ordered labeled tree. The primary parser is
SQLGlot~\citep{sqlglot}. Every SQL expression becomes a node labeled by its expression class,
such as \texttt{SELECT}, \texttt{WHERE}, \texttt{GROUP}, \texttt{SUM}, or
\texttt{WINDOW}. Each populated expression argument is represented by an
explicit role node, such as \texttt{ARG:WHERE} or \texttt{ARG:GROUP}, and
child expressions preserve the original SQL tree order. Lists of expressions
preserve their order as multiple children. Scalar values are represented by
typed labels rather than raw database-specific values. If SQLGlot cannot parse
a dialect-specific skeleton, the implementation uses a deterministic
parenthesis-aware token tree as a conservative fallback and records the
backend used for that pair.

\subsubsection{TSED computation}
\label{app:tsed_computation}

For a source tree $T_s$ and a generated tree $T_g$, we compute the ordered
tree-edit distance with APTED~\citep{song-etal-2024-revisiting}. The allowed
operations are node deletion, node insertion, and node rename. All three
operations have fixed unit cost:
\[
c_{\mathrm{delete}}=c_{\mathrm{insert}}=c_{\mathrm{rename}}=1.
\]
The costs are not tuned on the benchmark data. Let
$d_{\mathrm{APTED}}(T_s,T_g)$ be the minimum total cost over all valid edit
sequences, and let $n_s$ and $n_g$ be the numbers of nodes in the two trees.
The normalized edit distance and TSED similarity are:
\[
d_{\mathrm{norm}}(T_s,T_g)
 = \min\left(1,\frac{d_{\mathrm{APTED}}(T_s,T_g)}
 {\max(n_s,n_g)}\right),
\qquad
\mathrm{TSED}(T_s,T_g)
 = \max\left(0,1-d_{\mathrm{norm}}(T_s,T_g)\right).
\]
Thus, identical normalized skeletons have TSED $=1$, and increasingly
different ordered SQL structures receive values closer to $0$. The
normalization by the larger tree size makes scores comparable for trees of
different lengths. TSED measures structural relatedness of the normalized
SQL trees; it does not establish semantic equivalence, equality of query
answers, or equality of database execution plans.

\textbf{Exact computation and caching.}
The full cross-product contains $286\times44=12{,}584$ possible shape pairs.
We parse each source and generated skeleton once and persist every evaluated
pair in a resumable SQLite cache. To reduce runtime without changing the
answer, we use the node-count bound
\[
\mathrm{TSED}(T_s,T_g)
\leq \frac{\min(n_s,n_g)}{\max(n_s,n_g)}.
\]
This follows because at least $|n_s-n_g|$ insertions or deletions are
necessary when the tree sizes differ. A pair is skipped only when this upper
bound cannot improve the current best match for either direction. The
remaining pairs are evaluated with the exact APTED computation. In the full
run, $3{,}225$ pair distances were computed exactly, and the other were
skipped by this bound; no approximate distance was substituted for a skipped
pair.

\subsubsection{Directional aggregation and results}
\label{app:tsed_aggregation}

Let $S$ be the set of unique source shapes and $G$ the set of unique
generated shapes. We report two directional nearest-neighbor summaries:
\[
\mathrm{S\mathord{\to}G}
 = \frac{1}{|S|}\sum_{s\in S}\max_{g\in G}\mathrm{TSED}(s,g),
\qquad
\mathrm{G\mathord{\to}S}
 = \frac{1}{|G|}\sum_{g\in G}\max_{s\in S}\mathrm{TSED}(g,s).
\]
The first score measures source-space coverage: for every source shape, a
closest generated shape is selected. The second measures containment: for
every generated shape, a closest source shape is selected. We summarize them
with the descriptive bidirectional TSED-Chamfer score
\[
\mathrm{Chamfer}_{\mathrm{TSED}}
 = \frac{1}{2}\left(\mathrm{S\mathord{\to}G}
 + \mathrm{G\mathord{\to}S}\right).
\]
This symmetric summary is not merged with query-centric fidelity because the
two experiments evaluate different objects.

\textbf{Interpretation and limitations.}
For the full generated workload, Source $\to$ Generated is $0.478$ and
Generated $\to$ Source is $0.702$. At threshold $\mathrm{TSED}\geq0.5$,
$96/286$ source shapes have a generated counterpart, while $45/49$ generated
shapes have a source counterpart. The asymmetry indicates that the generated
workload is mostly inside the structural envelope represented by the sources,
but it covers only part of the broader source space. The largest coverage gap
comes from long TPC-DS queries with common-table expressions, nested
subqueries, window functions, and set operations that are not well represented
by the current single-table template inventory.

The result is frequency-independent after shape deduplication: a shape
appearing many times does not receive more weight in the directional means.
This makes the audit complementary to frequency-sensitive workload metrics.
It also means that the score should not be interpreted as a fidelity score for
the synthetic tables themselves. Documentation sources with one or two
parseable queries, incomplete SQL snippets, and source-side parse failures are
reported in the accompanying audit files rather than silently converted to
zero similarity. The complete directional matrices, pair cache, parse audit,
and both comparison universes are frozen in the corresponding
\texttt{Bench2Tem} experiment bundle.

\clearpage
\section{Discussion and Future Works}
\label{app:discussion_future}
\label{sec:discussion-limitations}

\subsection{Discussion and Limitation}
\textbf{Query-centric fidelity and distance-based fidelity.}
Query-centric fidelity is not intended to replace distance-based fidelity. The two views measure different aspects of synthetic data quality. Distance-based metrics summarize global resemblance over columns or distributions, while query-centric fidelity asks whether the synthetic table preserves the answers to analytical operations that downstream users are likely to run. Our results show why both views are needed. This makes query-centric fidelity a complementary evaluation layer rather than a substitute for statistical similarity. In practice, the right evaluation depends on the downstream use case. Users who need mock data for coarse schema exploration may care more about global resemblance, while users who need data system testing or analytics on synthetic data may need stronger query-centric fidelity.

\textbf{Toward query-aware synthetic-data generation.}
\BenchmarkName{} also suggests a model-development direction: future tabular generative models can use query families as training, validation, or model-selection signals. The family-level breakdown identifies where current models lose analytical structure, rather than only reporting an aggregate query score. For example, high-cardinality results point to the need for better discrete support preservation. Tail and rarity results point to the need for better rare-region generation. Local conditional results point to the need for models that preserve filtered subpopulation structure. These targets are concrete because they correspond to executable SQL queries and measurable query answers. A query-aware generative model could therefore optimize not only global distributional resemblance, but also take these analytical queries as part of the loss function or reward signals so that the generated table can support these operations.

\textbf{Multi-table query-centric evaluation.}
The current benchmark is designed to support the single-table setting, where reusable analytical query patterns can be grounded across heterogeneous schemas through schema-aware realization. A natural next step is multi-table evaluation. In that setting, synthetic data must preserve not only column distributions and single-table query answers, but also joins, foreign-key structure, join selectivity, fanout behavior, and query answers across related tables. Multi-table query-centric fidelity would extend the same principle used in \BenchmarkName{}: synthetic data should preserve the analytical operations that downstream users are likely to run. This extension would be especially important for analytical questions which cross table boundaries.

\textbf{Privacy vs. fidelity.}
Query-centric fidelity also has a natural connection to privacy evaluation, where synthetic-data releases are often judged through privacy-utility tradeoffs, differential-privacy mechanisms, or threat-specific leakage analyses~\citep{tao2021dpbenchmark,hollig2025utilityprivacy,pdpc2024synthetic,tran2025quantifying}. The same structural assessors that measure whether a synthetic table preserves analytical behavior can reveal how much rare support, local subgroup structure, or high-cardinality information survives generation. Query-centric fidelity is not a formal privacy metric, and it is not a substitute for membership-inference, attribute-inference, or differential-privacy analyses. Instead, it highlights a fidelity-privacy tension that users must interpret for their specific settings. In particular, privacy risk is often contextual: some releases may tolerate accurate aggregates, while others may treat specific query outputs as sensitive information. When query outputs are the objects users want to protect, query-centric evaluation provides a useful diagnostic because the same query structure can be used to measure how much protected signal leaks through the synthetic table. A synthetic table that preserves rare values or high-cardinality keys may be more useful for analytics, but such preservation should be read together with privacy reports and threat models.

\textbf{LLM-assisted SQL grounding.}
\BenchmarkName{} uses an LLM only inside the constrained template-to-SQL realization step: the profiling, binding, and validation are deterministic, and the validation rejects invalid SQL, incompatible realizations, and queries that fail execution. Our stability study suggests that the resulting conclusions are broadly stable across regenerated query sets, but LLM-assisted SQL grounding remains a source of benchmark-instrument variance. The current study reruns the same grounding pipeline, but it does not yet compare multiple SQL realization models or multiple prompting policies. Future work could audit SQL grounding across different LLMs, compare family-level and model-level rankings under these alternative realizations, and include additional human or rule-based checks for ambiguous schema bindings. 

\vspace{-0.5em}

\end{document}